\documentclass[12pt]{report}
\usepackage{hyperref}
\usepackage{natbib}
\usepackage[section]{placeins}
\makeatletter
\AtBeginDocument{%
  \expandafter\renewcommand\expandafter\subsection\expandafter{%
    \expandafter\@fb@secFB\subsection
  }%
}
\makeatother
\usepackage{booktabs}
\usepackage{tablefootnote}
\usepackage{footnote}
\usepackage{footmisc}
\usepackage{float,lscape}
\usepackage{multirow}
\usepackage{url}
\usepackage[utf8]{inputenc}
\DeclareUnicodeCharacter{03B8}{\ensuremath{\theta}}
\usepackage{physics}
\usepackage[english]{babel}
\usepackage{array}
\usepackage{amsmath}
\usepackage{simplewick}
\usepackage{changepage}  
\usepackage{lipsum}
\usepackage{graphicx}
\usepackage{fancyhdr}
\usepackage{graphicx} 
\usepackage{placeins}
\usepackage{dsfont}
\usepackage{float}
\usepackage{caption}
\usepackage{subcaption}
\usepackage{amssymb,cancel}
\usepackage{mathtools}
\usepackage{mathrsfs}  
\usepackage{amsfonts}
\usepackage{color}
\usepackage{bm}
\newcommand{\uvec}[1]{\mathbf{\hat{\textbf{#1}}}}
\usepackage{array}
\newcolumntype{P}[1]{>{\centering\arraybackslash}p{#1}}
\usepackage[margin=1in]{geometry}
\usepackage{tikz}
\usepackage{calrsfs}
\usepackage{geometry}
\setcitestyle{notesep={; }}
\usetikzlibrary{arrows,automata,positioning,shapes.geometric}

\makeatletter

\newcommand{\Rmnum}[1]{\expandafter\@slowromancap\romannumeral #1@}
\makeatother
\newlength{\offsetpage}
{\end{adjustwidth}}

\begin{document}
\begin{titlepage}
\begin{center}
\begin{figure}[H]
\centering
\includegraphics[scale=0.1]{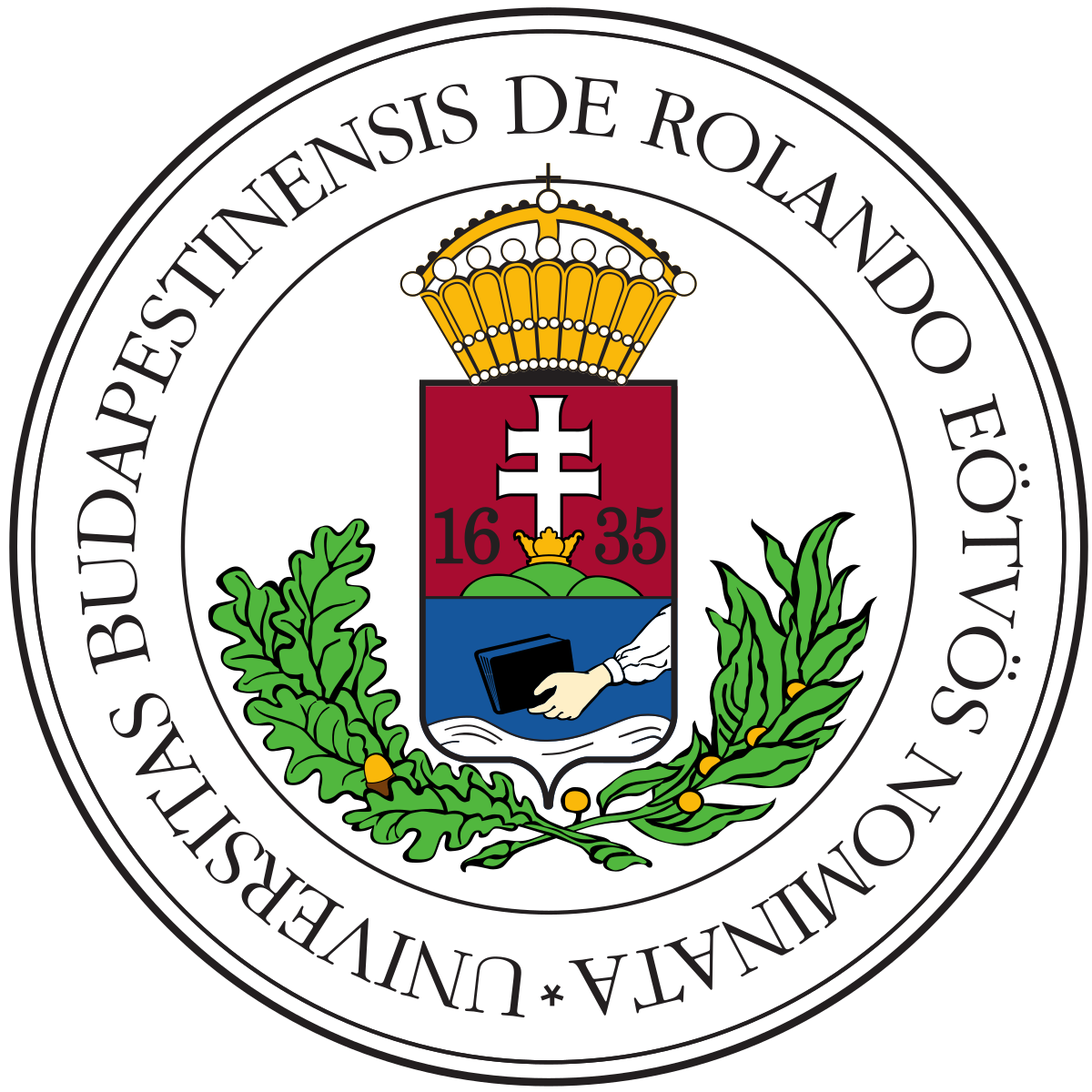}
\end{figure}
\textbf{{\large 
Eötvös Loránd University\\
Faculty of Science\\
Institute of Physics and Astronomy
}}

\vspace{1in}
\line(1,0){500}\\
\huge{\bfseries Propagation of
Interplanetary Shocks 
in the Heliosphere}\\
\line(1,0){500}\\
\textbf{\large MSc in Physics Thesis}\\
\vspace{0.2in}

\centering
\begin{tabular}{c}
    \textbf{\Large Munkhjargal LKHAGVADORJ}\\
\end{tabular}
\vspace{0.2in}

\centering
\begin{tabular}{r l}
    \textbf{\small Supervisor:} & \textbf{\large Gábor FACSKÓ, PhD}\\
    & {\small Department of Space Physics and Space Technology }\\
    & {\small Wigner Research Centre for Physics}\\
    \vspace{0.05in} & \\
    \textbf{\small Co-supervisor:} & \textbf{\large Krisztina Éva GABÁNYI, PhD}\\
    & {\small Department of Astronomy, Eötvös Loránd University}\\
\end{tabular}

\vfill 

\centering
\textbf{\small Budapest, May 30, 2023}
\end{center}
\end{titlepage}

\renewcommand{\abstractname}{Acknowledgements}
\begin{abstract}
 I would like to begin by expressing my deepest gratitude to my supervisor, Dr. Gábor FACSKÓ. His unwavering support, insightful guidance, and constant encouragement throughout my research journey have been invaluable. I am deeply appreciative of his mentorship.
 
 I am also grateful to Dr. Krisztina Éva Gabányi for being my internal supervisor and for all her valuable suggestions, advice, and support.
 
 My sincere thanks also go to Dr. Andrea OPITZ. Her professional advice and thought-provoking suggestions significantly enhanced the quality of my thesis work. 

 My appreciation also goes out to Dr. Péter KOVÁCS. His insight and deep knowledge of the field helped me immensely. 
 
My appreciation extends to the Department of Space Physics and Space Technology at the Wigner Research Centre for Physics. Their support and acceptance of me as a Wigner trainee have provided an enriching environment in which I could grow as a researcher. Being a part of this dynamic team has been an honor.

 I would like to thank all my professors at the Institute of Physics, Eötvös Loránd University, for imparting the knowledge and skills that have been instrumental in my academic journey. Their commitment to my education is something I hold in high regard.
 
 I would also like to acknowledge NASA CDAWeb, the STEREO$-$A and B IMPACT, PLASTIC, the Wind MFI, SWE, ACE MAG, SWEPAM, and the Cluster FGM, CIS, EFW teams for providing data, comprehensive Heliospheric Shock Database (ipshocks.fi) developed and hosted at University of Helsinki for the main data selection process, and TREPS, designed and developed by the French Plasma Physics Data Centre (CDPP), for coordinate transformations for this study.    
 
I am immensely grateful for the Stipendium Hungaricum scholarship. The opportunities it presented have played a pivotal role in my academic and personal development. The experiences I gained through this scholarship have been transformative, and for that, I am deeply thankful.

 Finally, I am indebted to extend my boundless appreciation
 towards my parents, for their love, unwavering support, and blessings have been instrumental in shaping my journey. I pay homage to my father, Lkhagvadorj ZEVEG, who, despite no longer being with us in this world, continues to inspire and guide me with cherished memories. His legacy remains a beacon in my life. My profound appreciation is directed towards my mother, Onon SANTUNDEV, whose ceaseless support, love, and warm wishes have been my constant source of strength. 
\end{abstract}
\newcommand{\keywords}[1]{\textbf{Keywords:} #1}
\renewcommand{\abstractname}{Abstract}
\begin{abstract}
    Interplanetary shocks are one of the crucial dynamic processes in the Heliosphere. They accelerate particles into a high energy, generate plasma waves, and could potentially trigger geomagnetic storms in the terrestrial magnetosphere disturbing significantly our technological infrastructures. 
       In this study, two IP shock events are selected to study the temporal variations of the shock parameters using magnetometer and ion plasma measurements of the STEREO$-$A and B, the Wind, Cluster fleet, and the ACE spacecraft. The shock normal vectors are determined using the minimum variance analysis (MVA) and the magnetic coplanarity methods (CP). During the May 07 event, the shock parameters and the shock normal direction are consistent. The shock surface appears to be tilted almost the same degree as the Parker spiral, and the driver could be a CIR. 
       During the April 23 event, the shock parameters do not change significantly except for the shock $\theta_{Bn}$ angle, however, the shape of the IP shock appears to be twisted along the transverse direction to the Sun-Earth line as well. The driver of this rippled shock is SIRs/CIRs as well. Being a fast-reverse shock caused this irregularity in shape.\\


\end{abstract}
\newpage
\tableofcontents
\newpage
\chapter{Introduction}
The solar corona is hotter than the photosphere, the chromosphere, and the transient layers beneath it. As a result, the high temperatures ionize atoms, creating a plasma of free-moving electrons and ions, known as the solar wind. Historically, \citep{parker1958dynamics} predicted the existence of the solar wind and coined the term "solar wind". He deducted it based on German astronomer Ludwig Bierman's observation of how the comet tail always points away from the Sun \citep{biermann201325}. The existence of the solar wind was confirmed by the Mariner 2 spacecraft \citep{snyder1965interplanetary}. The solar wind is a collisionless plasma, and it flows at both supersonic and super-Alfvénic speed, meaning they exceed the Alfvén speed, which is the speed of magnetohydrodynamic waves in a plasma. A shock wave is where a fluid changes from supersonic to subsonic speed. Therefore, the fast-moving solar wind tends to create a shock on its journey. Hence, interplanetary (IP) shocks are common through the heliosphere, which is a bubble-like region of space surrounding the Sun and extending far beyond the orbits of the planets and is filled with the solar wind. There are a few varieties of shocks such as planetary bow shocks, shocks that are risen due to the stream interaction regions (SIR), which is called co-rotation interaction region (CIR) when extending beyond 1 AU, and coronal mass ejection (CME) driven shocks. IP shocks are one of the main and efficient accelerators of energetic particles \citep{tsurutani1985acceleration, KEITH20211}. These accelerated particles can cause disturbances to the geomagnetic field and are hazardous to astronauts and satellites. (IP) shocks driven by (CMEs) precede large geomagnetic storms \citep{gonzalez1994geomagnetic}. Large geomagnetic storms can damage oil and gas pipelines and interfere with electrical power infrastructures. GPS navigation and high-frequency radio communications are also affected by ionosphere changes brought on by geomagnetic storms \citep{cid2014extreme} and can cause internet disruptions around the world for many months \citep{jyothi2021solar}. 
Therefore, IP shocks are important in determining and understanding space weather.

The main goal of this thesis is to study and determine parameters such as IP shock normals, upstream and downstream plasma parameters (magnetic field, density, temperature, velocity), and how they vary in their temporal evolution.
 There are several methods for determining the shock normal vector \citep{paschmann1998analysis}. In this thesis, the minimum variance analysis (MVA) and the magnetic coplanarity method (CP) are used. These two methods are primarily utilized because they require solely magnetic field data. The data are from NASA's twin Solar Terrestrial Relations Observatory Ahead (STEREO$-$A) and Behind (STEREO$-$B) \citep{kaiser2008stereo}, the Wind \citep{ogilvie1997wind}, and  Advanced Composition Explorer (ACE) spacecraft \cite{stone1998advanced}, and ESA's four identical Cluster constellation satellites--Cluster-1 (C1), Cluster-2 (C1), Cluster-3 (C3), and Cluster-4 (C4) \citep{escoubet1997cluster}. The temporal resolution of the magnetometers of the heliospheric (and the Cluster) spacecraft is significantly higher than the plasma instruments because the variations are quite slow in the heliosphere. So, any agreement between the two methods indicates relatively accurate shock normal vectors \citep{facsko2008statistical, facsko2009global, facsko2010study}. 
Here, two events are studied, one is May 07, 2007, and the other is April 23. The data selection year, 2007, is special because location wise it was the year when twin STEREO$-$A and B spacecraft were closer to the Sun-Earth line until their gradual separation from each other in the following years. Later, the spatial separation is so high that it is hard to distinguish the spatial and temporal changes. Hence, shocks during this period are proper to study shock propagation and their temporal developments in the case of using these spacecraft.
 Chapter 2 introduces the basics of plasma physics and magnetohydrodynamics, which are the governing equations of these heliosphere phenomena. Chapter 3 discusses a brief description of the Sun, the solar wind, and the interplanetary magnetic field. Chapter 4 is about instrumentation, database, and methods. Chapter 5 presents the results and discusses them. Chapter 6 is the summary and conclusions.  
\chapter{Basics of Magnetohydrodynamics}
\section{Plasma}
The term plasma for this state of matter was coined by Irvin Langmuir after its similarity with the blood plasma carrying white and red cells \citep{tonks1967birth}. A plasma is a set of charged particles made up of an equal number of free carriers for positive and negative charges. Having nearly the same number of opposite charges ensures that the plasma appears quasi-neutral from the outside. Free particle carriers mean the particles inside a plasma must have enough kinetic (thermal) energy to overcome the potential energies of their nearest neighbors, which means a plasma is a hot and ionized gas.
 There are the three basic criteria for a plasma \citep[Chapters 1-4]{baumjohann2012basic}.\\
The first criterion is a test-charged particle is clouded by its opposite-charged particles, canceling the electric field of the test particle. This is known as the Debye shielding and its so$-$called Debye$-$lenght, $\lambda_{D}$, is defined as follows:
\begin{align}\label{eq:Debyalength}
    \lambda_D = \sqrt{\frac{\epsilon_0 k_B T_e}{n_e e^2}},
\end{align}
where $\epsilon_0$ is the free space permittivity, $k_B$ the Boltzmann constant, $T_e = T_i$ the electron and ion temperature, $n_e$ the electron plasma density, and $e$ electric charge. To ensure the quasineutrality of a plasma, the system length $L$ must be greater than the Debye length
\begin{align}
    L>>\lambda_D
\end{align}
The second criterion is since the Debye shielding results from the collective behavior of particles inside the Debye sphere with the radius $\lambda_D$, the Debye sphere must contain enough particles
\begin{align}
    \Lambda = n_e \lambda^3_D >> 1
\end{align}
which indicates the mean potential energy of particles between their nearest neighbors must be less than their mean individual energies.\\
The third criterion is the collision time scale $\tau$ of the system is greater than the inverse of the plasma $\omega_p^{-1}$, electron $\Omega_e^{-1}$, and ion $\Omega_i^{-1}$ cyclotron gyrofrequencies.
\begin{align}\label{eq:time}
    \tau >> \omega_p^{-1}, \Omega_i^{-1}, \Omega_e^{-1}
\end{align}
By solving each particle, plasma dynamics can be described, but this approach is too difficult and inefficient. Therefore, there are certain approximations, depending on the corresponding problems. Magnetohydrodynamics is one such approximation that instead of taking account of individual particles, the plasma is assumed as a magnetized fluid.

\section{The single fluid MHD}\label{sec:thesinglefluidmhd}
In this section, the magnetohydrodynamics (MHD) equations are briefly introduced without too much detail and derivations.
The following formulizations are based on \citep{baumjohann2012basic}[page 138-158], \citep{IdealMHD}, \citep{freidberg2014ideal} and \citep{SinglefluidMHD}.
Magnetohydrodynamics (MHD) was developed by Hannes Alfvén \citep{davidson2002introduction}. MHD equations are the result of coupling the Navier-Stokes equations (the fluid equations) to the Maxwell equations. In the MHD, the plasma is treated as a single fluid with macroscopic parameters such as average density, temperature, and velocity.  
Since plasma consists of generally two species of particles, namely electrons, and ions, the different species should be handled together. Hence, the single fluid approximation is utilized.
The single fluid variables are defined as follows:\\
Mass density, considering the mass $m_e$ of an electron is significantly lower than the mass $m_i$ of an ion, $m_e<<m_i$:
\begin{align}
\rho_m = n_e m_e + n_i m_i \approx m_i n
\end{align}
Momentum density:
\begin{align}
    \rho_m \mathbf{v} = n_e m_e \mathbf{u_e} + n_i m_i \mathbf{u_i} \approx \rho_m \mathbf{u_i}
\end{align}
where $\mathbf{v}$, $\mathbf{u_e}$, $\mathbf{u_i}$ are plasma, electron, ion velocities respectively, and  $\rho$ is the plasma density, $n_e$, $n_i$ are the electron and ion number densities.

Current density:
\begin{align}
    \mathbf{J} = n_e q_e \mathbf{u}_e + n_i q_i \mathbf{u}_i = en(\mathbf{u}_i-\mathbf{u}_e)
\end{align}
where $q_e$ and $q_i$ are electron and ion charges.

Total pressure:
\begin{align}
    \mathbf{P}=\mathbf{P}_e+\mathbf{P}_i
\end{align}
By using the single fluid variables, the single fluid MHD equations
can be defined as follows:\\
The continuity equation for the mass density:
\begin{align}\label{eq:mass}
    \frac{\partial \rho_m}{\partial t}+\nabla\cdot (\rho_m \Vec{v}) = 0
\end{align}
The momentum equation:
\begin{align}
\frac{\partial (\rho_m\mathbf{v})}{\partial t}+\nabla\cdot (\rho_m  \mathbf{v}\mathbf{v}) = -\nabla\cdot \mathbf{P}+\rho_e \mathbf{E}+\mathbf{J}\cross \mathbf{B}
\end{align}
where $\rho_e$ denotes the charge density, and $\mathbf{E}$, $\mathbf{B}$ electric and magnetic fields.\\
The generalized Ohm's law:
\begin{align}
    \mathbf{E}+ \mathbf{v}\cross \mathbf{B} = \eta \mathbf{J}+\frac{1}{ne}\mathbf{J}\cross \mathbf{B}-\frac{1}{ne}\nabla\cdot \mathbf{P}_e+\frac{m_e}{ne^2}\frac{\partial \mathbf{j}}{\partial t},
\end{align}
where $\eta$ is magnetic diffusivity , and $\eta \mathbf{J}$ is the resistive term.\\
Ampère's law:
\begin{align}
    \nabla\cross\mathbf{B}=\mu_0 \mathbf{J}+\mu_0\epsilon_0\frac{\partial \mathbf{E}}{\partial t}
\end{align}
where $\mu_0$ is the vacuum magnetic permeability, $\epsilon_0$ is vacuum permittivity

Faraday's law:
\begin{align}
    \nabla\cross \mathbf{E} = -\frac{\partial \mathbf{B}}{\partial t}
\end{align}
And finally, the magnetic field divergence constraint:
\begin{align}
    \nabla\cdot \mathbf{B}=0
\end{align}
The MHD approximations are valid when the characteristic speed of the system is much slower than the speed of light, 
\begin{align}
    \mathbf{v}<<c
\end{align}
the characteristic frequency, $\omega$, must be smaller than the ion cyclotron frequency $\omega_i$
\begin{align}
    \omega<<\omega_i
\end{align}
the characteristic scale length $L$ of the system must be longer than the mean free path $r_{gi}$ of ion gyroradius
\begin{align}
    L>r_{gi}
\end{align}
the characteristic scale times must be larger than the collision times as stated in equation \ref{eq:time}.
\section{MHD wave modes}\label{sec:mhdwaves}
\begin{figure}[H]
\centering
\includegraphics[width=0.6\textwidth]{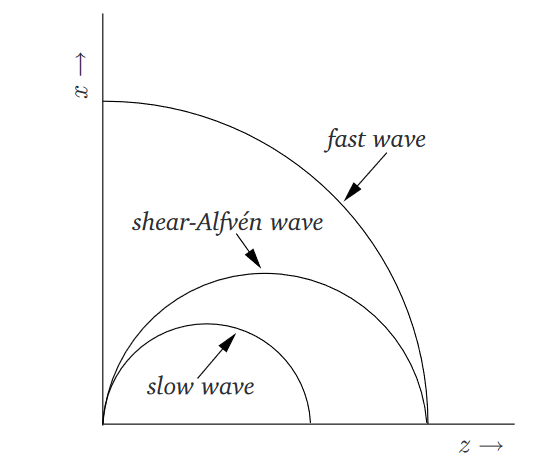}
\caption{Phase velocities of the three MHD waves. \protect\citep[Figure~is from][Figure~7.1]{fitzpatrick2014plasma}}
\label{fig:threewaves}
\end{figure}

Plasma is considered a highly conductive fluid, which consists of charged particles.
 Therefore, MHD waves in plasma fundamentally arise from two distinct wave speeds: the sound speed of a fluid and the Alfvén speed, which is due to the magnetic field pressure and tension. Their combination gives rise to magnetosonic waves. Thus, in MHD there are three wave modes--namely slow magnetosonic, the shear-Alfvénic and the fast magnetosonic waves \citep{fitzpatrick2014plasma} in addition to the sound wave, as seen in Figure~\ref{fig:threewaves}. 
The derivation of the waves can be obtained from the linearized MHD dispersion relation.\\
The sound wave is due to a compressible fluid and the wave is longitudinal. The sound speed is defined as:
\begin{align}\label{eq:soundspeed}
C_s=\sqrt{\frac{\gamma k_B T}{m_i}}
\end{align}
where $\gamma$ is the polytropic index and in space plasma, it is in the range $1<\gamma<5/3$ \citep{livadiotis2021relationship}, $k_B$ is the Boltzmann constant, $T$ is temperature, $m_i$ is mass of a particle species.
  The Mach number, a ratio of flow to the (thermal) sound speed
\begin{align}
M_s=\sqrt{\frac{m_i v^2}{\gamma k_B T}}
\end{align}
The shear Alfvénic wave is incompressible and transverse.
 The Alfvén speed is defined as follows:
\begin{align}\label{eq:alfvenicspeed}
V_A=\sqrt{\frac{B^2}{\mu_0 \rho}}
\end{align}
here $\frac{B^2}{\mu_0}$ is the magnetic pressure and $\rho$ is density.\\
Similarly, the magnetic Mach number, which is defined as a ratio between the flow speed $V_{flow}$ and the Alfvénic speed $V_A$, is defined as follows:
\begin{align}
M_A=\frac{V_{flow}}{V_A}
\end{align}
The magnetosonic waves are as follows:
\begin{align}\label{eq:truemagnetosonic}
    C^2_{ms}=\frac{1}{2}(C^2_s+V^2_A)\bigg [1\pm \bigg(1-4\frac{cos^2\theta}{b^2}\bigg)\bigg],
\end{align}
where $b$ term is as follows:
\begin{align}\label{eq:bchunk}
b=\frac{C_s}{V_A}+\frac{V_A}{C_s}\geq 2
\end{align}
The equation \ref{eq:truemagnetosonic} has two terms. The term containing (+) is the fast magnetosonic wave while the one containing (-) is the slow magnetosonic wave.\\
When $V_A >> V_s$ or $V_s >> V_A$ in \ref{eq:bchunk}, as well as the wave propagation direction, is nearly perpendicular to the magnetic field $(cos^2\theta<<1)$, the slow magnetosonic wave speed becomes as follows:
\begin{align}\label{eq:slowms}
    C^2_{ms}=cos^2\theta \frac{C_s^2 V_A^2}{(c^2_s+V^2_A)}
\end{align}
while the fast magnetosonic wave is simplified as follows:
\begin{align}\label{eq:magetosonic}
    C_{ms}=\sqrt{C_s^2+V_A^2}
\end{align}
from \ref{eq:magetosonic}, the fast magnetosonic Mach number is defined as
\begin{align}
    M_{ms}=\frac{V_{flow}}{C_{ms}}
\end{align}
\section{MHD discontinuities}
When plasmas of different properties collide, they reach equilibria, resulting in the boundaries separating neighboring plasma regions \citep[Chapter 8]{baumjohann2012basic}. These boundary regions are called discontinuities, and in astrophysics, heliopause and magnetopause are examples of these discontinuities. 
Through discontinuities, the field and plasma parameters change abruptly, but these sudden changes satisfy a few conditions, namely the Rankine-Hugoniot jump conditions.
 To derive the jump conditions, it is suitable to integrate the conservation laws across the discontinuity. Therefore, it is better to write the single fluid MHD equations \ref{sec:thesinglefluidmhd} in conservative form, assuming that the two sides of the discontinuity satisfy an ideal single-fluid MHD. Following some notations and derivations of \citep[Chapter 8]{baumjohann2012basic}, the ideal single-fluid MHD equations in conservative forms are defined below:
\begin{equation} \label{eq:momentum}
\frac{\partial (\rho\mathbf{v})}{\partial t}+\nabla\cdot \Bigg[ (\rho\mathbf{vv})+\left(\mathbf{P}+\frac{B^2}{2\mu_0}\mathbf{I}\right)-\frac{1}{\mu_0} (\mathbf{BB})\Bigg]=0,
\end{equation}
where $\mathbf{P}$ the plasma pressure, B the magnetic field magnitude,$\mathbf{B}$ the magnetic field vector, $\mu_0$ the vacuum permeability, and $\mathbf{I}$ the identity tensor.\\
The induction equation:
\begin{equation}\label{eq:induction}
\frac{\partial \mathbf{B}}{\partial t}=\nabla\cross(\mathbf{v}\cross \mathbf{B})
\end{equation}
The divergence of the magnetic field:
\begin{equation}
\nabla\cdot \mathbf{B}=0
\end{equation}
The energy conservation equation:
\begin{equation} \label{eq:energy}
\nabla\cdot\Bigg\lbrace \rho\mathbf{v} \left[\frac{1}{2}v^2+w+\frac{1}{\rho}\left(p+\frac{B^2}{\mu_0}\right)\right]-\frac{1}{\mu_0}(\mathbf{v}\cdot \mathbf{B})\mathbf{B}\Bigg\rbrace=0,
\end{equation}
where for slowly variable fields $\mu_0 j=\nabla\cross \mathbf{B}$ and the ideal Ohm's law $E=-\nabla\cross \mathbf{B}$ are implemented as well as neglecting charges $\rho\mathbf{E}=0$, and $w=\frac{c_v P}{\rho k_B}$ is the internal enthalpy.
\\
For completion, the equation of state is set:
\begin{equation} \label{eq:eos}
\mathbf{P}=p\mathbf{I},
\end{equation} 
where $p$ is the scalar pressure.
Choosing a co-moving reference frame with discontinuity, a steady-state situation is assumed that all the time-dependent terms are canceled, leaving the flux terms in the conservation laws. 
A discontinuity causes the plasma parallel to the discontinuity to stay the same while the plasma perpendicular to the discontinuity changes. For these reasons, a two-dimensional function S(x)=0 can be used to characterize the discontinuity surface, and the normal vector to the discontinuity, $\mathbf{n}$, is defined as follows
\begin{align}\label{eq:normal}
\mathbf{n}=-\frac{\nabla S}{|\nabla S|}
\end{align}  
To the direction of the $\mathbf{n}$ the vector derivative has only one component $\nabla=\mathbf{n}(\frac{\partial}{\partial n})$. 
After these considerations, integrating over the discontinuity for the flux terms has to be done. Remembering only two perpendicular sides of the discontinuity contribute twice to the integration, a quantity X crossing $S$ becomes
\begin{align}
\oint_s\frac{\partial X}{\partial n}dn=2\int_1^2\frac{\partial X}{\partial n}dn=2(X_2-X_1)=2[X],
\end{align}
where parenthesis $[X]$ indicates the jump of a quantity X.
Now replacing the conservation laws of the ideal one-fluid MHD with the jump conditions, \textbf{the Rankine-Hugoniot} conditions are defined as follows:
\begin{equation}
\mathbf{n}\cdot[\rho_m\mathbf{v}]=0
\end{equation}
\begin{equation}
\mathbf{n}\cdot[\rho_m\mathbf{vv}]+\mathbf{n}\bigg[ p+\frac{B^2}{2\mu_0}\bigg]-\frac{1}{\mu_0}\mathbf{n}\cdot [\mathbf{BB}]=0
\end{equation}
\begin{equation}
[\mathbf{n}\cross \mathbf{v}\cross \mathbf{B}]=0
\end{equation}
\begin{equation}
\mathbf{n}\cdot[\mathbf{B}]=0
\end{equation}
\begin{equation}\label{eq:nenergy}
\Bigg [ \rho_m \mathbf{n}\cdot\mathbf{v}\Bigg\lbrace \frac{v^2}{2}+w+\frac{1}{\rho}\left(p+\frac{B^2}{\mu_0}\right)\Bigg\rbrace-\frac{1}{\mu_0}(\mathbf{v}\cdot \mathbf{B})\mathbf{n}\cdot \mathbf{B}\Bigg]=0
\end{equation}
From the equations above, the normal component of the magnetic field is continuous across all discontinuities, which leads to its jump condition vanishing.
\begin{equation}\label{eq:Bn}
[B_n]=0
\end{equation}
Also normal direction mass flow is continuous: 
\begin{equation}\label{eq:nvn}
[\rho_m v_n]=0
\end{equation}
Splitting the fields between the normal and tangential components, the remaining jump conditions are derived:
\begin{equation}\label{eq:totalpressure}
\rho_m v_n[v_n]=-\Bigg [p+\frac{B^2}{2\mu_0}\Bigg]
\end{equation}
\begin{equation}\label{eq:tangentialvelocity}
\rho_m v_n[\mathbf{v_t}]=\frac{B_n}{\mu_0}[\mathbf{B_t}]
\end{equation}
\begin{equation}\label{eq:magneticfieldwithtangentialvelocity}
B_n[v_t]=[v_n\mathbf{B_t}]
\end{equation}
where subscript $t$ and $n$ denote normal and tangential components respectively. The equations \ref{eq:Bn} and \ref{eq:nvn} demonstrate that the normal components of the magnetic field, as well as the mass flow, are constant across the discontinuity.
\textbf{The Rankine-Hugoniot} conditions provide the four types of MHD discontinuities \citep[Chapter-8]{baumjohann2012basic}, namely contact discontinuity, tangential discontinuity, rotational discontinuity, and shocks as the values are defined in Table \ref{tab:discontinuitytable}.\\
\begin{table}[H]
\centering
\begin{tabular}{ |c|c|c|c|  }
 \hline
 \multicolumn{4}{|c|}{\centering The properties of MHD discontinuities}\\
 \hline
Discontinuity type & Normal Mass flux & Normal Magnetic field & Density \\
 \hline
Contact  & $v_n=0$ & $B_n \neq 0$ & $\rho \neq 0$\\
 \hline
Tangential  & $v_n=0$ & $B_n = 0$ & $\rho \neq 0$\\
 \hline
Rotational & $v_n \neq 0$ & $B_n \neq 0$ & $\rho=0$\\
 \hline
Shocks & $v_n \neq 0$ & $B_n \neq 0$ or $B_n = 0$  & $\rho \neq 0$\\
 \hline
 
\end{tabular}
\caption{Some properties of MHD discontinuities. \citep[The values are from][]{oliveira2015study}}
\label{tab:discontinuitytable}
\end{table}

\textbf{Contact discontinuity} is determined by the condition when there is no normal mass flow across discontinuities, which means from the Rankine-Hugoniot conditions $v_n=0$. When $B_n\neq 0$ or $[B_n]=0$, the only quantity that experiences a change across the discontinuity is the density, $[\rho]\neq 0$. Due to these constraints, the plasma on the two sides of the discontinuity are attached and connected by the normal component of the magnetic field, As a result, they flow together at the same tangential speed. This is known as the contact discontinuity. 

\textbf{Tangential discontinuity}
is when $B_n=0$ in addition to $v_n=0$, only non-trivially satisfied jump condition is the total pressure balance in the equation \ref{eq:totalpressure}, which indicates that a discontinuity is created between two plasma with total pressure balance on both sides and no mass and magnetic fluxes are crossing across the discontinuity while all other quantities are changing. This is the tangential discontinuity. 

\textbf{Rotational discontinuity} is formed when there is a continuous normal flow velocity $[v_n]=0$ with a non-vanishing $B_n\neq 0$ such that from the equations \ref{eq:tangentialvelocity} and \ref{eq:magneticfieldwithtangentialvelocity} the tangential velocity and the magnetic field can only change together, especially the equation \ref{eq:magneticfieldwithtangentialvelocity} indicates they rotate together when crossing the discontinuity.

\textbf{Shocks}\label{shocks}
Unlike the previous three discontinuities, shocks are irreversible in that the entropy increases \citep{goedbloed2019magnetohydrodynamics}. And one of the main distinctive characteristics of shocks is normal fluxes of the Rankine-Hugionot conditions take non-zero values, $\rho v_n\neq 0$, and the density $\rho$ is discontinuous.

Consequently of the three wave modes, there are corresponding three shocks -- fast shock (FS), intermediate shock (IS), and slow shock (SS) \citep{tsurutani2011review}. When the fast wave speed is greater than the upstream magnetosonic speed \ref{eq:magetosonic}, the fast shocks (FSs) are formed. Similarly, when the shear-Alfvén wave speed is greater than the upstream Alfvén velocity \ref{eq:alfvenicspeed}, the intermediate shocks (ISs) are formed, and when the slow wave speed is greater than the upstream (thermal) sound speed \ref{eq:soundspeed}, the slow shocks (SSs) are formed with their respective Mach numbers greater than at least 1.

\begin{figure}[H]
\centering
\includegraphics[width=0.6\textwidth]{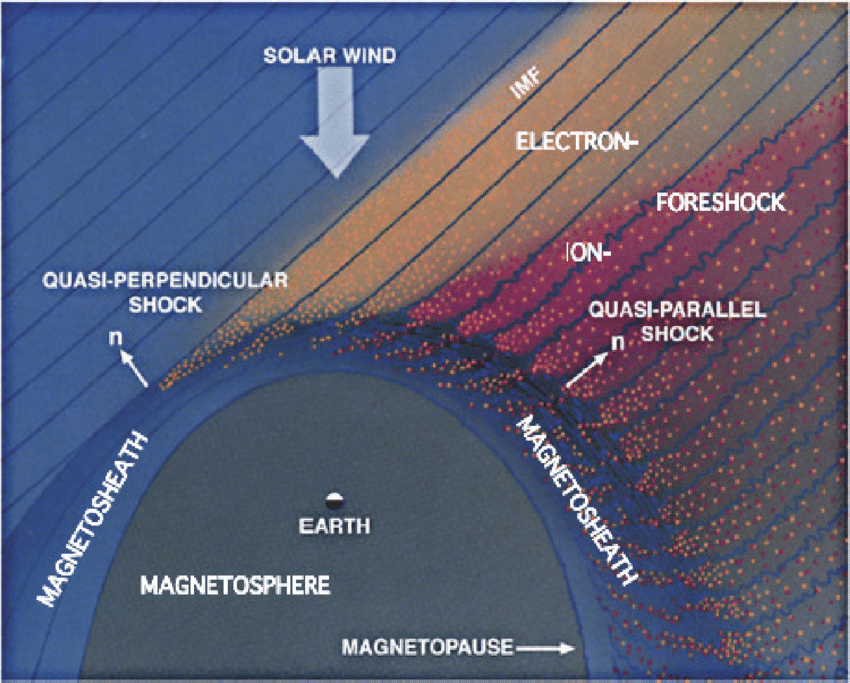}
\caption{The solar wind interaction with the Earth's bow shock.\protect\citep[Figure~is from][Figure~2-5]{oliveira2015study}}
\label{fig:differentshock}
\end{figure}

The three MHD waves \ref{sec:mhdwaves} are anisotropic, which means their speeds depend on the angle between wave propagation direction and the upstream (unshocked) magnetic field.
 An illustration of this is the Parker spiral propagation of the solar wind such that the interplanetary magnetic field (IMF) hits, for example, the Earth from the morning side, it creates parallel and perpendicular shocks concerning the shock normal and the upstream magnetic field $\mathbf{B}$ as shown in Figure~\ref{fig:differentshock}   
Hence, depending on the shock normal angle $\theta_{B_n}$, the shocks can be geometrically classified as \textbf{parallel}, $\theta_{B_n}=0^{\circ}$, \textbf{perpendicular}, $\theta_{B_n}=90^{\circ}$, \textbf{oblique}, $0^{\circ}<\theta_{B_n}<90^{\circ}$, \textbf{quasi-parallel}, $0^{\circ}<\theta_{B_n}<45^{\circ}$, and \textbf{quasi-perpendicular}, $45^{\circ}<\theta_{B_n}<90^{\circ}$  \citep{CHAO1984641, johlander2022quasi}. 
\chapter{Solar and heliospheric physics}
 The Sun is the main source of energy for all lives on the earth as well as the main defining object of the solar system dynamics. Like all other stars, the Sun has both inner complex dynamics and influence on interplanetary space. In this chapter, for the sake of completeness and to explain the origin of these tremendous dynamics, the inner structure of the Sun as well as some important heliospheric activities are briefly discussed.
\section{The Sun}
\begin{figure}[H]
\centering
\includegraphics[width=0.6\textwidth]{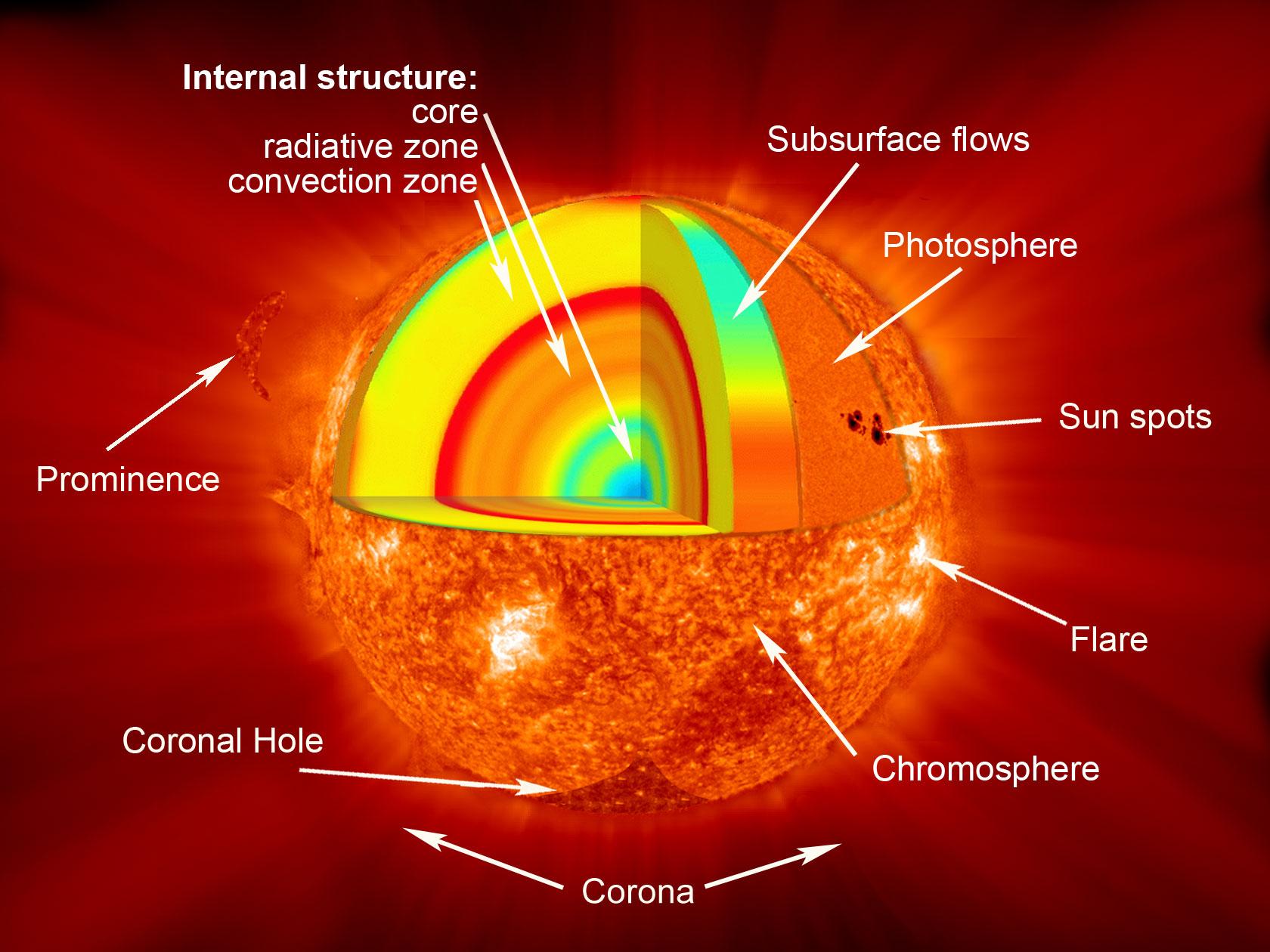}
\caption{The layers of the Sun \protect\citep[Credit:][]{NASASunLayers2017}}
\label{fig:layers}
\end{figure}

Our star, the Sun, is a G2-V type star \citep{aschwanden2014encyclopedia}. 
All stars are in a balance of opposing forces between outward radiation fueled by nuclear fusion and inward pressure dictated by gravitational force.
 The radiative force is the product of the fusion of hydrogen atoms, which constitutes about 73 $\%$, into helium atoms, which constitute 25 $\%$ of the Sun \citep{Basu2007HelioseismologyAS}. This process is a thermonuclear reaction via turning a proton into a neutron. \\
Structurally, the Sun has inner zones and outer zones. The former consists of the core, radiative zone, interface layer (tachocline), and convection zone. The latter consists of the photosphere, the chromosphere, the transition region, and the corona \citep{NASASunLayers2017} as shown in Figure~\ref{fig:layers}.  
The core of the Sun is 25$\%$ of the solar radius \citep{garcia2007tracking}, the density is about $150g/cm^3$ \citep{basu2009fresh} contains 34$\%$ of the solar mass even though only 0.8$\%$ of the total volume, and it is the hottest part of the Sun that a temperature of 15 million Kelvins \citep{TheSolarInterior}. This is the zone where the thermonuclear reaction is going.
 For the radiative zone, it starts from the edge of the core to the interface layer (tachocline), and much of the energy generated in the core is carried away by photons though photons take a million years to reach the next layer due to the dense material of the region \citep{chemin2022introductory}. 
 
The very next layer, the interface layer (tacholine) is much of our interest because it is believed that the solar magnetic dynamo originates in this very thin layer \citep{tobias2007solar} that magnetic field lines can become stronger due to the changes in the fluid flow velocities crossing this layer. However, recent radio observations of brown dwarf stars show that, despite not having a tachocline layer, they can have similar magnetic strength and activities just like the Sun, which indicates the convection zone may be solely responsible for the solar dynamo \citep{route2016discovery}.

 Next, the convection zone is the outermost layer of the inner zones and it extends about 200,000 km from the depth and the temperature is approximately 2 million Kelvins \citep{christensen1991depth}, which makes the zone relatively cooler for heavy ions to keep their electrons so that the zone is opaque to the radiation \citep{ortuno2012endoreversible}, which in turn traps the heat and eventually the fluid becomes unstable and start convecting. As a result, this layer is very turbulent  \citep{brummell1995turbulent}, which in addition to the rotational motion creates electric currents and magnetic fields, and the gas pressure is much more dominant than the magnetic pressure in this region, and because of this the magnetic field is dragged and twisted by the fluid, which then propagates passing through the photosphere, chromosphere, the transition region up to the corona and creates a multitude of activities on the surface of the Sun called solar activity. 
 
 In the photosphere, visible darker areas called sunspots appear due to localized strong magnetic regions, which inhibit some of the heat from reaching the surface and this makes the areas cooler than the other parts of the Sun \citep{babcock1955sun}. Because of the mentioned strong magnetic nature, the magnetic field lines around the sunspot often twist and cross, causing a tension of energy that bursts in an explosion called a solar flare \citep{chakraborty2022brief}.

\begin{figure}[H]
\centering
\includegraphics[width=0.8\textwidth]{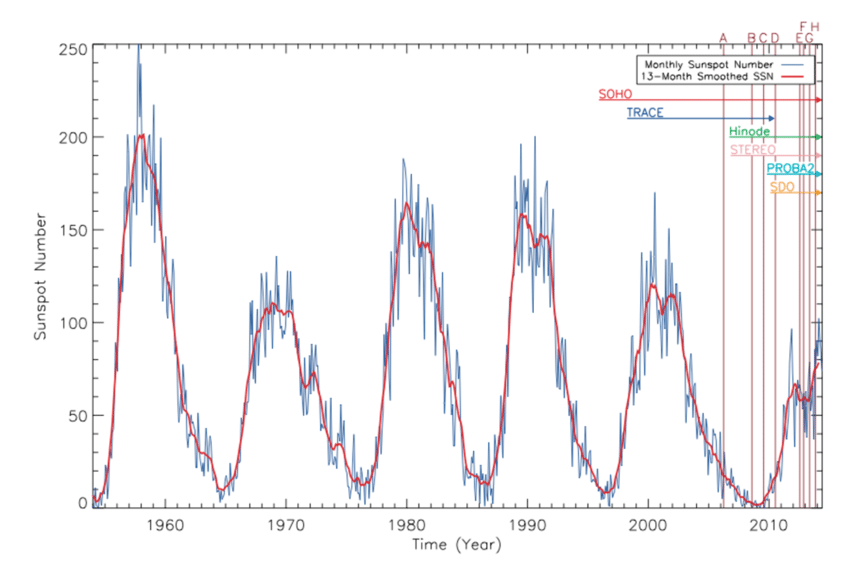}
\caption{The sunspot cycle over the last several decades \protect\citep[Figure~is from][Figure~8]{article}}
\label{fig:sunspot}
\end{figure}

The corona is the hot and ionized outermost layer of the Sun and it is millions of Kelvins greater than that of the surface of the Sun \citep{aschwanden2006physics}.
 The solar magnetic field confines much of the coronal plasma, but some of the plasma spread with a supersonic speed into interplanetary space, which is called the solar wind. The solar magnetic field forms so-called flux tubes. These ropes are tensed by the differential rotation of the Sun. Therefore, store plenty of energy. If a certain configuration appears (X-shape) the magnetic structure reorganizes. Hence, a huge amount of energy is released. The name of this 20 million K-degree flash is a solar flare. It emits plenty of energized protons and other ions in the heliosphere and very often, but not always a huge amount of solar plasma ejects to the IP space. The process is called coronal mass ejections (CME). However, CME could occur without a flare too. These phenomena are the main drivers of space weather and the conditions of the terrestrial cosmic environment \citep{gabor2022space}. The solar activity is cyclic, which is called the solar cycle, and it is approximately 11 years \citep{SDO}, see Figure~\ref{fig:sunspot}. Each cycle the magnetic field of the Sun flips, and the periods, in which, the sunspots are the greater in number are called solar maximum and the lower in number are called solar minimum.  
\section{The Solar Wind and the Interplanetary Magnetic Field}
\begin{figure}[H]
\centering
\includegraphics[width=0.6\textwidth]{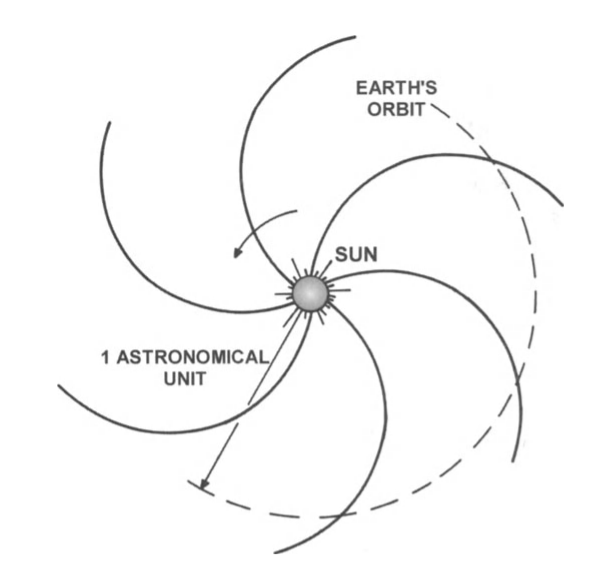}
\caption{Schematic representation of the Archimedes spiral structure of the interplanetary magnetic field \protect\citep[Figure~is from][Figure~3]{bittencourt2004fundamentals}}
\label{fig:parker}
\end{figure}

The Sun ejects highly energized and ionized charged particles continuously in all directions, which is the solar wind. 
The solar wind is a collisionless plasma that consists of equal amounts of protons and electrons with the addition of negligible 3 to $6\%$  helium \citep{neugebauer1981observations}. As a result, the solar wind is a quasi-neutral plasma whose characteristics can be defined by magnetohydrodynamics (MHD). 
In MHD, Alfvén theorem states that in a fluid with high electric conductivity the magnetic field line is frozen in it and moves along with it \citep{alfven1942existence}. Since the solar wind is one such fluid the Sun's magnetic field lines are frozen in the flow that they are forced to propagate with the solar wind \citep{roberts2007alfven}, forming the interplanetary magnetic field (IMF). 
Depending on the origination point whether in the coronal holes or the equatorial belt of the Sun, the solar wind is classified as the fast solar wind with a velocity of 750 km/s and the slow solar wind with a velocity of 400 km/s respectively \citep{feldman2005sources}. The coronal holes are areas that appear dark in X-ray images because these areas are much cooler and less dense than other regions \citep{cranmer2009coronal} and the magnetic field around these areas does not loop back down but extends into the interplanetary space \citep{parker1959extension}, so the plasma can easily flow out, creating the fast solar wind. The coronal holes can appear anywhere on the coronal areas during the solar maximum while they usually appear on northern or southern poles of the sun during the solar minimum \citep{mccomas2003three}.
\begin{figure}[H]
\centering
\includegraphics[width=0.6\textwidth]{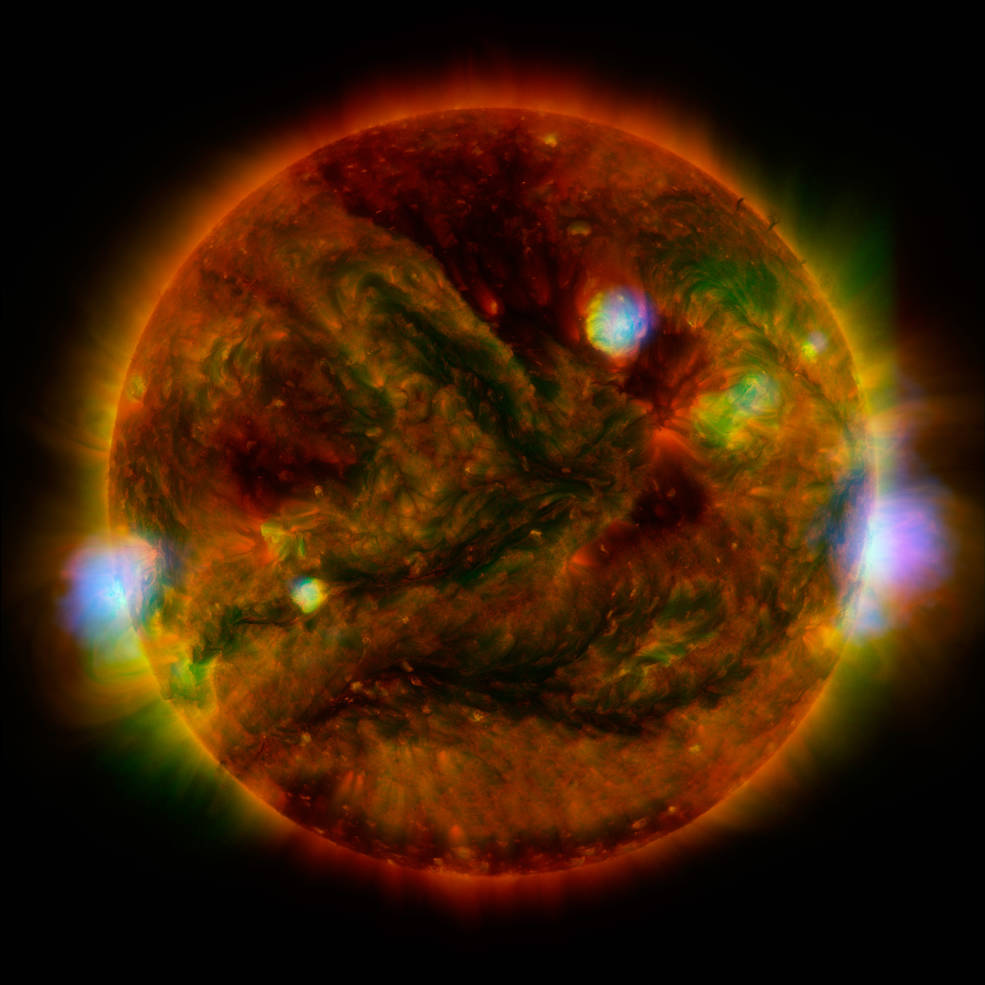}
\caption{Combined X-ray image of the Sun's active regions observed from several telescopes. High-energy X-rays from NASA's Nuclear Spectroscopic Telescope Array (NuSTAR) are shown in blue, low-energy X-rays from Japan's Hinode spacecraft are green, and extreme ultraviolet light from NASA's Solar Dynamics Observatory (SDO) is yellow and red \protect\citep[Credit:][]{sunxray}}
\end{figure}
The expansion of the solar wind is a radial propagation away from the Sun that it is diluted and cools down on its journey. As seen from some observations the density of the solar wind decreases approximately as $r^{-2}$ while the temperature decrease is less significant about a factor of 20 \citep{baumjohann2012basic}. During the expansion of the radial flow, the Sun rotates, which is 27 days on average, and because the magnetic field lines are anchored in the sun, the radial flow looks like an Archimedean spiral called the Parker Spiral. At 1 AU the spiral makes $45^0$ to the Earth-Sun line, see Figure~\ref{fig:parker}.
The solar wind travels throughout the solar system and defines the heliosphere the region in space whose frontier is impeded by the interstellar medium. Consequently, the heliosphere is a giant bubble whose center is the Sun and protects the solar system from interstellar radiations and cosmic rays. 
The size of the heliosphere is about 121 AU \citep{cowen2013voyager}. The solar wind decelerates when it flows outward through the Solar System. At a certain point, the flow becomes subsonic and forms a shock. Its name is termination shock \citep{jokipii2013heliospheric}. Then the flow continues its journey outward and interacts with the interstellar material. It is under debate whether the speed of the Solar System is super- or sub-sonic. A bow shock or bow wave forms before the heliosphere, respectively. Furthermore, due to the interaction, the solar wind becomes turbulent and this region is called heliosheath which is in between the termination shock and the heliopause \citep{burlaga2015voyager}. 
\section{Interplanetary shocks}
The average interplanetary magnetic field strength is around 6 nT at 1 AU \citep{lowrie_fichtner_2020} and considering the average speed of the solar wind, the solar wind is both supersonic and super-Alfvénic, which means it is super-magnetosonic. 
As a result of this nature of the solar wind, when it collides with celestial objects such as planets, moons, asteroids, and comets or its slow-flowing part, shocks are formed. The main characteristic of shocks that can be found in interplanetary space to interstellar regions is a denser state in contrast to the medium in which they propagate due to the shock formation.
 Furthermore, if the celestial objects are magnetized, then the shock formations create interesting interaction regions, and their physics looks very exciting. 
 
\begin{figure}[!t]
\centering
\includegraphics[width=0.6\textwidth]{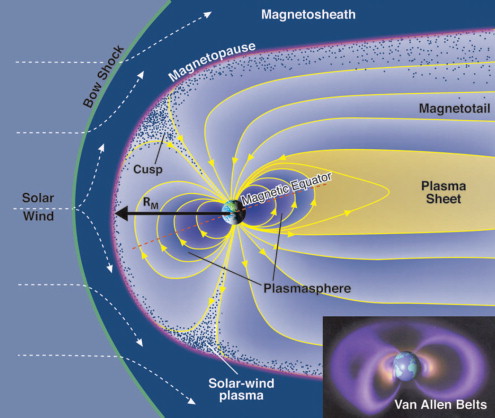}
\caption{Illustration of the Earth's magnetosphere and its interaction with the solar wind. \protect\citep[Figure~is from][Figure~1]{KIVELSON2007519}}
\label{fig:bowshock}
\end{figure}

One such fascinating magnetized object is our planet, the Earth. The Earth has its magnetic field, known as the magnetosphere, which originated from its inner core via the dynamo effect \citep{GILBERT2003355}.   
The Earth's magnetosphere extends 60000 kilometers in the sunward direction while a million kilometers in the anti-sunward direction \citep{lakhina2009overview}. Such a big extension is a common characteristic of all the magnetized planets, and as a result of this extension, the cross-section of a planet is increased by a large factor. For example for the Earth, the factor is 150 \citep{baumjohann2012basic}.
 The magnetic field frozen in the supersonic and super-Alfvénic solar wind plasma cannot enter the magnetosphere, the region where the terrestrial magnetosphere dominates. As a result, a special standing wave, the bow shock forms \citep{BAUMJOHANN200777}. (\ref{fig:bowshock}). 
Consequence of the slowing down of the solar wind plasma, the kinetic energy of some of the particles is converted into thermal energy, which occurs behind the bow shock, and this region is called the magnetosheath. The boundary region between two magnetic field lines is called the magnetopause. Bow shock formation is common to all the planets and celestial objects with or without magnetospheres \citep{articlea}.
\begin{figure}[H]
\centering
\includegraphics[width=0.6\textwidth]{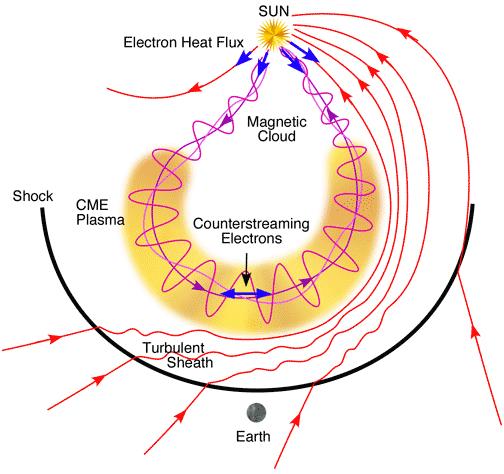}
\caption{Illustration of a CME event. \protect\citep[Credit:][]{cme}}
\label{fig:cme}
\end{figure}
\raggedbottom
In addition, there are several other shocks, the previously mentioned termination shock, coronal mass ejection (CME) driven shock, and a co-rotating interaction region (CIR) driven shock. 
When a CME event occurs, it moves faster than the background solar wind flow, resulting in a shock wave. On the created shock waves charged particles accelerate. So usually CMEs are one of the main causes of the solar energetic particles (SEPs) \citep{cane2006introduction}, see Figure~\ref{fig:cme}.

\begin{figure}[H]
\centering
\includegraphics[width=0.6\textwidth]{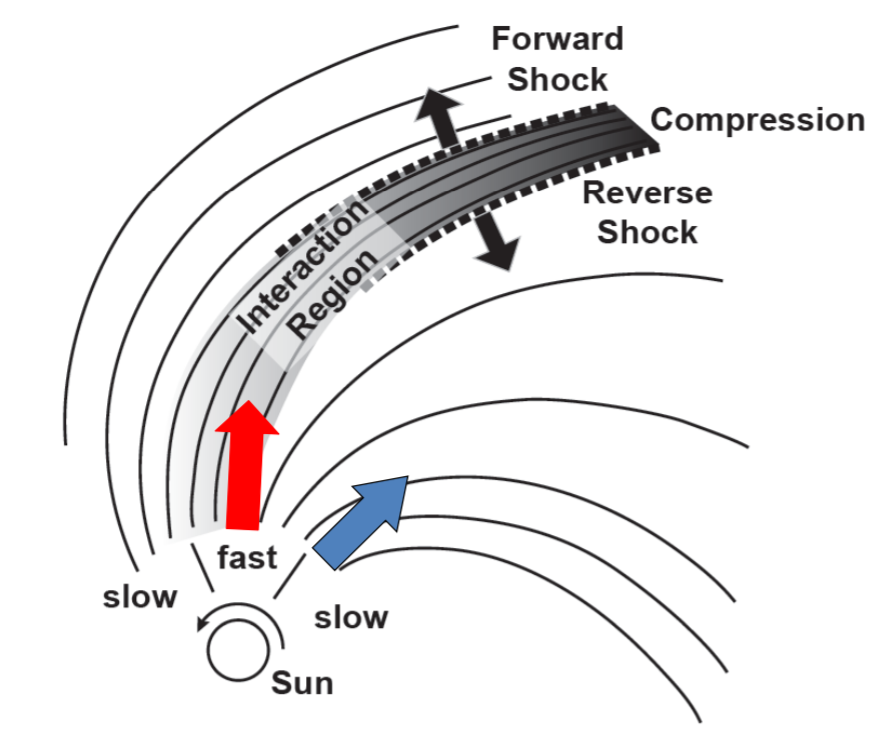}
\caption{Co-rotating interaction region. \protect\citep[Figure~is from][]{cir}}
\label{fig:cir}
\end{figure}

Similarly to the shocks caused by CMEs, When the fast-moving solar wind flow catches the slow-moving solar wind flow, the so-called co-rotating interaction region is formed, and if the pressure gradient gets sufficiently large and the speed difference surpasses the local speed shocks can arise \citep{heber1999corotating}, Figure~\ref{fig:cir}.  
\subsection{Classifications of Interplanetary (IP) shocks}
\begin{figure}[H]
\centering
\includegraphics[width=0.4\textwidth]{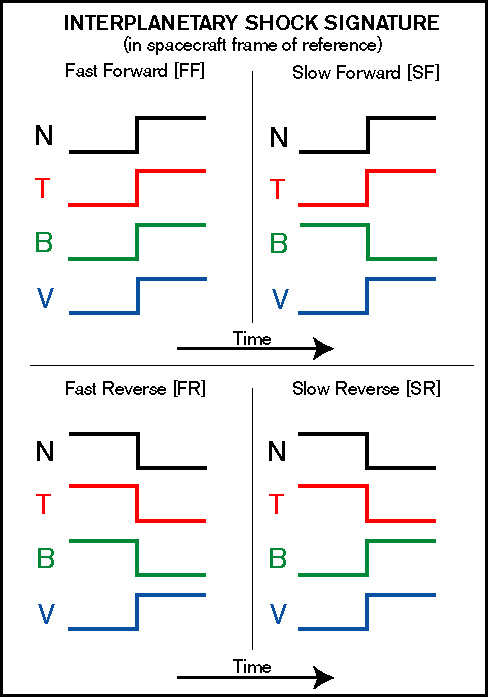}
\caption{Categorizations of IP shocks. N, T, B, and V denote number density, the proton temperature, the magnitude of the magnetic field, and speed respectively. \protect\citep[Figure~is from][]{NASAipshock}}
\label{fig:ipshocks}
\end{figure}
Armed with the above shock definitions and categorizations knowledge, now we can finally be able to classify IP shocks. The IP shocks can be classified based on their travel directions concerning the solar wind frame of reference. If the shock is moving away from its driver, here the Sun is referred but drivers can be detailed such as ICMEs, CIRs, etc, the shock is called forward shock (FS), and if it is moving toward its driver, it is called reverse shock (RS). Adding the previous definitions of fast and slow shocks, IP shocks are usually categorized as fast forward  (FF), fast reverse  (FS), slow forward (SF) and slow reverse (SR) shocks \citep{berdichevsky2000interplanetary}, see Figure~\ref{fig:ipshocks}.
 As you can see from Figure~\ref{fig:ipshocks}, the solar wind plasma parameters -- number density N, the proton temperature T, the magnitude of the magnetic field B, and the bulk speed V parameters increase dramatically from upstream (unshocked) to downstream (shocked) regions in fast forward (FF) shocks while the parameters except for the bulk speed decrease in fast reverse (FR) shocks\footnote{ In a sense of the reverse upstream to downstream in fast reverse (FR) shocks, the parameters except for the bulk speed increase from reversed upstream to downstream}. In the case of slow forward (SF) shocks, the parameters except for the magnitude of the magnetic field increase from upstream to downstream whereas in the case of slow reverse (SR) shocks, the number density N and the proton temperature T decrease while the magnitude of magnetic field B and the bulk speed V increase\footnote{Again in the sense of the reversed upstream to downstream in slow reverse (SR) shocks, the number density N and the proton temperature T increase while the magnitude of magnetic field B and the bulk speed V decrease}.

Within 1 AU, the most frequent IP shocks are fast forward (FF) shocks \citep{richter1985review}
\newpage
\chapter{Analysis methods}
\section{Instruments}
In this thesis, Solar Terrestrial Relations Observatory (STEREO) A and B, Wind, Advanced Composition Explorer (ACE), and the Cluster spacecraft magnetic field and ion plasma data are used. In the following subsections, the basic descriptions of each spacecraft, its orbits, and its instruments are mentioned. 
\subsection{The STEREO mission}
\begin{figure}[H]
\centering
\includegraphics[width=0.8\textwidth]{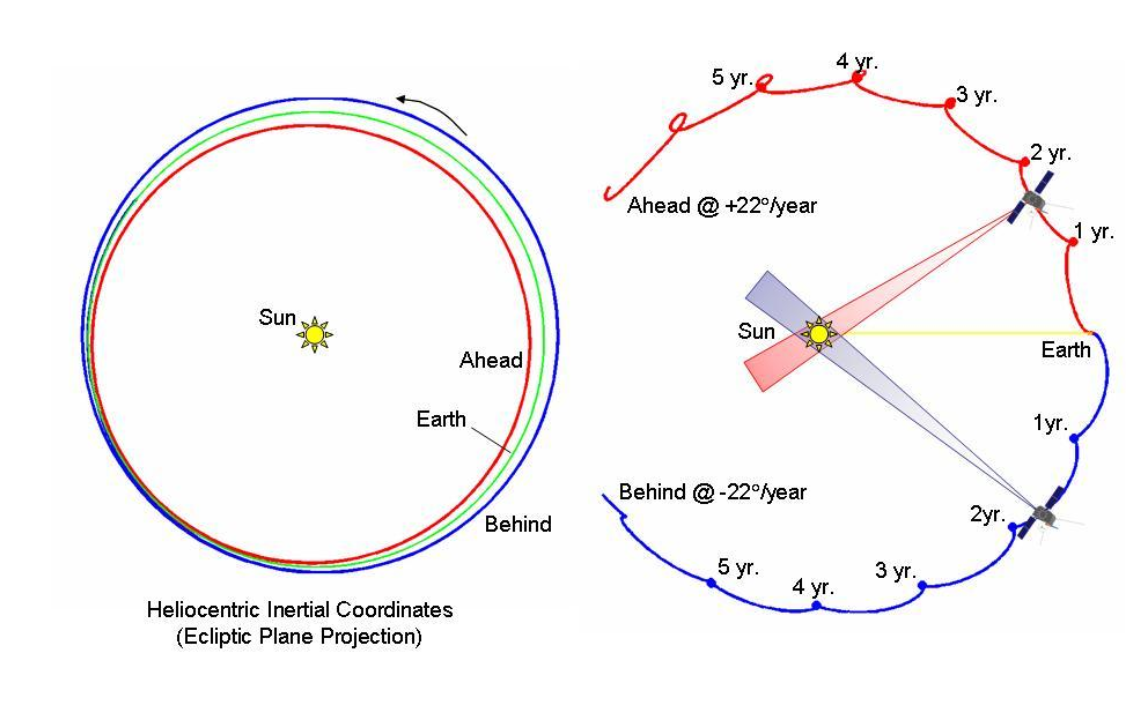}
\caption{Left panel shows the orbits of STEREO$-$A and B relative to that of Earth. The right panel shows the degree of separation of two spacecraft in years from their first launch period. \protect\citep[Figure~is from][Figure~2]{KAISER20051483}}
\label{fig:stereoAB}
\end{figure}

\begin{figure}[H]
\centering
\includegraphics[width=0.6\textwidth]{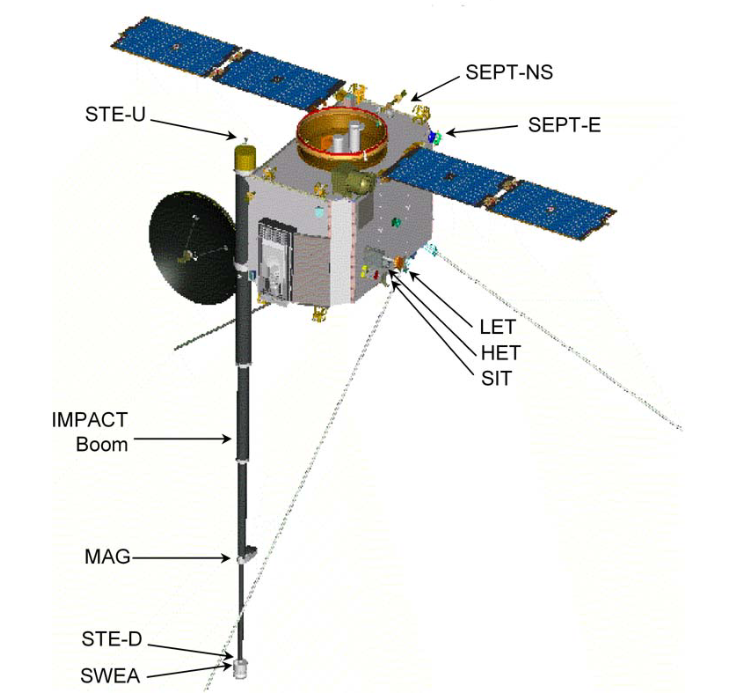}
\caption{Relative placement of the IMPACT boom on the STEREO \protect\citep[Figure~is from][Figure~1]{acuna2008stereo}}
\label{fig:stereoIMPACTboom}
\end{figure}

The twin Solar Terrestrial Relations Observatory (STEREO) A and B spacecraft were launched on October 26, 2006, from Kennedy Space Center \citep{kaiser2008stereo}. In heliospheric orbit at 1 AU, STEREO$-$A (Ahead) leads while STEREO$-$B (Behind) trails Earth. The two spacecraft separate at $44^{\circ}$ from each other annually, see Figure~\ref{fig:stereoAB}. This formation allows the stereoscopic image of the Sun from the Sun-Earth axis angle in 3D. The purposes of the mission are to study the cause and initiation mechanisms of coronal mass ejection (CME) and their propagation through the heliosphere, discover the sites and procedures of solar energetic particles in the corona as well as the interplanetary medium, and construct a three dimensional and time-dependent model of the parameters of the solar wind \citep{KAISER20051483}. 
The two spacecraft are each equipped with the complement of four scientific instruments, particularly two instruments and two instrument suites, with a total of 13 instruments on each spacecraft. The complements of the four instruments are as follows:\begin {itemize}
    \item 1. Sun-Earth Connection Coronal and Heliospheric
Investigation (SECCHI)
    \item 2. STEREO/WAVES (S/WAVES)
    \item 3. In situ Measurements of Particles and CME Transients (IMPACT) 
    \item 4. Plasma and Suprathermal Ion Composition
(PLASTIC)
\end{itemize}
The SECCHI suite of instruments has two white light coronagraphs, an extreme ultraviolet imager, and two heliospheric white light imagers for tracking CMEs \citep{howard2008sun}. S/WAVES instrument utilizes radio waves to trace locations of the CME-driven shocks and the 3-D open field lines for the particles created by solar flares \citep{bougeret2008s}. The PLASTIC instrument measures proton, the composition of heavy ions, and alpha particles in the solar wind plasma \citep{galvin2008plasma}. The IMPACT suite of instruments consists of seven instruments, and three of them are located on the 6-meter boom as shown in Figure~\ref{fig:stereoIMPACTboom} while the others are in the main hull of the spacecraft \citep{acuna2008stereo}. The IMPACT measures protons, heavy ions, and electrons, and the MAG magnetometer sensor in it measure the in situ magnetic fields in a range of $\pm 512 nT$ with 0.1 nT accuracy \citep{kaiser2007stereo}.  
Unfortunately, multiple hardware issues affecting control of the spacecraft resulted in the loss of contact with STEREO$-$B on October 1, 2014, \citep{NASASTEREO-B}.
\subsection{Wind}
\begin{figure}[H]
\centering
\includegraphics[width=0.6\textwidth]{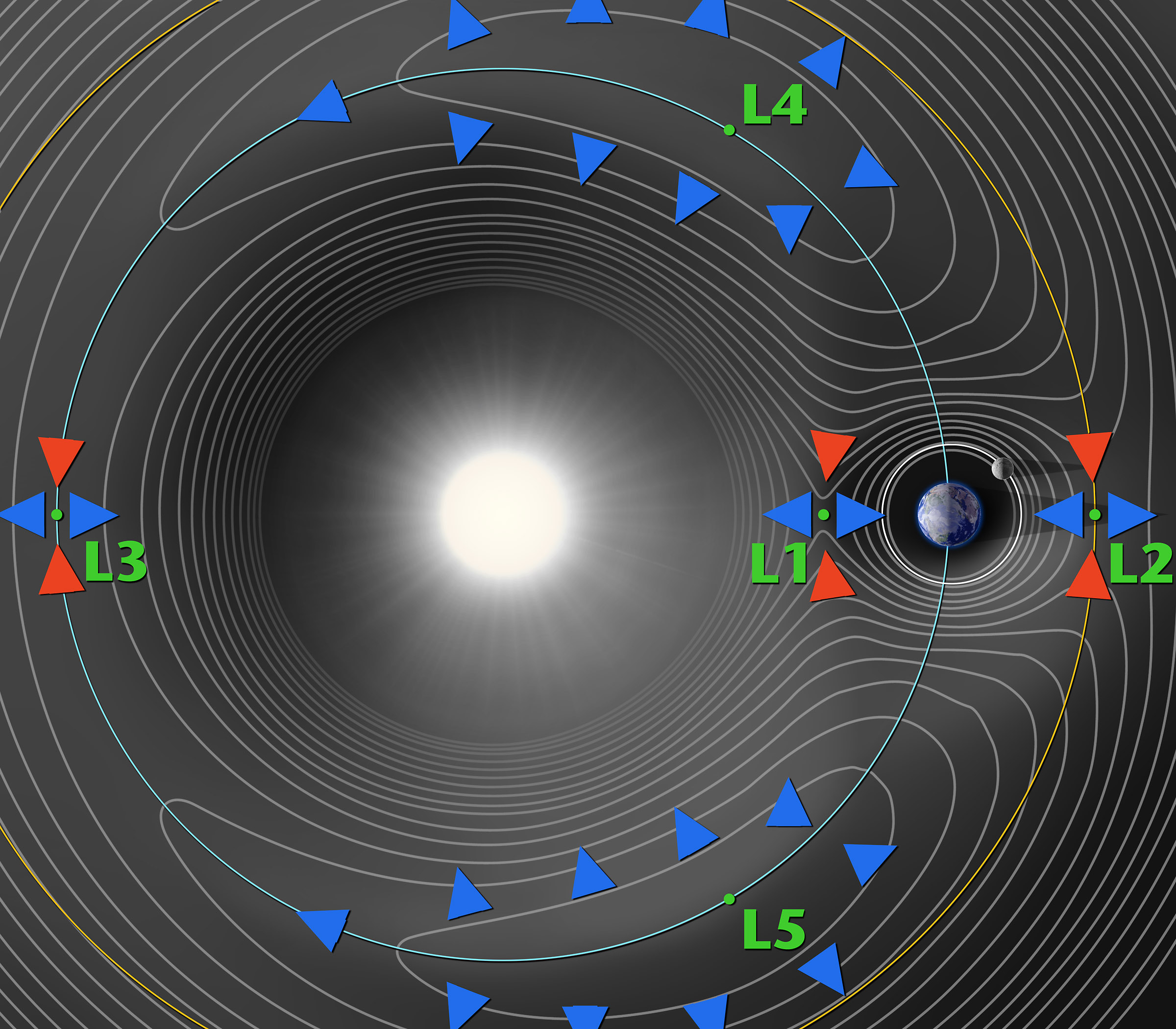}
\caption{Lagrangian points in the Sun-Earth system. \protect\citep[Credit:][]{NASALagranianpoints}}
\label{fig:lagrangianpoints}
\end{figure}

\begin{figure}[H]
\centering
\includegraphics[width=0.6\textwidth]{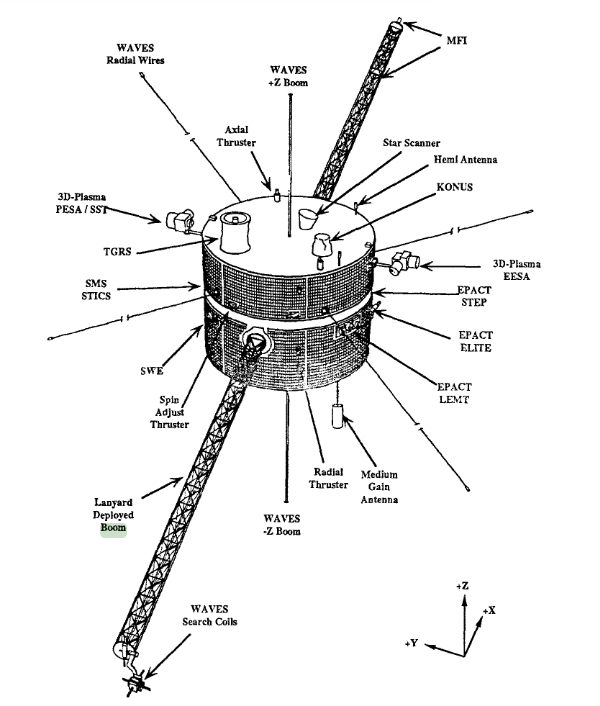}
\caption{Configuration of the Wind instruments. \protect\citep[Figure~is from][Figure~1]{harten1995design}}
\label{fig:windboom}
\end{figure}

NASA's Wind spacecraft was launched on November 1, 1994, \citep{wilson2021quarter}. The Wind was initially planned sent to $L_1$ Lagrangian point but was delayed to study the magnetosphere and lunar environment. Following a sequence of orbital adjustments, the Wind spacecraft was positioned in a Lissajous orbit close to the $L_1$ Lagrange point in early 2004 for studying the incoming solar wind on the verge of impacting Earth's magnetosphere \citep{WINDSpacecraft}, see Figure~\ref{fig:lagrangianpoints}.
The object of the mission is to study solar wind plasma, magnetic field, and solar and cosmic energetic particles. The spacecraft is equipped with the eight instruments such as Solar Wind Experiment \citep[SWE][]{ogilvie1995swe}, 3-D plasma \citep[3DP][]{wilson2011notes}, Magnetic Field Investigation \citep[MFI][]{lepping1995wind}, Solar wind/mass suprathermal ion composition studies \citep[SMS][]{gloeckler1995solar}, Energetic Particles: Acceleration, Composition Transport \citep[EPACT][]{von1995energetic}, Radio and Plasma Wave Experiment \citep[WAVES][]{bougeret1995waves}, \citep[KONUS][]{mazets1981recent} and Transient Gamma-Ray Spectrometer \citep[TGRS][]{owens1991transient}, see Figure~\ref{fig:windboom}. From these instruments, MFI and SWE are of interest to this thesis. The MFI consists of two magnetometers at the 12-meter boom, its measurement capability is 4$\,$nT, 65536$\,$nT, and measures vector magnetic field up in a time resolution of 22 or 11 vectors per second for the calibrated high-resolution data and primary science data is in time resolutions of three seconds, one minute and one hour \citep{WINDDataSources}. 
The SWE measures the solar wind key parameters such as velocity, density, and temperature.  
\subsection{ACE}
\begin{figure}[H]
\centering
\includegraphics[width=0.5\textwidth]{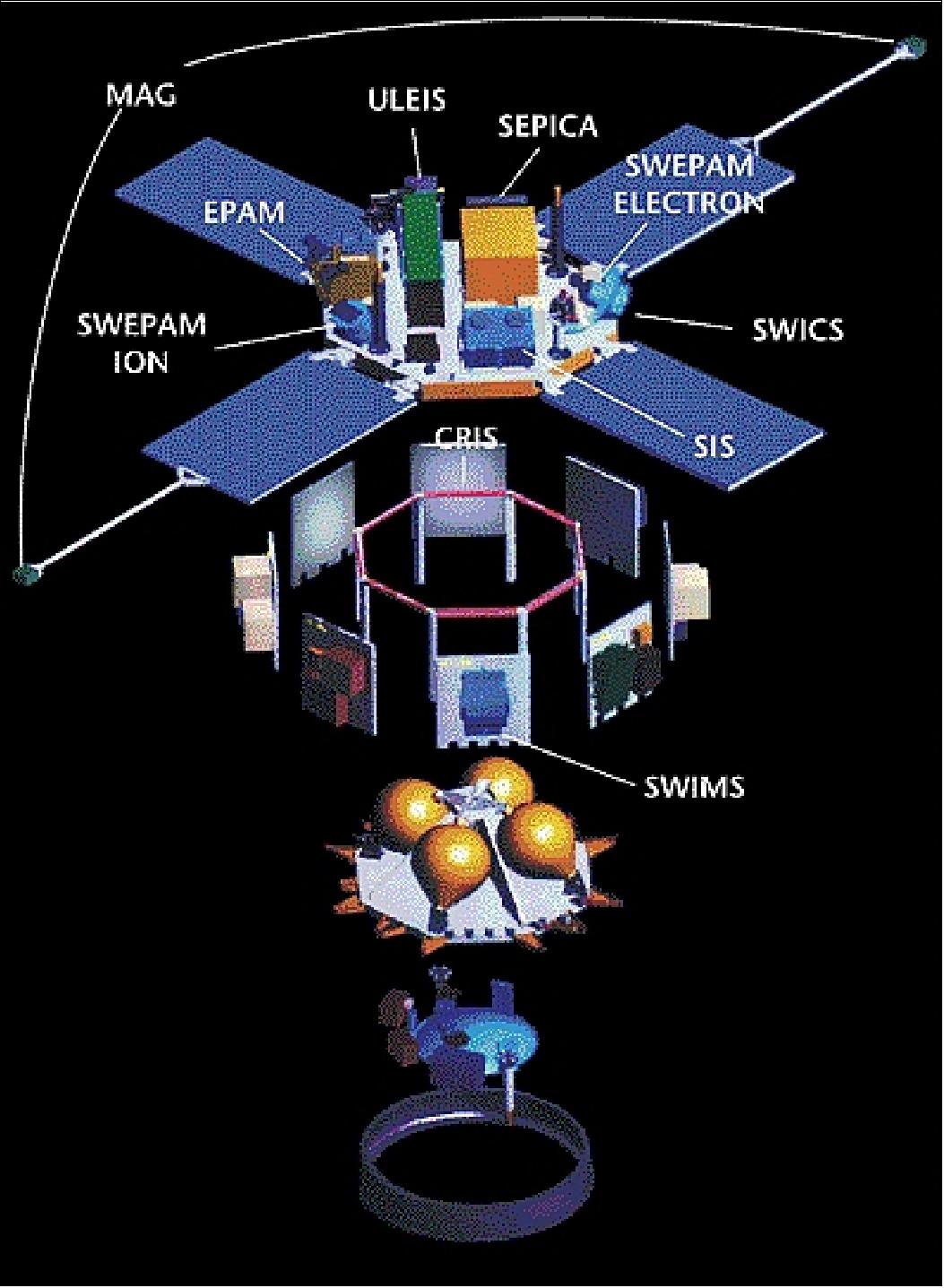}
\caption{Instruments of ACE spacecraft. \protect\citep[Figure~is from][Figure~6]{stone1998advanced}}
\label{fig:ace_spacecraft}
\end{figure}

NASA's ACE spacecraft was launched on August 25, 1997, \citep{stone1998advanced}. The spacecraft is located at $L_1$ Lagrangian point same as the Wind spacecraft, see Figure~\ref{fig:lagrangianpoints}. The general purposes of ACE are to gather and study particles originating from the sun, interplanetary or interstellar mediums, and the galaxy as well as to investigate the solar wind structures such as ICMEs and magnetic clouds. 
The spacecraft is equipped with nine primary scientific instruments and one engineering instrument such as Cosmic-Ray Isotope Spectrometer \citep[CRIS][]{stone1998cosmic}, Electron, Proton, and Alpha-particle Monitor (EPAM), Magnetometer \citep[MAG][]{smith1998ace}, Real-Time Solar Wind \citep[RTSW][]{zwickl1998noaa}, Solar Energetic Particle Ionic Charge Analyzer \citep[SEPICA][]{mobius86solar}, Solar Isotope Spectrometer \citep[SIS][]{stone1998solar}, Solar Wind Electron, Proton and Alpha Monitor \citep[SWEPAM][]{mccomas1998solar}, Solar Wind Ion Composition Spectrometer (SWICS) and Solar Wind Ion Mass Spectrometer \citep[SWIMS][]{gloeckler1998investigation} and Ultra-Low-Energy Isotope Spectrometer \citep[ULEIS][]{mason1998ultra}, see Figure~\ref{fig:ace_spacecraft}.
 The MAG consists of twin triaxial flux-gate magnetometers such that magnetometer sensors have between 3 and 6 vectors $s^{-1}$ resolutions for continuous observation of the interplanetary magnetic field \citep{smith1998ace}.
\subsection{Cluster}
\begin{figure}[!t]
\centering
\includegraphics[width=0.6\textwidth]{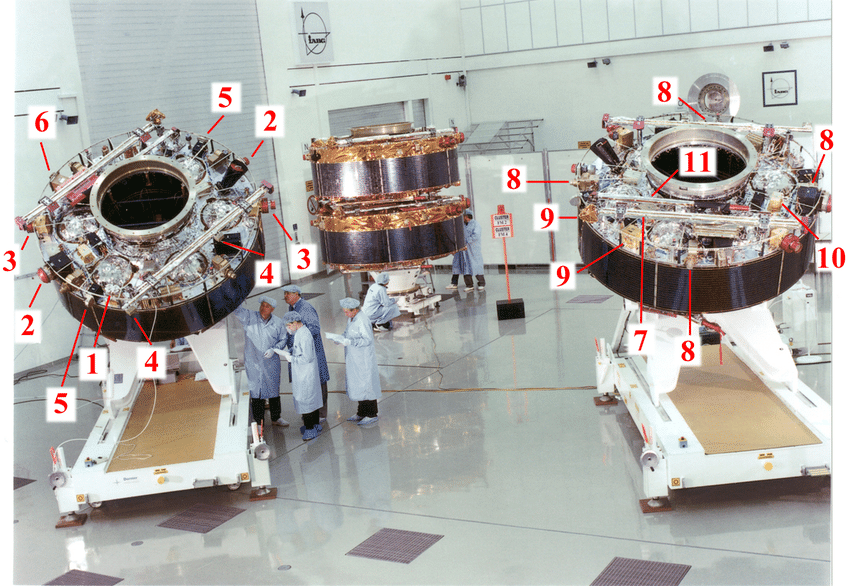}
\caption{Instruments on Cluster satellites. ASPOC
(1), CIS (2), EDI (3), FGM (4), PEACE (5), RAPID (6), DWP (7), EFW (8), STAFF (9), WBD (10), WHISPER (11) \protect\citep[Figure~is from][Figure~3]{escoubet2001cluster}}
\label{fig:clusterinstruments}
\end{figure}

ESA's Cluster constellations consist of 4 satellites, which were launched on 16 July and 9 August 2000 \citep{escoubet2001cluster}. The Cluster satellites orbit in a tetrahedral formation around Earth. The orbits feature perigees close to 4 Earth radii ($R_E$) and apogees approximately 19.6 $R_E$ away \citep{zhang2010case}. The primary objectives of the Cluster mission involve examining small-scale plasma formations and macroscopic turbulences in three dimensions in crucial plasma areas, including solar wind, bow shock, magnetopause, polar cusps, magnetotail, and auroral regions \citep{escoubet2001cluster}. The four satellites are each equipped with 11 instruments such as Active Spacecraft Potential Control \citep[ASPOC][]{torkar2016active}, Ion Composition \citep[CIS][]{reme1997cluster}, Electron Drift Instrument \citep[EDI][]{haaland2007high}, Fluxgate Magnetometer \citep[FGM][]{balogh1997cluster}, Plasma Electron And Current Experiment \citep[PEACE][]{johnstone1997peace}, Research with Adaptive Particle Imaging Detectors \citep[RAPID]{daly2010rapid}, Digital Wave processor \citep[DWP][]{woolliscroft1997digital}, Electric field and waves  \citep[EFW][]{gustafsson1997electric}, Spatio-Temporal Analysis of Field Fluctuations \citep[STAFF][]{cornilleau1997cluster}, Wide-band plasma wave \citep[WBD][]{gurnett1997wide}, and Waves of High frequency and Sounder for Probing of Electron Density \citep[WHISPER][]{decreau1997whisper} as shown in Figure~\ref{fig:clusterinstruments}. From the instruments, the magnetic and plasma parameters are up to our interest. Hence, the FGM magnetometer, CIS, and EFW instruments are emphasized. The Fluxgate Magnetometer (FGM) is composed of two tri-axial fluxgate magnetometers, which are installed on one of the two 5-meter radial booms.
 It measures in the dynamic range $\pm 65,536$nT. At the highest dynamic level, the resolution is  $\pm8$ nT, and the time resolution is 100 vectors per second \citep{wmo_fgm_2021}. CIS (Ion composition) instrument measures three-dimensional ion distribution, and it is composed of two distinct sensors: the Composition Distribution Function (CODIF) sensor and the Hot Ion Analyzer (HIA) sensor. The CIS experiment is not operational for Cluster-2 and the HIA sensor is switched off for Cluster-4 due to a problem with the high voltage of the electrostatic analyzer \citep{caa_cis_user_guide_2021}. Hence, for Cluster-2 and Cluster-4, the Electric Field and Wave (EFW) instrument measurement is useful. The EFW instrument measures the electric field fluctuations as well as the spacecraft potential, which is essential for studying plasma density and spacecraft charging. 
\section{Methods}
The following two methods are suitable for single-spacecraft measurements and both of them are based on the divergence of the magnetic field, $\nabla \cdot \mathbf{B} = 0$, constraint, and both of them use solely magnetic field data.
 The following formalization, descriptions, and derivations are based on \citep{sonnerup1998minimum} and \citep{paschmannshock}.
\subsection{Minimum Variance Analysis}
The Minimum variance analysis technique was first developed by \citep{sonnerup1967magnetopause}. 
It is based on the assumption that variations in the magnetic field would be observed when a single spacecraft passes through a 1-D\footnote{In reality, it is a 2-D or 3-D transition layer} current layer or wavefront. Remembering the divergence of the magnetic field constraint, $\nabla \cdot \mathbf{B} = 0$, such that the normal component of the magnetic field must remain constant. If such a normal direction can be found, then the variations in the magnetic field are zero or at the least has a minimum variance. Thus, $\uvec{n}$ can be determined by the minimization of the following equation: 
\begin{align}\label{eq:min}
    \sigma^2 = \frac{1}{M}\sum_{m=1}^M\bigg|(\mathbf{B}^{(m)}-\langle \mathbf{B}\rangle) \cdot \uvec{n}\bigg|^2,
\end{align}
where ${\mathbf{B}^{(m)}}, (m=1,2,3...M)$ is the magnetic field in a time series data and $\mathbf{B}$ is the average magnetic field.

\begin{figure}[H]
\centering
\includegraphics[width=0.6\textwidth]{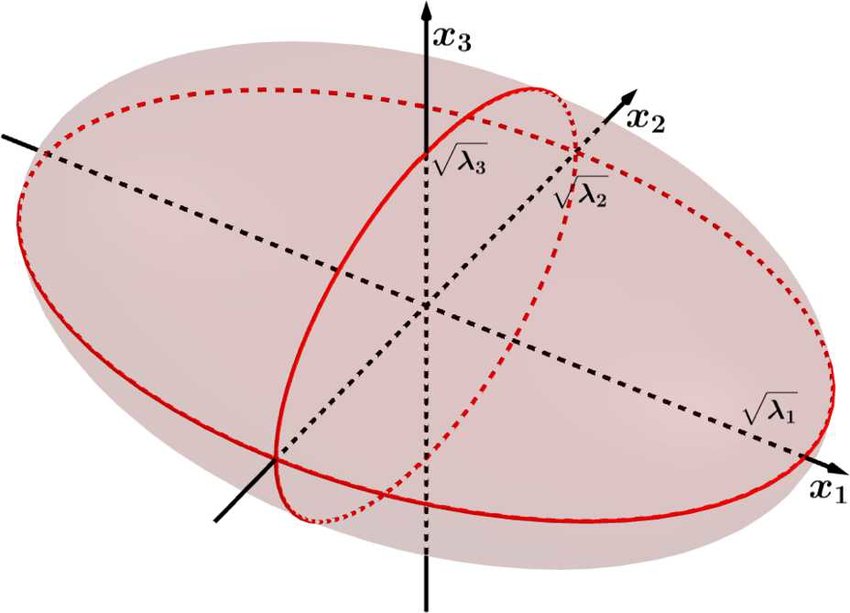}
\caption{The illustration of the variance space ellipsoid geometry \protect\citep[Figure~is from][Figure~1]{rosa2020new}}
\label{fig:mvaellipsoid}
\end{figure}

Taking the account of the normalization constraint $|\uvec{n}|^2=1$, on which minimization is conditioned, and introducing a Lagrange multiplier, $\lambda$, for utilizing the constraint, the solution of three homogeneous linear equations can be found. The three homogeneous equations are:
\begin{equation}\label{eq:homogeneouseq}
    \begin{aligned}
    \frac{\partial }{\partial n_X}\bigg(\sigma^2-\lambda|\uvec{n}|^2-1\bigg) = 0\\
    \frac{\partial }{\partial n_Y}\bigg(\sigma^2-\lambda|\uvec{n}|^2-1\bigg) = 0\\
    \frac{\partial }{\partial n_Z}\bigg(\sigma^2-\lambda|\uvec{n}|^2-1\bigg) = 0, 
    \end{aligned}
\end{equation}
where $\sigma^2$ is given by \ref{eq:min} and $\uvec{n}$ is expressed in its three components, which are along X, Y, and Z coordinates. 
After the differentiation procedures of \ref{eq:homogeneouseq} are done, the resulting sets of equations can be written in matrix form:
\begin{align}\label{eq:variancematrix}
    \sum_{\nu=1}^3 M_{\mu\nu} n_{\nu} = \lambda n_{\mu},
\end{align}
where the subscripts $\mu$ and $\nu$ indicate cartesian coordinates and $M_{\mu\nu}$ is the magnetic variance matrix: 
\begin{align}
    M_{\mu\nu} n_{\nu} = <B_{\mu} B_{\nu}> - <B_{\mu}><B_{\nu}> 
\end{align}
with $\lambda$ being the eigenvalues with three possible values $\lambda_1$, $\lambda_2$, and $\lambda_3$ with decreasing order. The eigenvalues have their corresponding eigenvectors $\mathbf{x_1}$, $\mathbf{x_2}$, and $\mathbf{x_3}$ of the matrix where they represent the directions of maximum, intermediate, and minimum variance of the magnetic field component. The eigenvectors corresponding to the smallest value of the $\lambda$ eigenvalues are the shock normal vectors.
 The eigenvalue ratios, especially intermediate to minimum eigenvalue ratio, must be greater than 2 or 3 to keep the variance space ellipsoid \citep{paschmann1998analysis} as shown in Figure~\ref{fig:mvaellipsoid} 
\subsection{The Magnetic Coplanarity method}
If two vectors that lie inside a coplanar surface can be determined, the normal to the coplanar surface can be found.

The coplanarity methods are based on the magnetic coplanarity theorem, which stated that both sides of the magnetic field vectors on the shock and the shock normal lie in the same plane. Similarly, the velocity on both sides or, in other words, the velocity jump through the shock also lie in the same plane. 
The upstream and downstream magnetic fields and velocities satisfy the Rankine-Hugionot conditions.
 So, there are multiple vectors lie in the shock plane, and the resulting constraint equations are as follows:
\begin{align}\label{eq:cp1}
    (\mathbf{B}_{down}-\mathbf{B}_{up})\cdot \uvec{n} = 0
\end{align}
\begin{align}\label{eq:cp2}
    (\mathbf{B}_{down}\cross \mathbf{B}_{up})\cdot \uvec{n} = 0
\end{align}
\begin{align}\label{eq:mixed1}
    (\mathbf{B}_{up}\cross (\mathbf{V}_{down} - \mathbf{V}_{up}))\cdot \uvec{n} = 0
\end{align}
\begin{align}\label{eq:mixed2}
    (\mathbf{B}_{down}\cross (\mathbf{V}_{down} - \mathbf{V}_{up}))\cdot \uvec{n} = 0
\end{align}
\begin{align}\label{eq:mixed3}
    ((\mathbf{B}_{up}-\mathbf{B}_{down})\cross (\mathbf{V}_{down} - \mathbf{V}_{up}))\cdot \uvec{n} = 0, 
\end{align}
where the indexes up and down denote upstream and downstream, respectively.
From \ref{eq:cp1} and \ref{eq:cp2} the magnetic coplanarity normal is defined:
\begin{align}\label{eq:coplanarity}
\uvec{n}_{MC}= \pm \frac{(\mathbf{B}_{down}\cross \mathbf{B}_{up})\cross (\mathbf{B}_{down}-\mathbf{B}_{up})}{\abs{(\mathbf{B}_{down}\cross \mathbf{B}_{up})\cross (\mathbf{B}_{down}-\mathbf{B}_{up})}},
\end{align}
where the signs are arbitrary. 

\begin{figure}[H]
\centering
\includegraphics[width=0.6\textwidth]{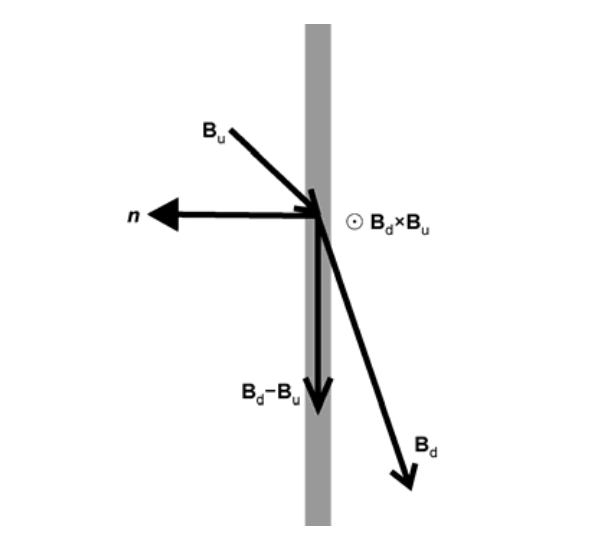}
\caption{The illustration of the magnetic coplanarity. The gray vertical line represents the shock layer. \citep[Figure~is from][Figure~1]{shan2013comparison}}
\label{fig:coplanaritymethod}
\end{figure}

The magnetic coplanarity schematic illustration is shown in Figure~\ref{fig:coplanaritymethod}.
Similarly, from \ref{eq:mixed1}, \ref{eq:mixed2} and \ref{eq:mixed3} the mixed modes for normals can be defined, respectively.
\begin{align}
    \uvec{n}_{mix1} = \pm \frac{(\mathbf{B}_{up}\cross (\mathbf{V}_{down}-\mathbf{V}_{up}))\cross (\mathbf{B}_{down}-\mathbf{B}_{up})}{\abs{(\mathbf{B}_{up}\cross (\mathbf{V}_{down}-\mathbf{V}_{up}))\cross (\mathbf{B}_{down}-\mathbf{B}_{up})}}
\end{align}
\begin{align}
    \uvec{n}_{mix2} = \pm \frac{(\mathbf{B}_{down}\cross (\mathbf{V}_{down}-\mathbf{V}_{up}))\cross (\mathbf{B}_{down}-\mathbf{B}_{up})}{\abs{(\mathbf{B}_{down}\cross (\mathbf{V}_{down}-\mathbf{V}_{up}))\cross (\mathbf{B}_{down}-\mathbf{B}_{up})}}
\end{align}
\begin{align}
    \uvec{n}_{mix3} = \pm \frac{((\mathbf{B}_{down}-\mathbf{B}_{up})\cross (\mathbf{V}_{down}-\mathbf{V}_{up}))\cross (\mathbf{B}_{down}-\mathbf{B}_{up})}{\abs{((\mathbf{B}_{down}-\mathbf{B}_{up})\cross (\mathbf{V}_{down}-\mathbf{V}_{up}))\cross (\mathbf{B}_{down}-\mathbf{B}_{up})}}
\end{align}
From these, the equation \ref{eq:coplanarity} is used and as denoted further on as (CP). 
The method of magnetic coplanarity is straightforward to implement and, as previously mentioned, it only necessitates the use of a magnetic field data \citep{paschmann1998analysis}.
\subsection{The Utilization of the Methods}
By comparing the two methods, the upstream and downstream time intervals of the magnetic field measurements are set. 
In this thesis work, as stated before the magnetic coplanarity method \ref{eq:coplanarity} is used out of the other coplanarity methods.

To accept the intervals, there are some criteria to be put in such as:
\begin{itemize}
  \item The angle between the vectors defined by the minimum variance analysis (MVA) and the magnetic coplanarity method must be less than  $15^{\circ}$ \citep{facsko2008statistical, facsko2009global, facsko2010study}.
  \item The ratio between the intermediate eigenvalue and the smallest eigenvalue should be greater than 2 \citep{facsko2008statistical, facsko2009global, facsko2010study}, the same for more data points and greater than 10 for data points less than 50 \cite{sonnerup1967magnetopause} or the ratio between the smallest eigenvalue to the intermediate eigenvalue should be smaller than 1/3 \citep{shan2013comparison}, which is in reverse means greater than 3. 
\end{itemize}
\subsection{Estimating the solar wind parameters}\label{sec:estimatingparameters}
The following parameter estimations are based on this paper \citep{lumme2017database}
\paragraph{Shock criteria:}
The solar wind bulk speed jump should fulfill the following conditions:
\begin{align}\label{eq:criteria1}
    V_{down}-V_{up} \geq 20\quad\text{km/s for FF}\quad\text{and}\quad
    V_{up}-V_{down} \geq 20\quad\text{km/s for FR}
\end{align}
And downstream to upstream ratios should fulfill the following ratios of upstream and downstream the magnetic field, density, and temperature respectively:
\begin{align}\label{eq:criteria2}
    \frac{B_{down}}{B_{up}}\geq 1.2\quad
    \frac{N^{down}_p}{N^{up}_p}\geq 1.2\quad\text{and}\quad
    \frac{T^{down}_p}{T^{up}_p}\geq \frac{1}{1.2}
\end{align}
\paragraph{Shock theta:}
This is the angle between the normal vector $\uvec{n}$ and the upstream magnetic field lines:
\begin{align}\label{eq:shocktheta}
    \theta_{Bn} = cos^{-1}\bigg (\abs{\frac{\mathbf{B}_{up}\cdot\uvec{n}}{|\mathbf{B}_{up}||\uvec{n}|}}\bigg)
\end{align}
\paragraph{Shock speed:}
The shock speed in the spacecraft frame of reference:
\begin{align}
    V_{sh}=\bigg |\frac{N^{up}_p\mathbf{V}_{up}-N^{down}_p\mathbf{V}_{down}}{N^{down}_p - N^{up}_p}\cdot\uvec{n}\bigg|
\end{align}
\paragraph{Upstream sound speed:}
\begin{align}\label{eq:sound}
    C^{up}_s=\bigg < \sqrt{\gamma k_B \frac{T_p+T_e}{m_p}}\bigg >,
\end{align}
where $m_p$ is the proton mass, $T_p$ and $T_e$ are proton and electron temperature. In the solar wind, the electron temperature at 1 AU is assumed to be $\sim 140,000$ K \citep{newbury1996electron}.
\paragraph{Upstream Alfvén speed:}
\begin{align}\label{eq:Alfven}
    V^{up}_A = \bigg < \frac{B}{\sqrt{\mu_0 N_p m_p}}\bigg>,
\end{align}
where $\mu_0$ is the magnetic vacuum permeability, $N_p$ is the proton density and $m_p$ is the proton mass.
\paragraph{Upstream magnetosonic speed:}
The equation \ref{eq:magetosonic} is used for here
\paragraph{Upstream plasma beta:}
\begin{align}\label{eq:plasmabeta}
    \beta^{up} = \bigg < \frac{2\mu_0 k_B N_p(T_p+T_e)}{B^2}\bigg>
\end{align}
\paragraph{Alfvén Mach number:}
\begin{align}\label{eq:alfvenmach}
    M_A = \frac{|\mathbf{V}_{up}\cdot \uvec{n}\pm \mathbf{V}_{sh}|}{V^{up}_A},
\end{align}
where $V_{up}$ is the upstream velocity, $V_{sh}$ is shock speed and $V^{up}_A$ is the Alfvén speed. The reason for $\pm V_{sh}$ is a Galilean coordinate transformation to the shock rest frame. The sign $\pm$ is depend on FF shock, for which (-) or FR shock, for which (+).
\paragraph{Magnetosonic Mach number:}
Similarly to \ref{eq:alfvenmach} the Magnetosonic Mach number is defined as follows:
\begin{align}\label{eq:msmach}
    M_{ms} = \frac{|V_{up}\cdot \uvec{n}\pm V_{sh}|}{C_{ms}},
\end{align}
where $C^{up}_{ms}$ is the magnetosonic upstream speed.
\section{Geomagnetic activity index Kp}
Changes in solar activity and solar wind disturb Earth's magnetosphere and cause fluctuations. The Ground-based magnetometers observe these variations in the magnetosphere. These geomagnetic activities are expressed by various types of geomagnetic indices \citep{rostoker1972geomagnetic}. One such method is the planetary K index as known as the Kp index. \citep{bartels1949standardized} introduced this index and K-stands for "Kennziffer" (the German word for "characteristic number") and "p" denotes planetary, representing global magnetic activity. The Kp value is calculated as the average of the standardized K-indices, measured every three hours across the 13 designated Kp observatories \citep{matzka2021geomagnetic}. The National Oceanic and Atmospheric Administration (NOAA) uses the Kp-index to categorize geomagnetic storms on a scale known as the G-scale. This scale extends from minor storms, represented as G1, which correlates with Kp=5, to extreme storms, classified as G5, which corresponds to a Kp=9 \url{https://www.swpc.noaa.gov/noaa-scales-explanation}.
 For determining the Kp index on May 07 and April 23, 2007, I used data from \url{https://kp.gfz-potsdam.de}.
\section{Data acquiring process}
\begin{figure}[H]
\centering
\includegraphics[width=0.8\textwidth]{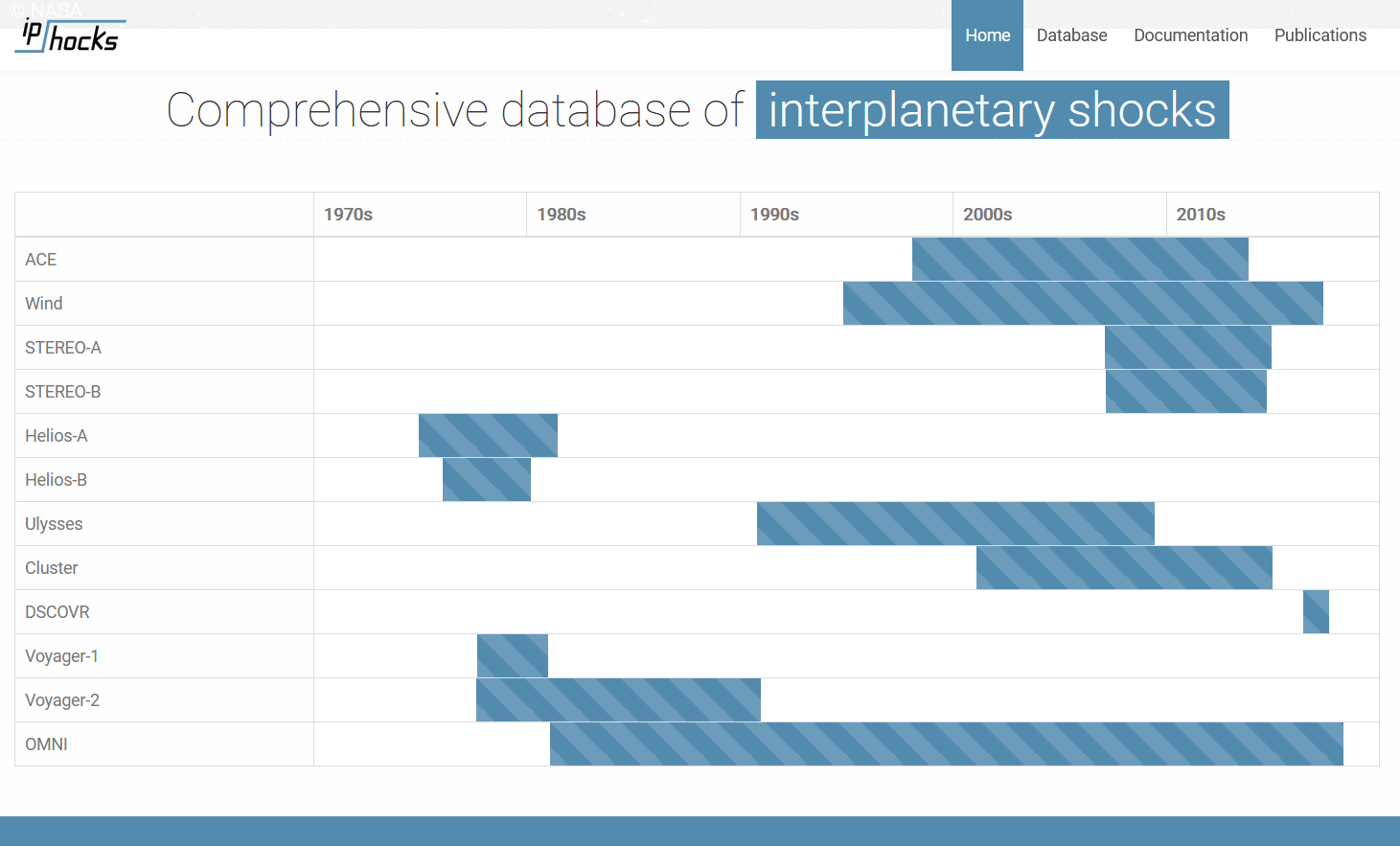}
\caption{Identified shocks detected by the spacecraft in the list}
\label{fig:ipshocksFinland}
\end{figure}

There are several steps for the data acquiring process.
 First of all, to determine the propagation of shocks, shocks that occurred in a single day are needed. For this purpose, shock candidates are chosen from the shock lists in the Database of Heliospheric Shock Waves maintained at the University of Helsinki \url{http://www.ipshocks.fi/} as shown in Figure~\ref{fig:ipshocksFinland}. 
For the event selection I chose the year 2007 for the reason that STEREO$-$A and STEREO$-$B were closer to each other as well as to the Sun-Earth line. Hence, the two events are from this year, particularly on May 07, 2007, and April 23, 2007. In the first event, May 07 of 2007, the selected spacecraft are STEREO$-$A, STEREO$-$B, Wind, and the four cluster satellites: Cluster-1 (C1), Cluster-2 (C2), Cluster-3 (C3), and Cluster-4 (C4), while for the second event, April 23, 2007, the spacecraft are STEREO$-$A and B, ACE, and Wind. 
After choosing the shock candidates, I downloaded the shock data from NASA's Coordinated Data Analysis Web (CDAWeb \url{https://cdaweb.gsfc.nasa.gov/}).
 I obtained the STEREO$-$A and B magnetic field observation and the plasma data from the STEREO In-situ Measurements of Particles and CME Transients magnetic field experiment \citep[IMPACT][]{luhmann2008stereo} with the time resolution of 100 ms and the STEREO PLAsma and SupraThermal Ion Composition \citep[PLASTIC][]{galvin2008plasma} respectively. The magnetic field and the plasma data of the STEREO$-$A and B are in the RTN\footnote{It is the spacecraft coordinate system and R is radially outward from the Sun, T is along the planetary orbital, and N is northward direction} coordinate system.
 I obtained the Wind magnetic and plasma data from the Magnetic Field Investigation instrument \citep[MFI][]{lepping1995wind} with a time resolution of 3 sec and The Solar Wind Experiment \citep[SWE][]{ogilvie1995swe} with a time resolution of 1 min respectively. The magnetic field and the plasma data of the Wind are in the Geocentric Solar Ecliptic System (GSE)\footnote{X-axis is pointing to the Sun from the Earth, Y-axis is in the ecliptic plane against the planetary motion and Z-axis is northward direction} coordinate system.
 I obtained the ACE magnetic and plasma data from the ACE Magnetic Field Experiment \citep[MAG][]{smith1998ace} with a time resolution of 1 sec and the Solar Wind Electron Proton Alpha Monitor \citep[SWEPAM][]{mccomas1998solar} with the time resolution of 64 sec, respectively. The magnetic field and the plasma data are in both the RTN and GSE coordinate systems.
 I acquired the magnetic data of Cluster satellites from the Cluster Fluxgate Magnetometer \citep[FGM][]{balogh1997cluster} with a time resolution of four sec for all the four Cluster satellites, ion data from the Cluster Ion Spectrometry \citep[CIS]{reme1997cluster} with a time resolution of 4 sec for the Cluster-1 and Cluster-3 satellites, and the spacecraft potential data from Electric Fields and Waves \citep[EFW][]{gustafsson1997electric} with a time resolution of 4 sec for the Cluster-2 and Cluster-4 satellites. All the Cluster data were in the GSE coordinate system.  

\begin{figure}[H]
\centering
\includegraphics[width=1.0\textwidth]{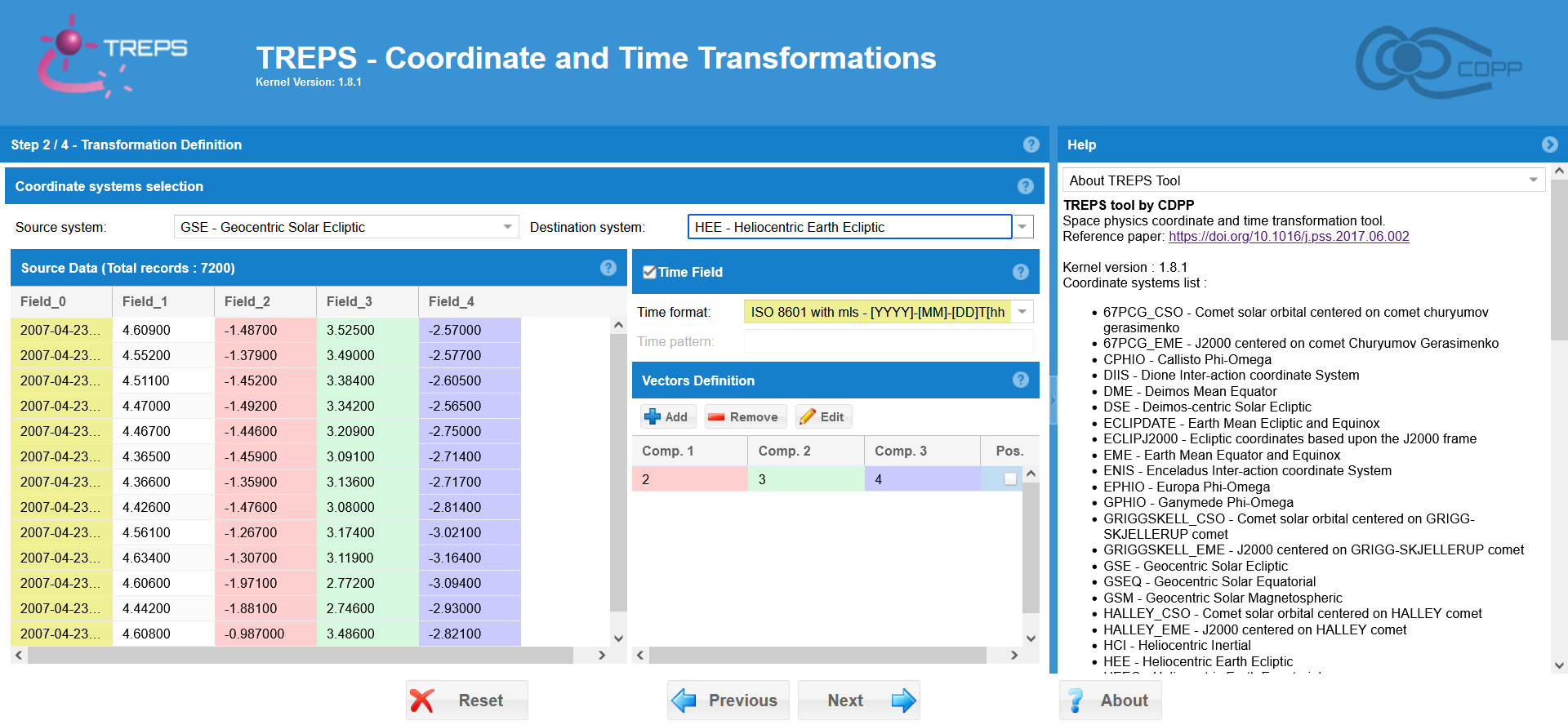}
\caption{TREPS interface, showing a transformation from the GSE to the HEE coordinate system. The green colored column denotes the time field and the other three colored columns denote X, Y, and Z vector components}
\label{fig:TREPS}
\end{figure}

After obtaining the data, I transformed the coordinate system transformation in such a way that all the coordinate systems are changed to the Heliocentric Earth Ecliptic (HEE)\footnote{In which the X-axis is toward the Earth from the Sun, Z-axis is northward direction. This system is fixed with respect to the Sun-Earth line} coordinate system. To transform from the RTN and the GSE coordinate systems to the HEE coordinate system, I used Transformation de REpères en Physique Spatiale \href{http://treps.irap.omp.eu/}{(TREPS)} online tool. The TREPS tool, which is developed by the French Plasma Physics Data Centre (CDPP), the national data center in France for the solar system plasmas, is based on SPICE (Spacecraft, Planet, Instrument, C-matrix, and Events) information system kernels created by NASA/NAIF. Its use is simple, with a 4-step process, which includes importing data, selecting the original coordinate system of the data and a coordinate system to transform, Figure~\ref{fig:TREPS}, choosing how the transformed vector should be exported, and finally, exporting the transformed file \citep{genot2018treps}.
\chapter{Discussion}
\section{Observations}
With the visual inspection of the magnetic field and plasma data, I identified jumps in magnetic field B, speed V, density N, and temperature T. Using these variations, I determined the times at which shocks occurred for each spacecraft data set in each event. From the plots, the shocks detected by the spacecraft in each event are believed to be propagation of the same shock due to their timing and the similarities of the magnetic field and plasma data profile with one another.
 The following case studies of two events are not orderly defined for one another. First, I analyzed the shock event on May 07, 2007, and then the event on April 23, 2007. Hence, the order follows this fashion.  
\subsection{The Event May 07, 2007}
\begin{figure}[htp]
\centering
\includegraphics[width=1.\textwidth]{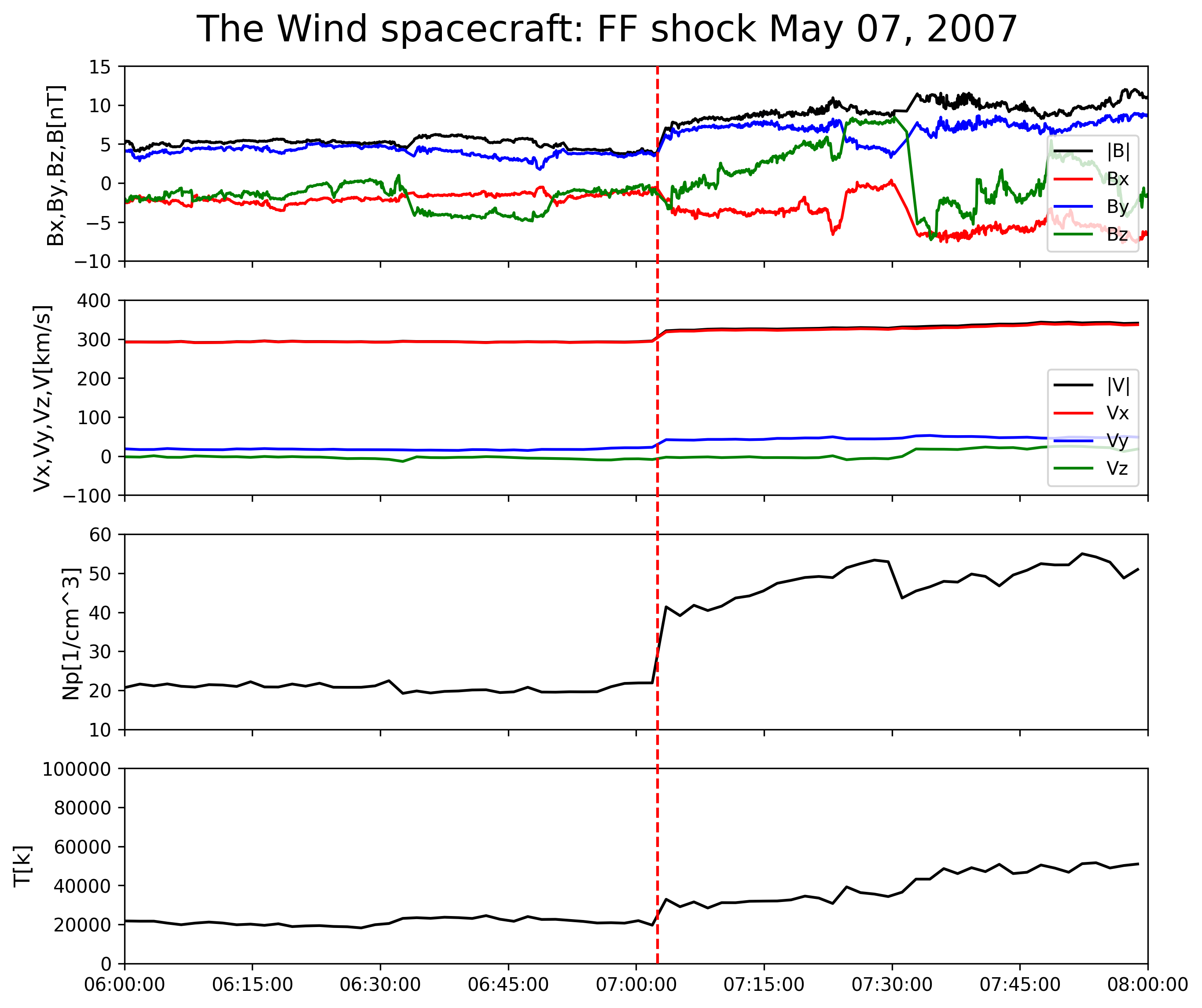}
\caption{The plot of the shock detected by the Wind spacecraft on May 07, 2007, at \textbf{07:02:30} UTC. FF stands for the fast forward shock, which means the shock is traveling away from its driver. The panels show from top to bottom, the magnetic field magnitude as well as its components, the total velocity, and its components, density, and temperature. The dashed red line represents the exact shock time. The duration of the plot is two hours.}
\label{fig:WindIP0507}
\end{figure}

\begin{figure}[htp]
\centering
\includegraphics[width=1.\textwidth]{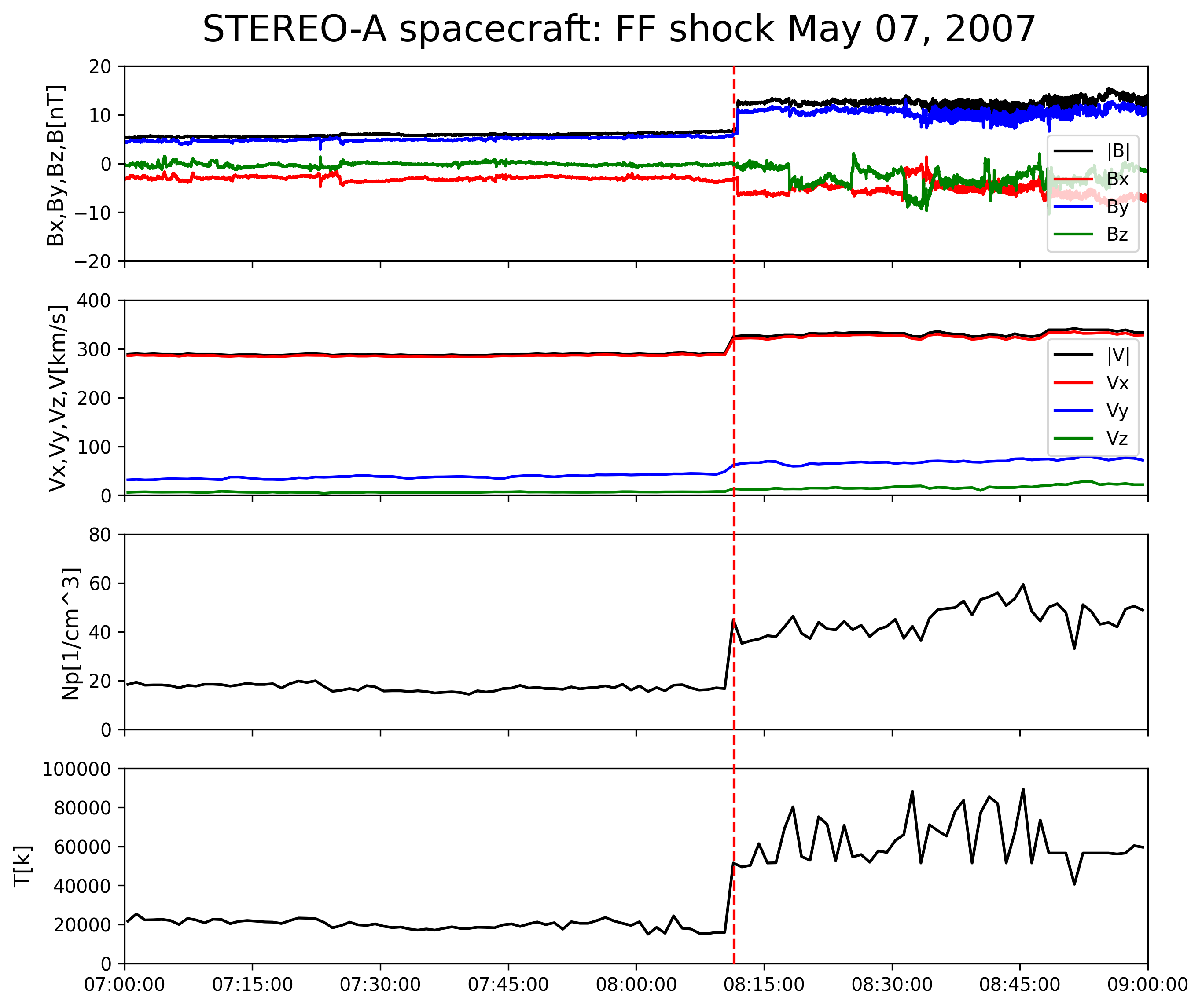}
\caption{The plot of the shock detected by the STEREO$-$A spacecraft on May 07, 2007, at \textbf{08:11:30} UTC. The symbols and details of the figures are the same as \ref{fig:WindIP0507}. The duration of the plot is two hours.}
\label{fig:staIP0507}
\end{figure}

\begin{figure}[htp]
\centering
\includegraphics[width=1.\textwidth]{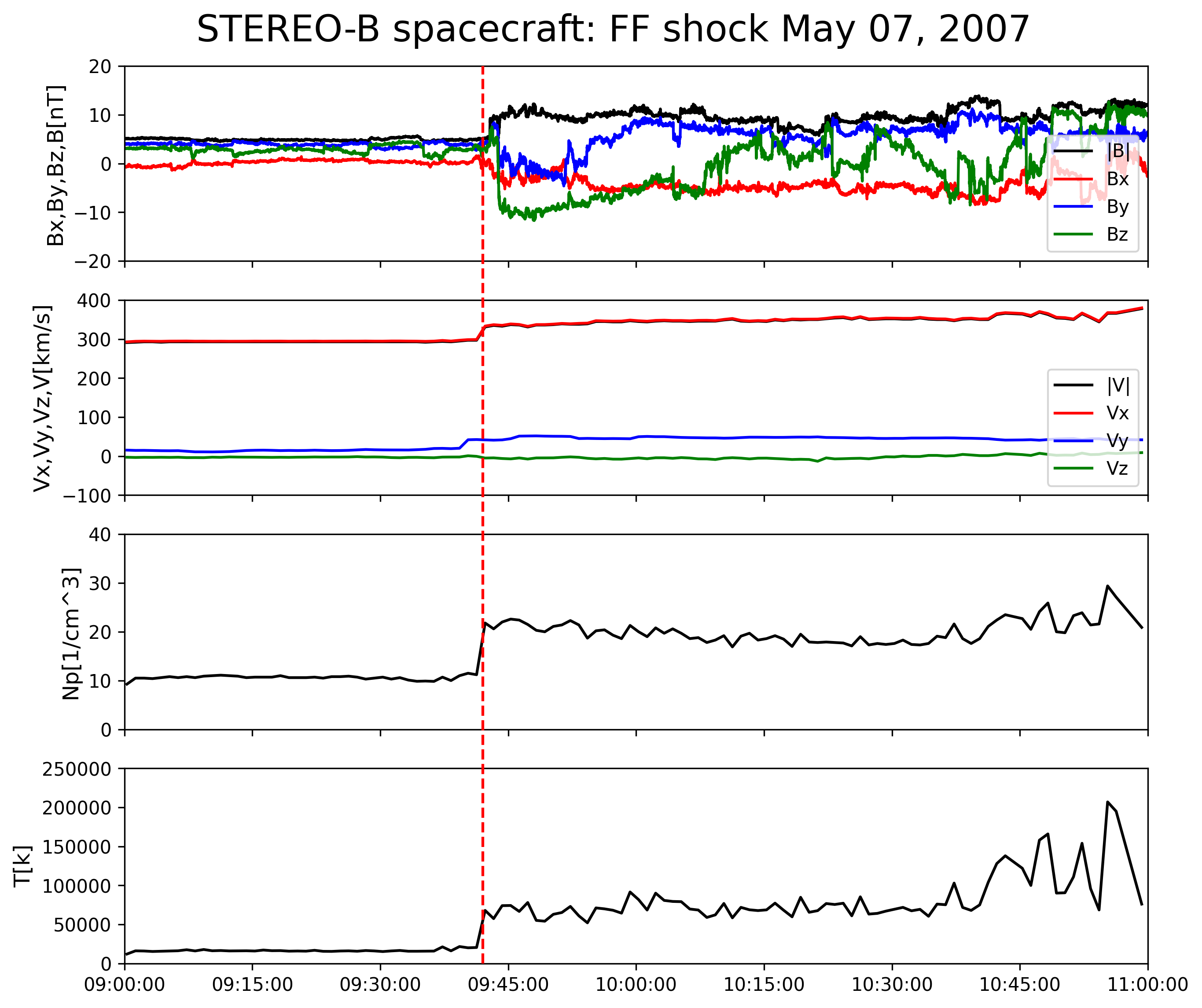}
\caption{The plot of the shock detected by the STEREO$-$B spacecraft on May 07, 2007, at \textbf{09:42:00} UTC. The symbols and details of the figures are the same as \ref{fig:WindIP0507}. The duration of the plot is two hours.}
\label{fig:stbIP0507}
\end{figure}

\begin{figure}[htp]
\centering
\includegraphics[width=1.\textwidth]{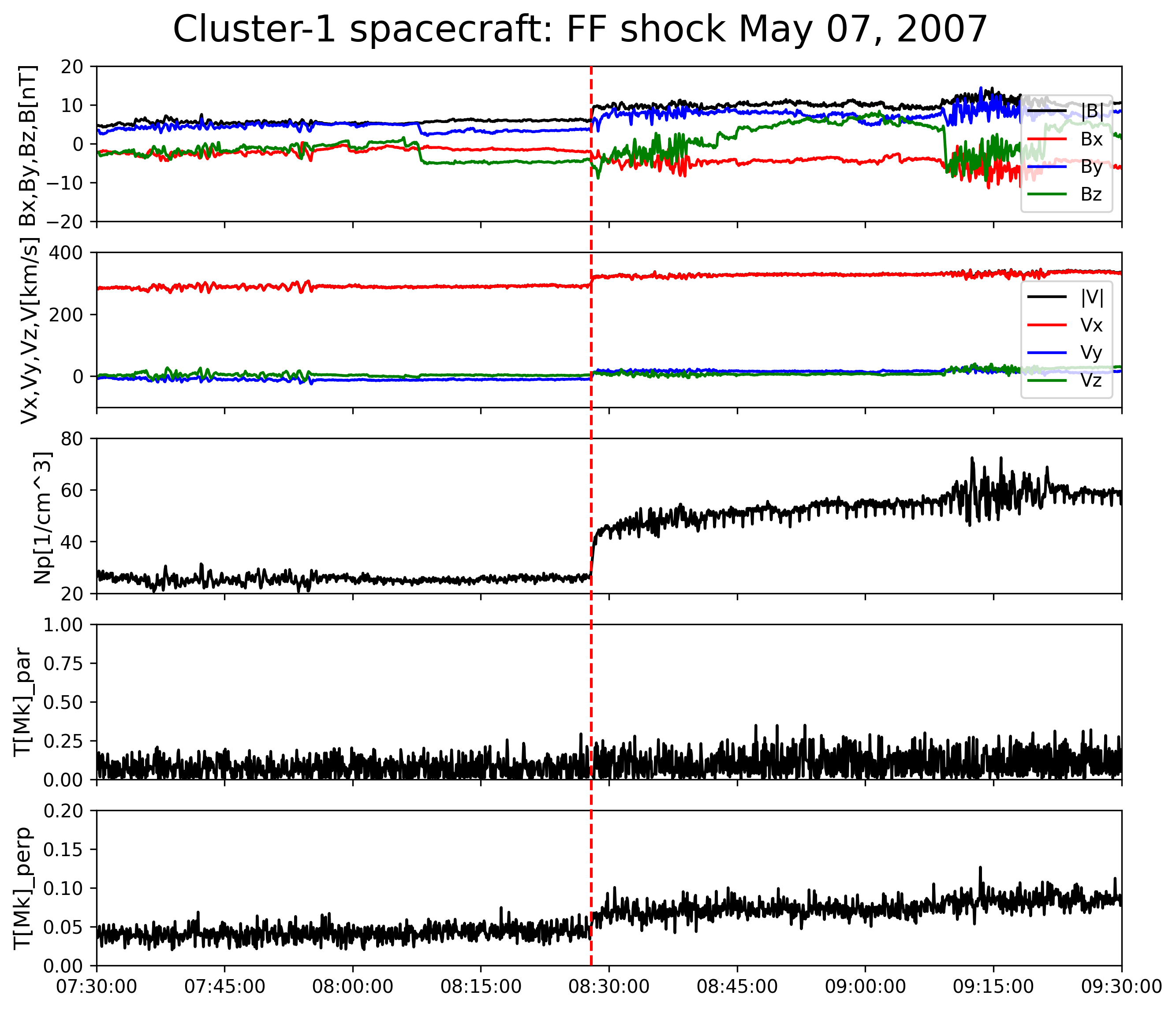}
\caption{The plot of the shock detected by the Cluster-1 spacecraft on May 07, 2007, at \textbf{08:27:55} UTC. The panels show from top to bottom, the magnetic field magnitude as well as its components, the total velocity, and its components, density, and parallel and perpendicular temperatures. The dashed red line represents the exact shock time. The duration of the plot is two hours.}
\label{fig:cl10507}
\end{figure}

\begin{figure}[htp]
\centering
\includegraphics[width=1.\textwidth]{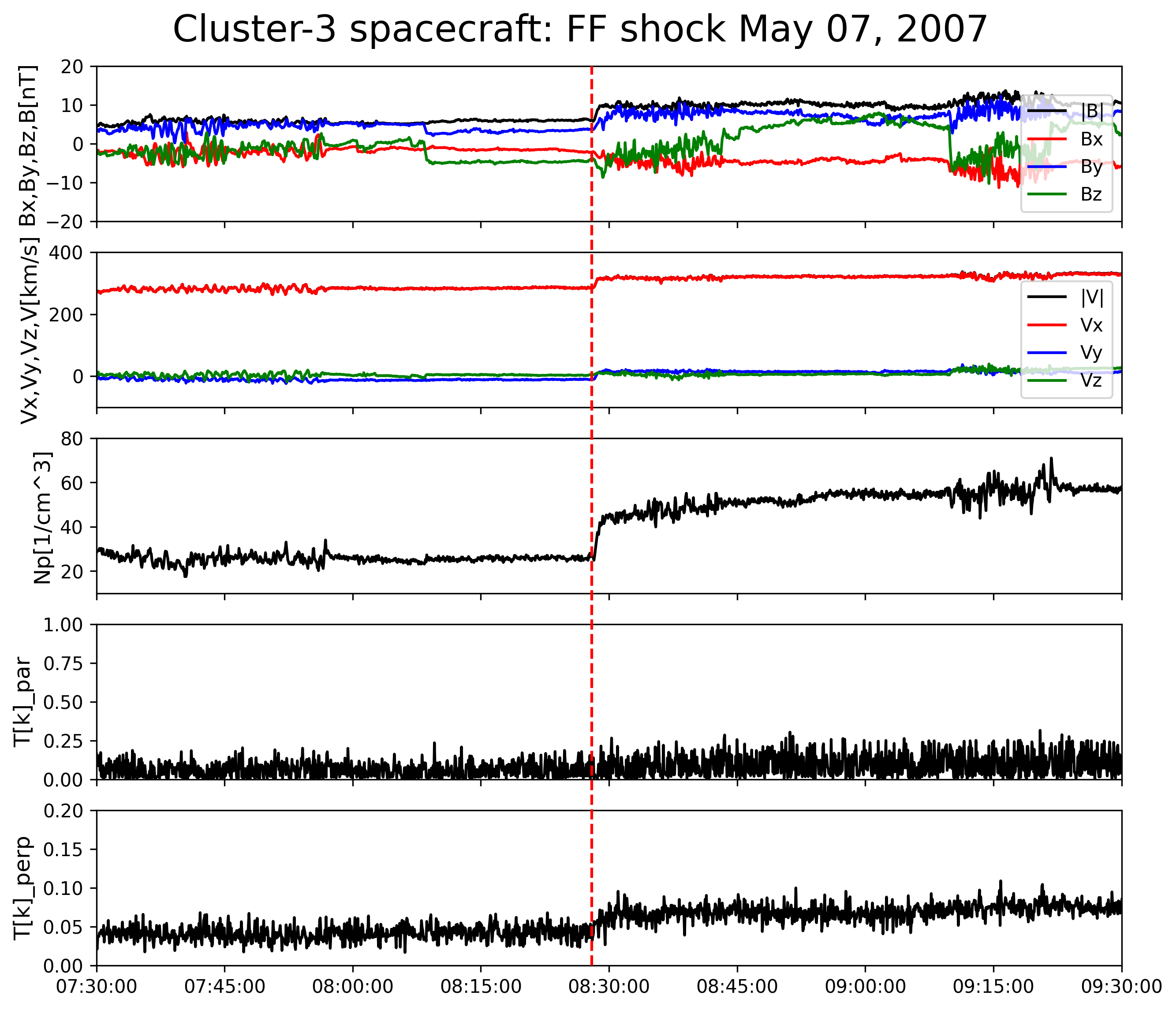}
\caption{The plot of the shock detected by the Cluster-3 spacecraft on May 07, 2007, at \textbf{08:28:00} UTC. The symbols and details of the figures are the same as \ref{fig:cl10507}. The duration of the plot is two hours.}
\label{fig:cl30507}
\end{figure}

On this day, the Wind spacecraft first detected the shock at \textbf{07:02:30} UTC. After that, the STEREO$-$A spacecraft detected the shock at \textbf{08:11:30} UTC, and then the STEREO$-$B spacecraft detected the shock at \textbf{09:42:00} UTC. The four Cluster satellites detect the shock as well. Cluster-1 spacecraft detected the shock at \textbf{08:27:55} UTC, and the Cluster-3 spacecraft detected the shock at \textbf{08:28:00} UTC. The magnetic field and the plasma plot are shown in Figure~\ref{fig:WindIP0507} for the Wind, in Figure~\ref{fig:staIP0507} for the STEREO$-$A, in Figure~\ref{fig:stbIP0507} for the STEREO$-$B, in Figure~\ref{fig:cl10507} for the Cluster-1, in Figure~\ref{fig:cl30507} for the Cluster-3. There is no indication that Cluster-2 and 4 detected a shock in the shock database list. But the four Cluster satellites are much closer together, so they must have detected the shock. Without plasma data, which is a problem for the Cluster-2 and 4 satellites, a shock cannot be confirmed based solely on the magnetic field data, but there is a way to obtain at least the plasma density parameter using the spacecraft potential, which is a resulting electric potential as the spacecraft travels through the plasma environment, it acquires an electric charge as a result of contact with charged particles. This charging process results in an electrical potential discrepancy between the spacecraft and the plasma around it, a phenomenon referred to as spacecraft potential. 

 The electron plasma density parameter is obtained by the empirical formula \ref{eq:empiricalformula}, as stated in \citep{WHISPERteam2022} and \citep{sandhu2016statistical}
\begin{align}\label{eq:empiricalformula}
    N_e = 200\cdot(-U_{EFW})^{-1.85}
\end{align}
So, comparing the magnetic field and the electron plasma density profile, the shock is determined.

\begin{figure}[htp]
\centering
\includegraphics[width=1.\textwidth]{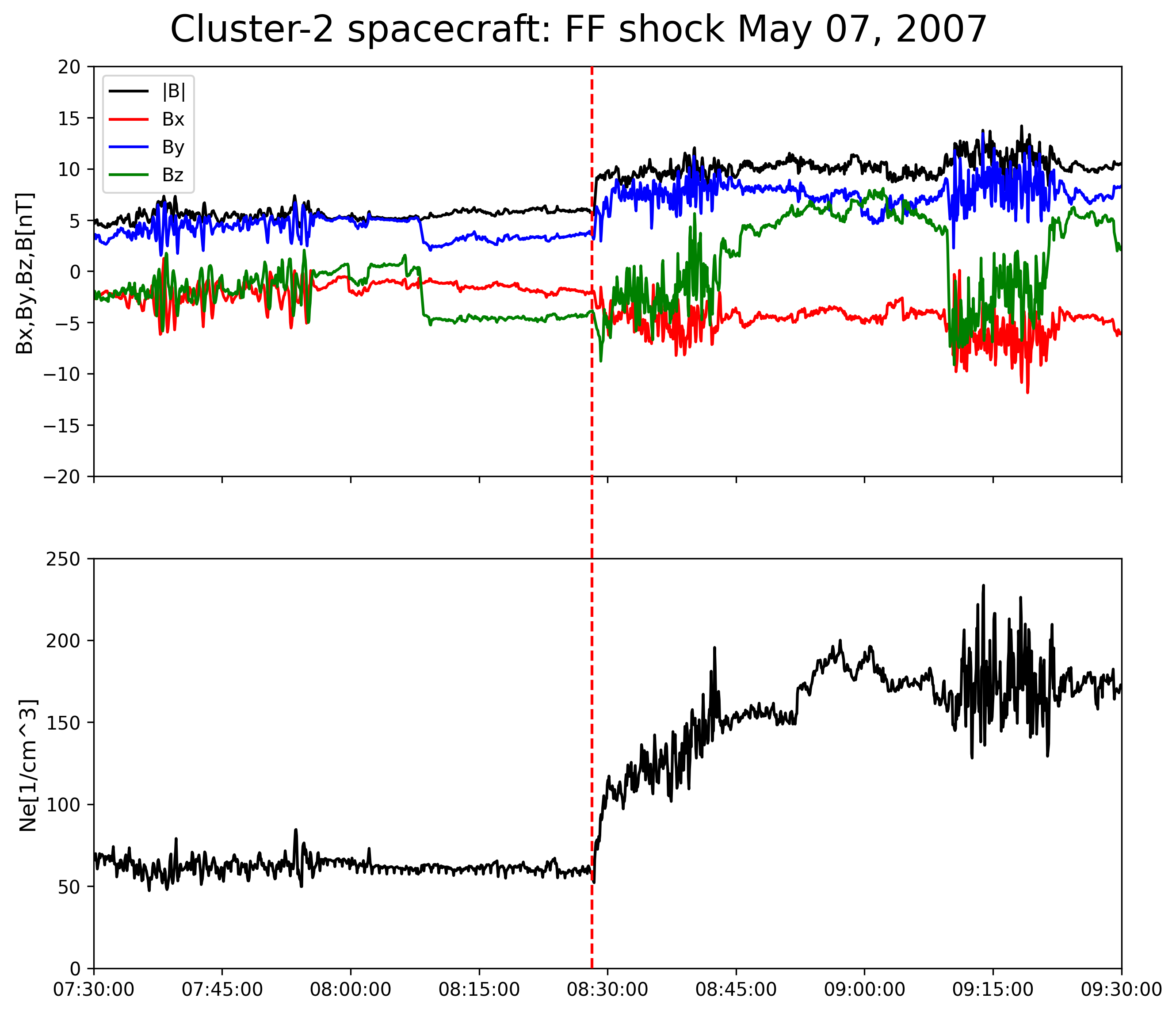}
\caption{The plot of the shock detected by the Cluster-2 spacecraft on May 07, 2007, at \textbf{08:28:10} UTC. The panels show from top to bottom, the magnetic field magnitude as well as its components, and the obtained electron density from the spacecraft potential. The dashed red line represents the exact shock time. The duration of the plot is two hours.}
\label{fig:cl20507}
\end{figure}

\begin{figure}[htp]
\centering
\includegraphics[width=1.\textwidth]{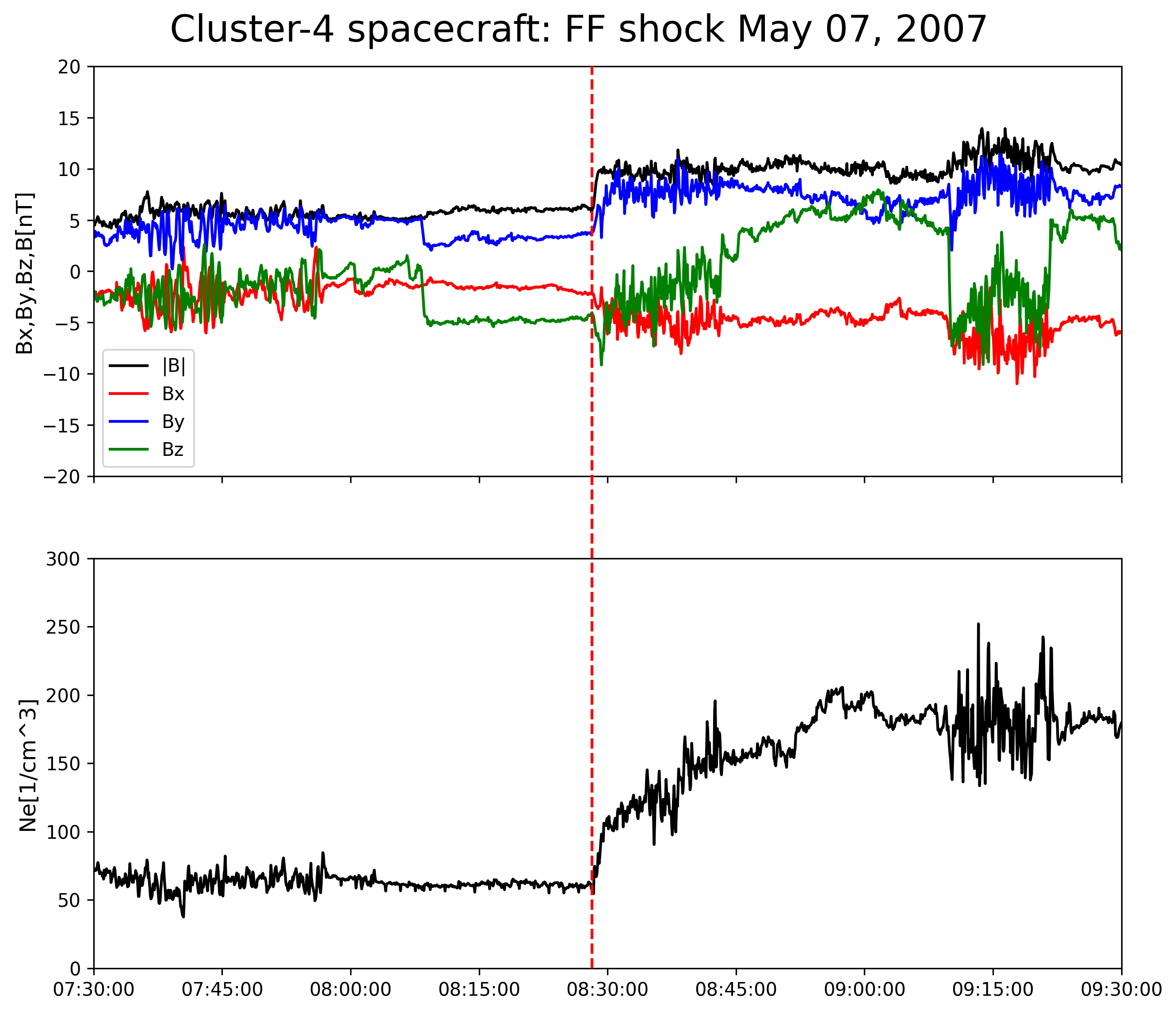}
\caption{The plot of the shock detected by the Cluster-4 spacecraft on May 07, 2007, at \textbf{08:28:10} UTC. The symbols and details of the figures are the same as \ref{fig:cl20507}. The duration of the plot is two hours.}
\label{fig:cl40507}
\end{figure}

The Cluster-2 spacecraft detected the shock at \textbf{08:28:10} UTC. Figure~\ref{fig:cl20507} shows the magnetic field and density plot. The Cluster-4 spacecraft detected the shock at \textbf{08:28:10} UTC. Figure~\ref{fig:cl40507} shows the magnetic field and density plot. Similarly, The electron density parameter is obtained by the empirical formula \ref{eq:empiricalformula}.

\subsection{The Event April 23, 2007}
\begin{figure}[htp]
\centering
\includegraphics[width=1.0\textwidth]{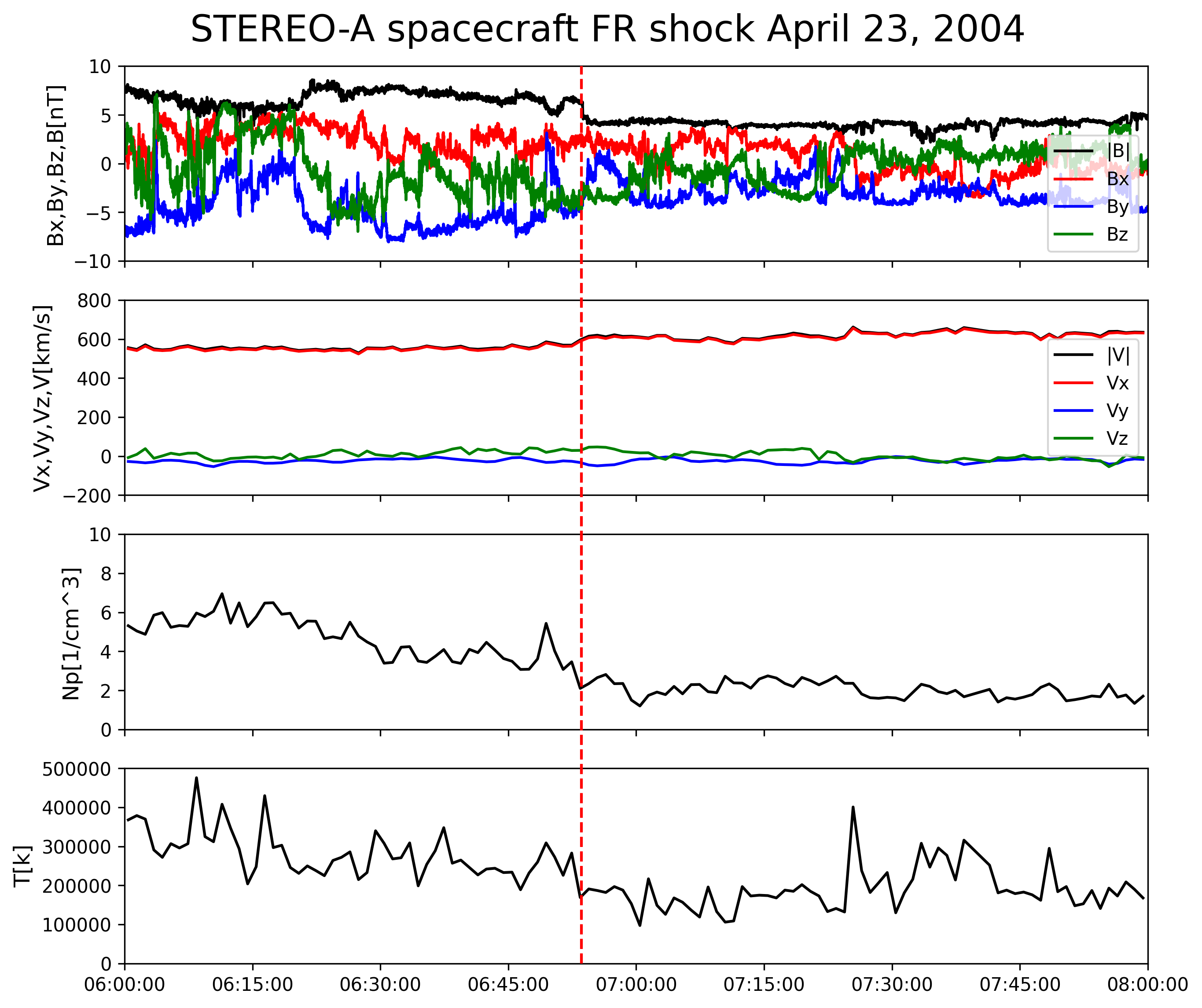}
\caption{The plot of the shock detected by the STEREO$-$A spacecraft on April 23, 2007, at \textbf{06:53:35} UTC. FR stands for the fast reverse shock, which means the shock is moving toward its driver. The panels show from top to bottom, the magnetic field magnitude as well as its components, the total velocity and its components, density, and temperature. The dashed red line represents the exact shock time. The duration of the plot is two hours}
\label{fig:staIP0423}
\end{figure}

\begin{figure}[htp]
\centering
\includegraphics[width=1.\textwidth]{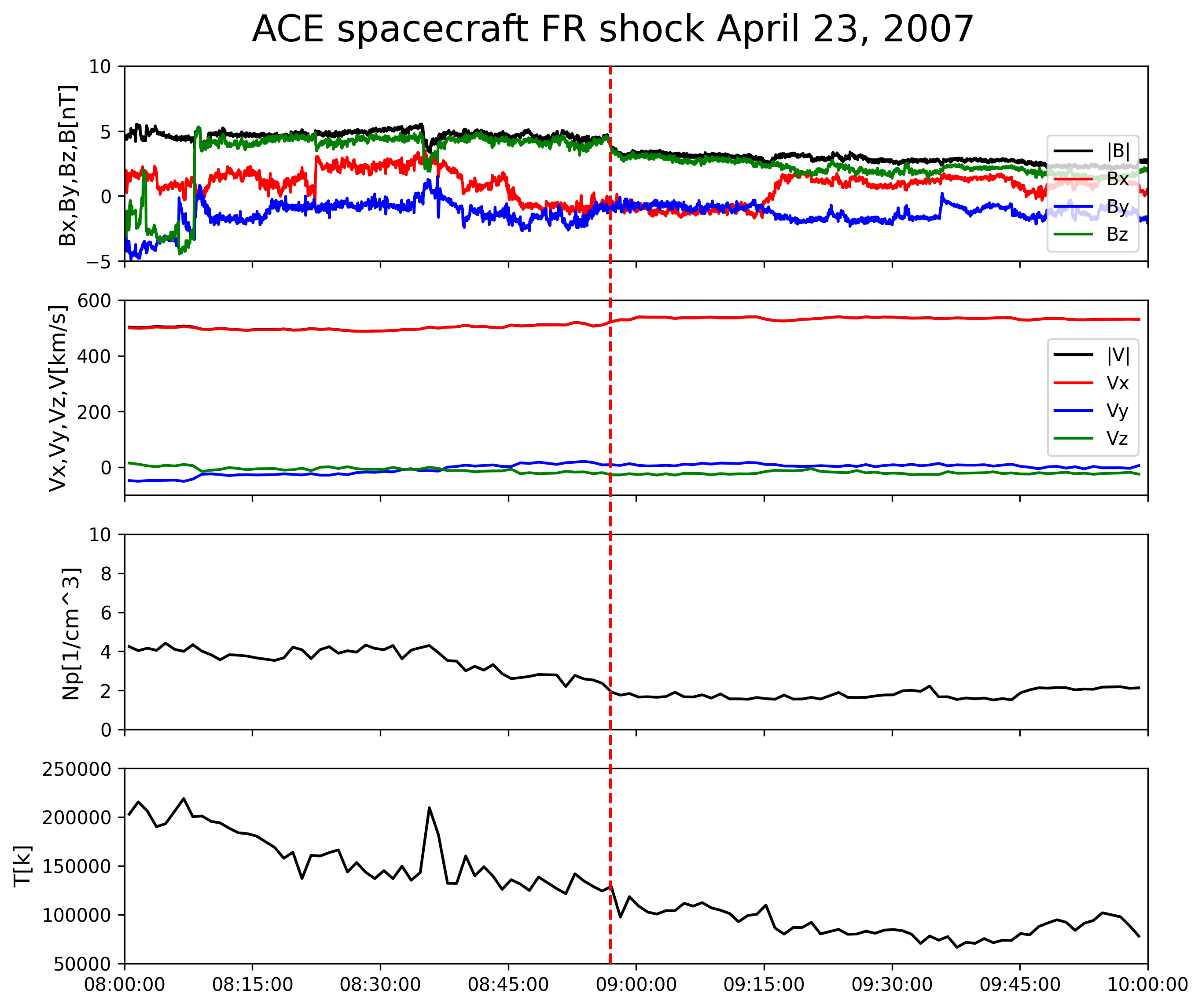}
\caption{The plot of the shock detected by the ACE spacecraft on April 23, 2007, at \textbf{08:57:00} UTC. The symbols and details of the figures are the same as \ref{fig:staIP0423}. The duration of the plot is two hours.}
\label{fig:aceIP0423}
\end{figure}

\begin{figure}[htp]
\centering
\includegraphics[width=1.\textwidth]{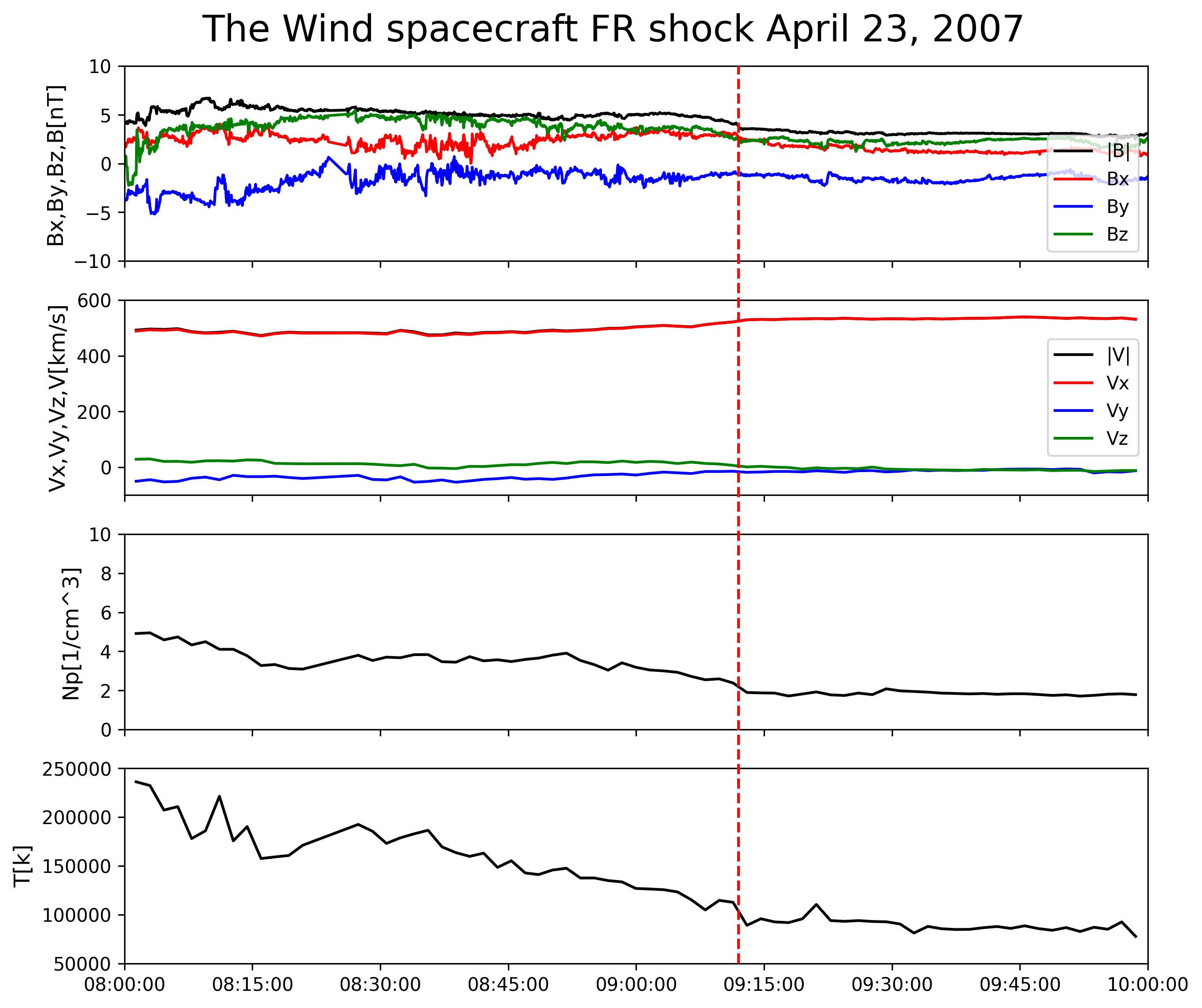}
\caption{The plot of the shock detected by the Wind spacecraft on April 23, 2007, at \textbf{09:12:00} UTC. The symbols and details of the figures are the same as \ref{fig:staIP0423}. The duration of the plot is two hours.}
\label{fig:WindIP0423}
\end{figure}

In this specific event STEREO$-$A spacecraft detected a shock first at \textbf{06:53:35} UTC, then the ACE spacecraft detected the shock at \textbf{08:57:00} UTC, and the Wind spacecraft detected the shock at \textbf{09:12:00} UTC. The magnetic field and the plasma data are shown in Figure~\ref{fig:staIP0423} for the STEREO$-$A, in Figure~\ref{fig:aceIP0423} for the ACE, and in Figure~\ref{fig:WindIP0423} for the Wind. 

\begin{figure}[htp]
\centering
\includegraphics[width=1.\textwidth]{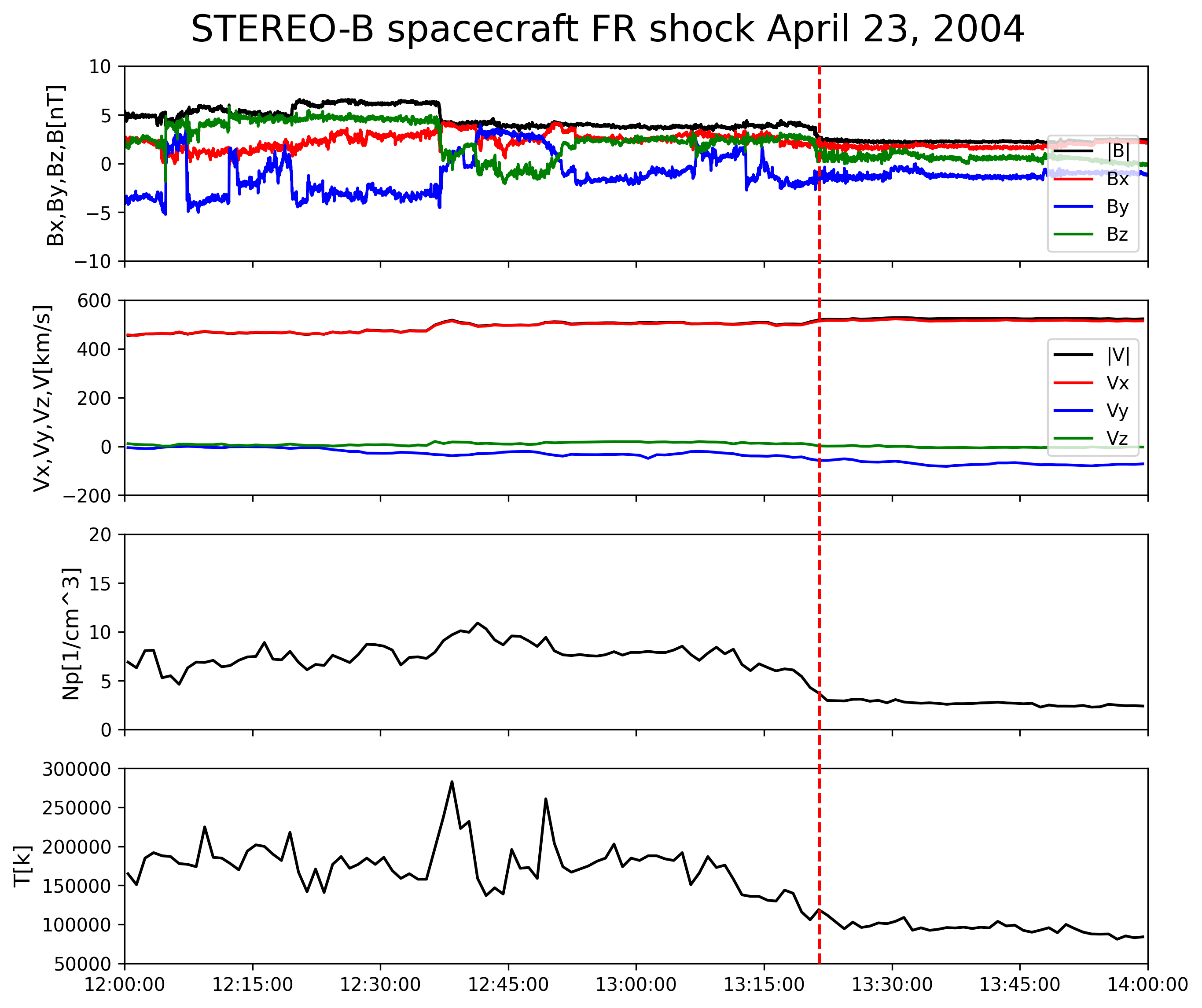}
\caption{The plot of the shock detected by the Wind spacecraft on April 23, 2007, around \textbf{13:21:30} UTC. The symbols and details of the figures are the same as \ref{fig:staIP0423}. The duration of the plot is two hours.}
\label{fig:stbIP0423}
\end{figure}

Since STEREO$-$A and B have identical instruments, STEREO$-$B must have detected the shock on that day even though there is no detected IP shock for STEREO$-$B in the shock lists database mentioned above. So, by using the average solar wind speed, 400 km/s, we concluded that the shock detection time for STEREO$-$B should be around 13:00 to 15:00. Considering this, a could-be shock signature from STEREO$-$B is found around \textbf{13:21:30} even though it is a faint signature. The magnetic field and the plasma plot are shown in Figure~\ref{fig:stbIP0423}

\section{Implementation of the methods on the events}
\subsection{The Event May 07, 2007}
\begin{figure}[H]
\centering
\includegraphics[width=0.8\textwidth]{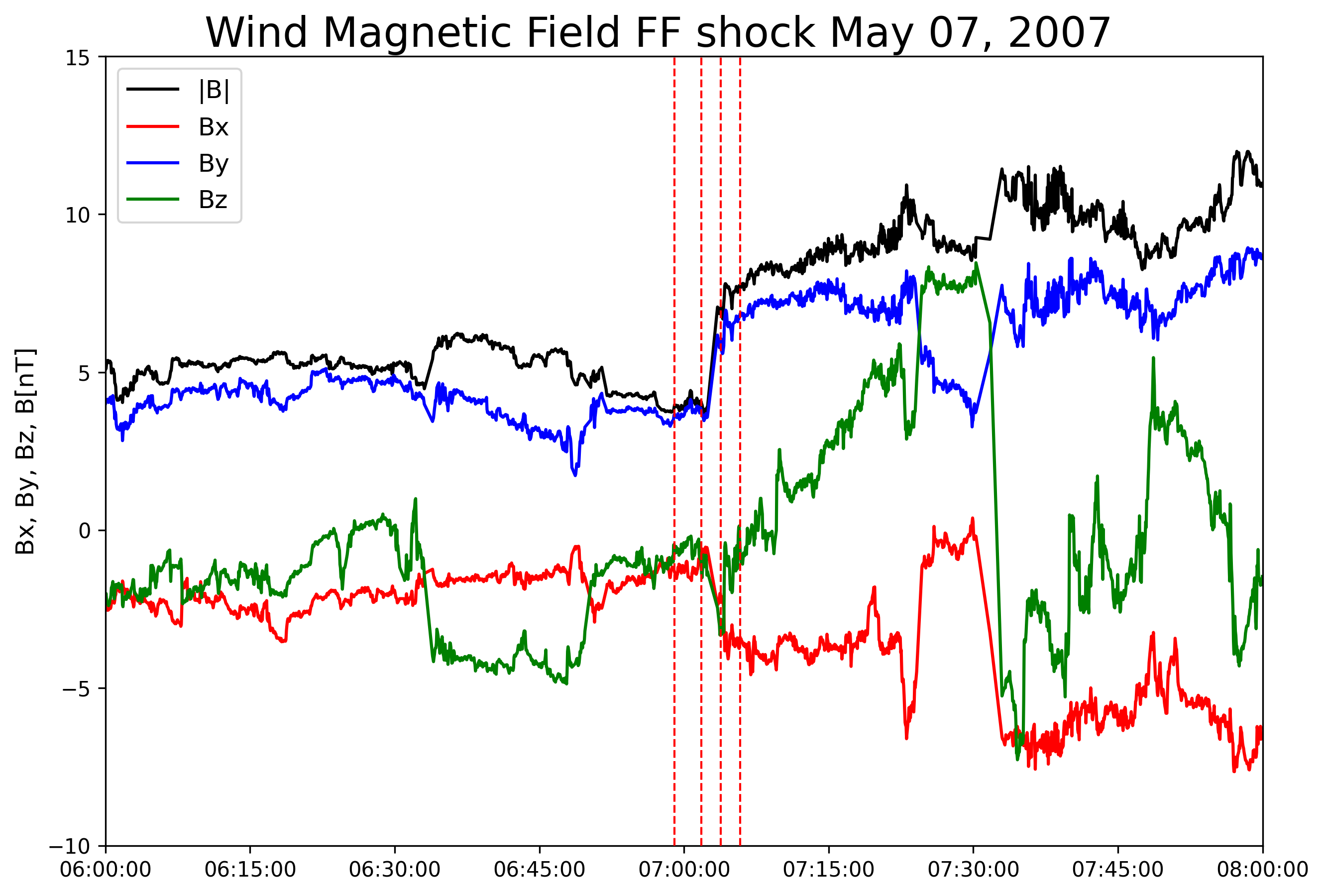}
\caption{The Wind magnetic field measurements. The upstream $\Delta t_{up}$ is between \textbf{(06:59:00 - 07:01:47)}, and the downstream $\Delta t_{down}$ is between \textbf{(07:03:50 - 07:05:50)}. FF stands for the fast forward shock, which means the shock is traveling away from its driver. The chosen intervals of the upstream and downstream magnetic field. The red dashed lines each represent the upstream starting time and ending time and the downstream starting time and ending time, respectively}
\label{fig:WindB}
\end{figure}

\begin{figure}[H]
\centering
\includegraphics[width=0.8\textwidth]{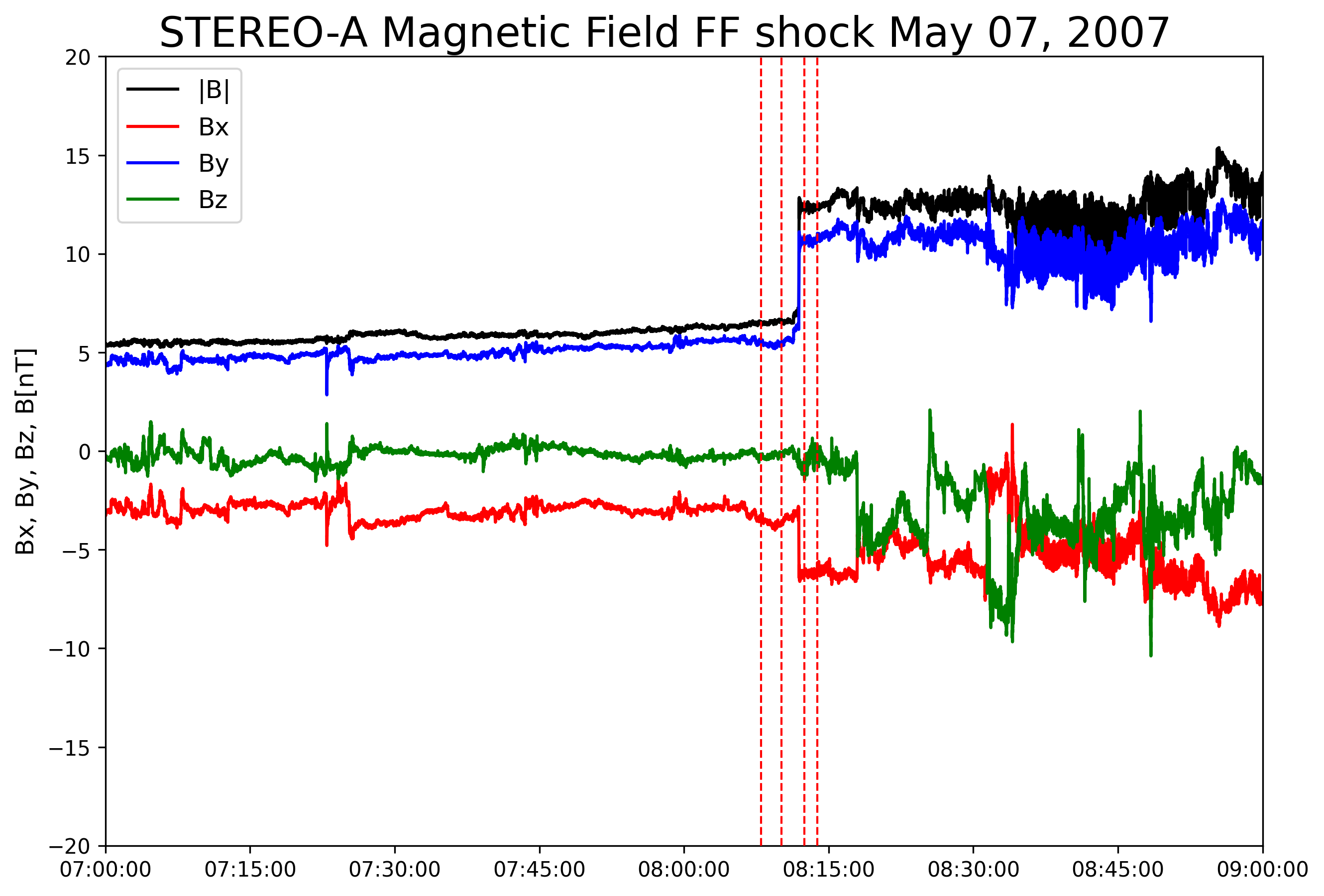}
\caption{The STEREO$-$A magnetic field measurements. The upstream $\Delta t_{up}$ is between \textbf{(08:08:00 - 08:10:05)}, and the downstream $\Delta t_{down}$ is between \textbf{(08:12:30 - 08:13:50)}. The symbols and details of the figures are the same as \ref{fig:WindB}}
\label{fig:staB}
\end{figure}

\begin{figure}[H]
\centering
\includegraphics[width=0.8\textwidth]{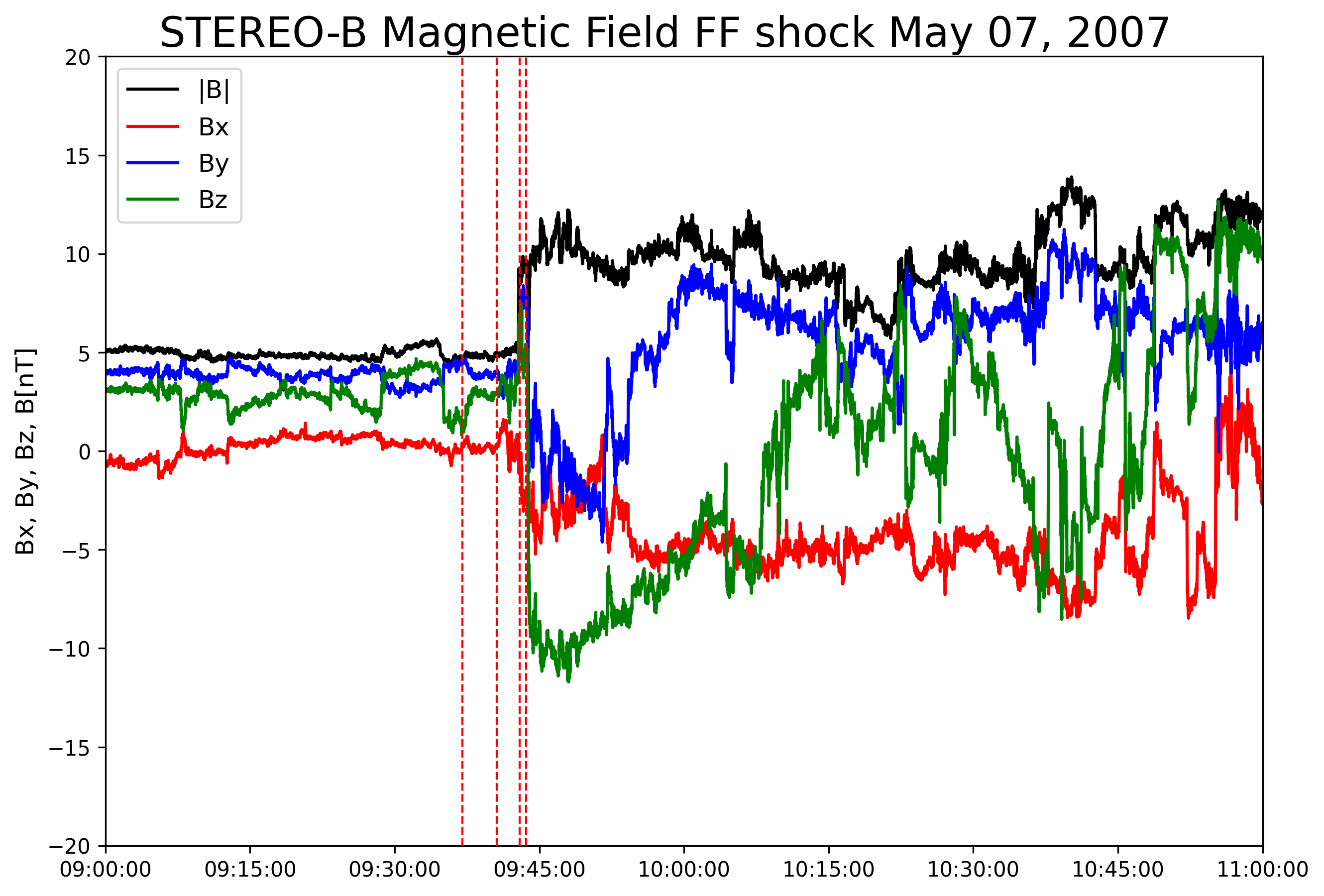}
\caption{The STEREO$-$B magnetic field measurements. The upstream $\Delta t_{up}$ is between \textbf{(09:37:00 - 09:40:35)}, and the downstream $\Delta t_{down}$ is between \textbf{(09:42:57 - 09:43:38)}. The symbols and details of the figures are the same as \ref{fig:WindB}}
\label{fig:stbB}
\end{figure}

\begin{figure}[H]
\centering
\includegraphics[width=0.8\textwidth]{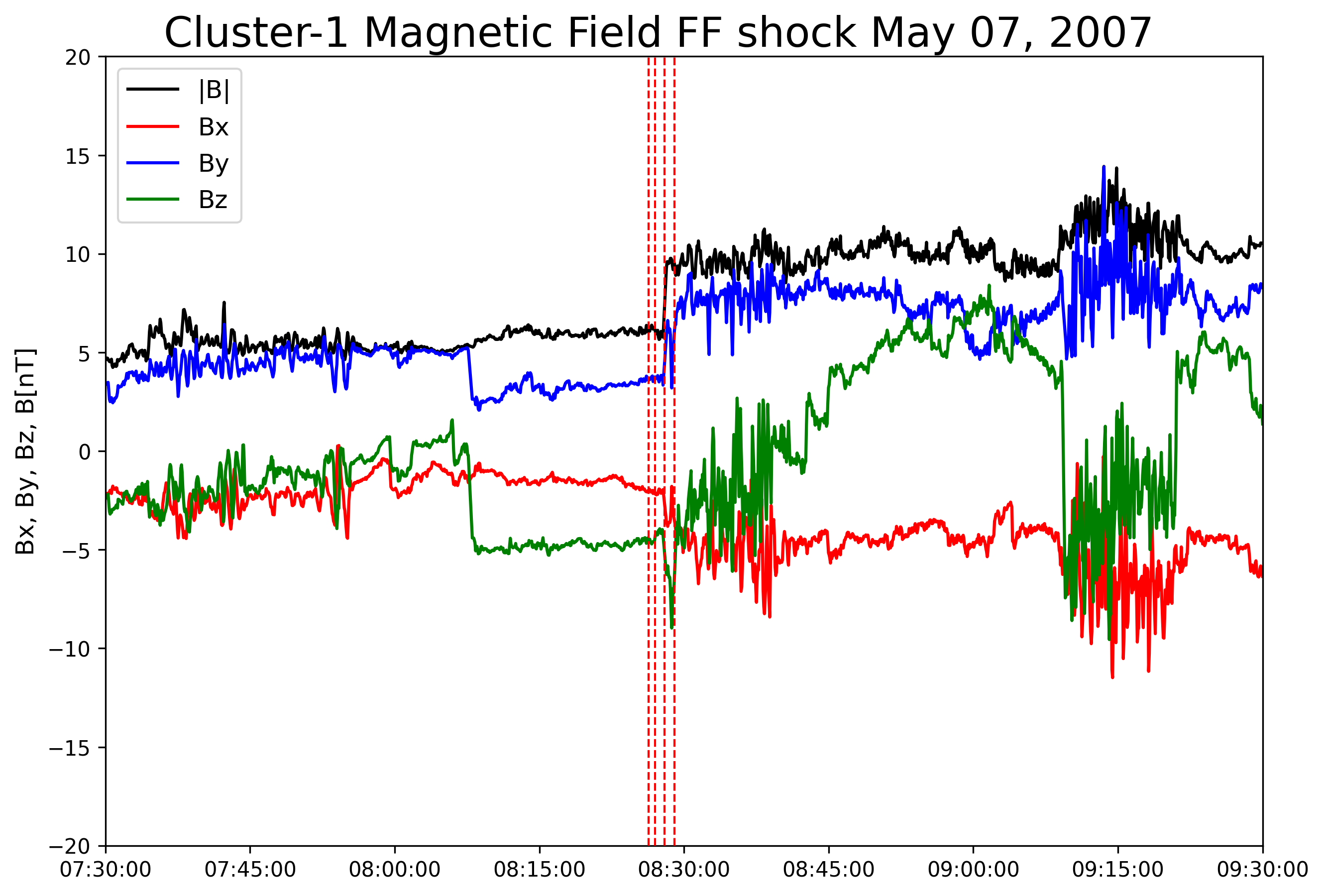}
\caption{The Cluster-1 B magnetic field measurements. The upstream $\Delta t_{up}$ is between \textbf{(08:26:10 - 08:27:00)}, and the downstream $\Delta t_{down}$ is between \textbf{(08:28:00 - 08:29:00)}. The symbols and details of the figures are the same as \ref{fig:WindB}}
\label{fig:cl1b}
\end{figure}

\begin{figure}[H]
\centering
\includegraphics[width=0.8\textwidth]{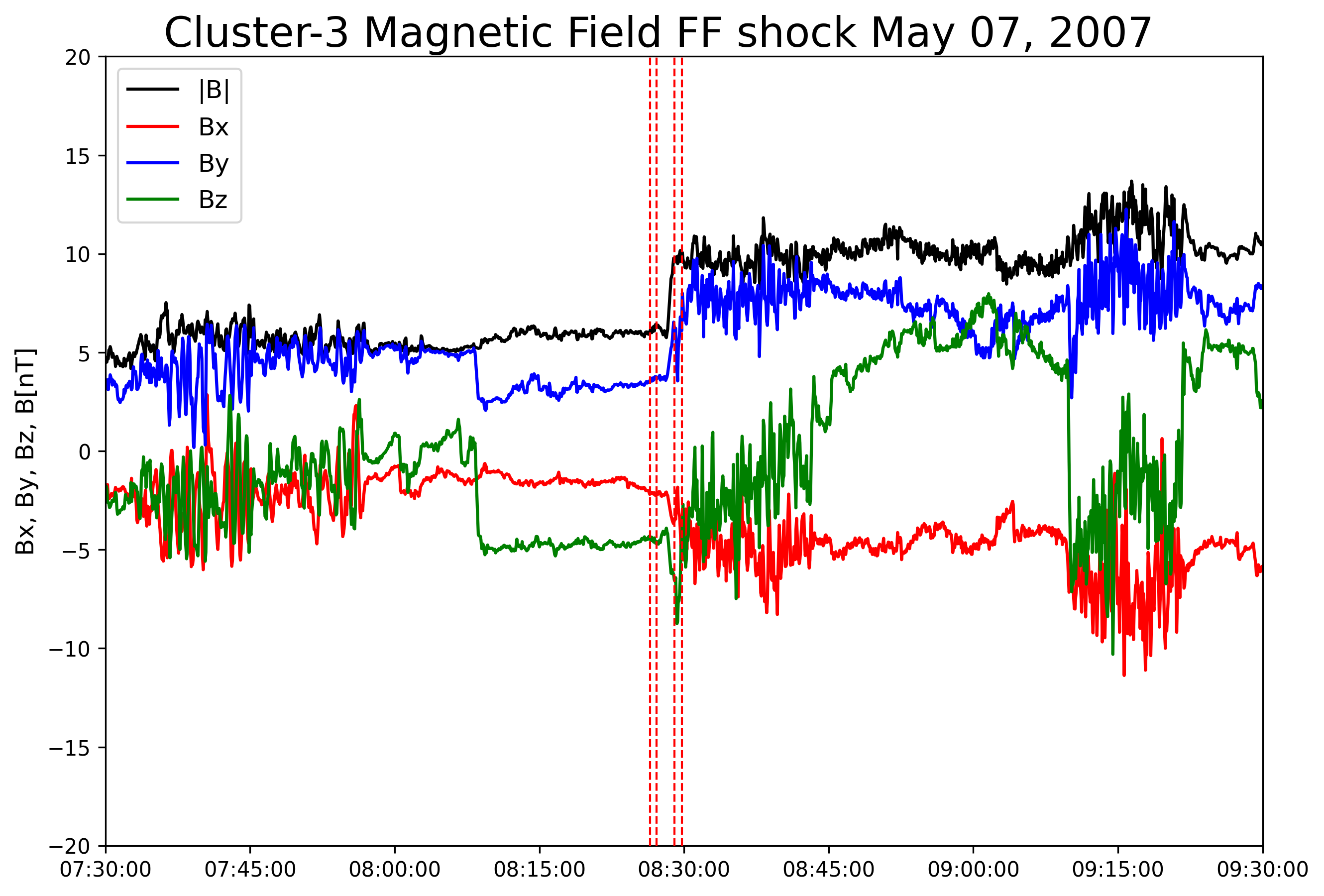}
\caption{The Cluster-3 B magnetic field measurements. The upstream $\Delta t_{up}$ is between \textbf{(08:26:30 - 08:27:10)}, and the downstream $\Delta t_{down}$ is between \textbf{(08:29:00 - 08:29:47)}. The symbols and details of the figures are the same as \ref{fig:WindB}}
\label{fig:cl3b}
\end{figure}

\begin{figure}[H]
\centering
\includegraphics[width=0.8\textwidth]{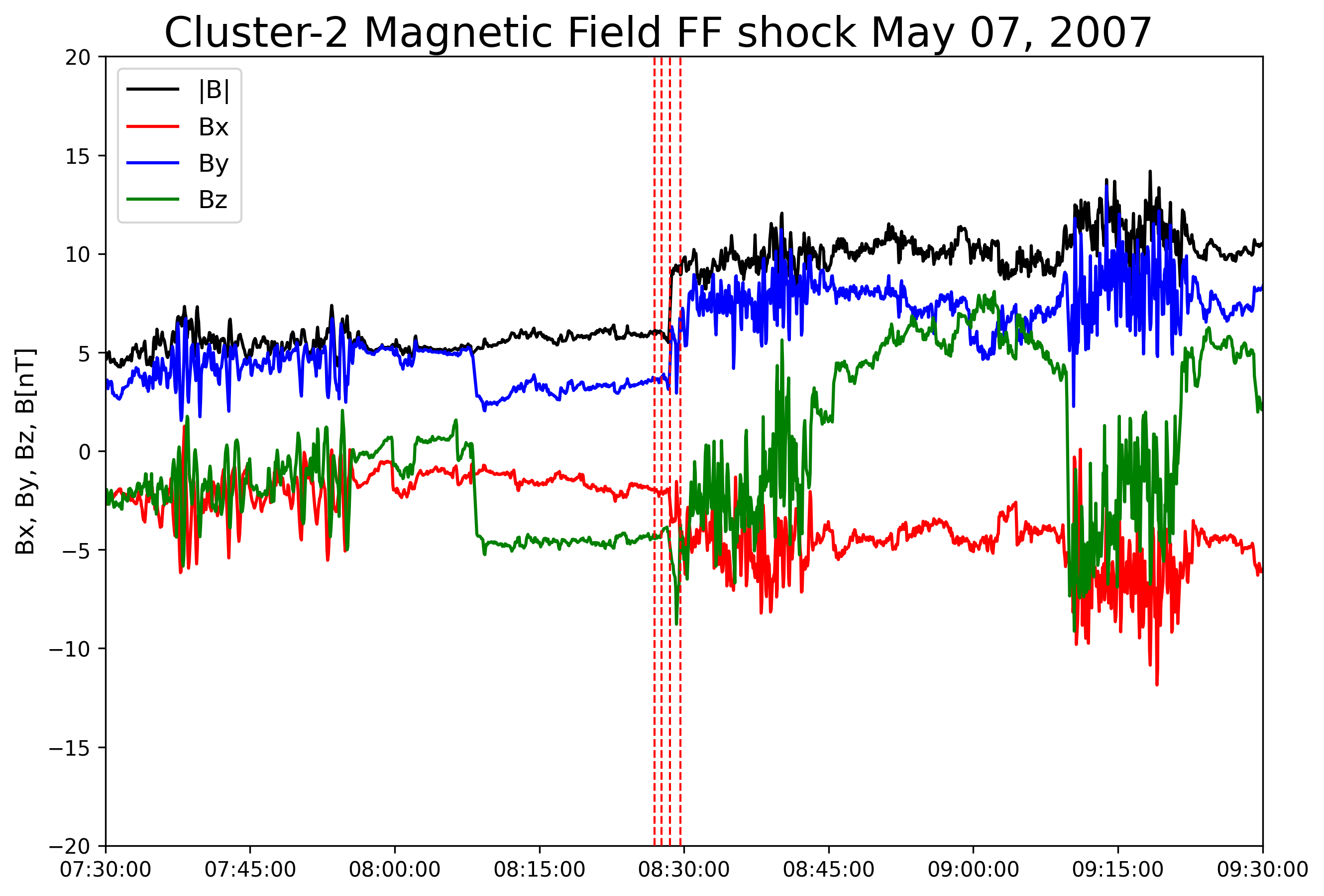}
\caption{The Cluster-2 B magnetic field measurements. The upstream $\Delta t_{up}$ is between \textbf{(08:26:56 - 08:27:40)}, and the downstream $\Delta t_{down}$ is between \textbf{(08:28:32 - 08:29:37)}. The symbols and details of the figures are the same as \ref{fig:WindB}}
\label{fig:cl2b}
\end{figure}

\begin{figure}[H]
\centering
\includegraphics[width=0.8\textwidth]{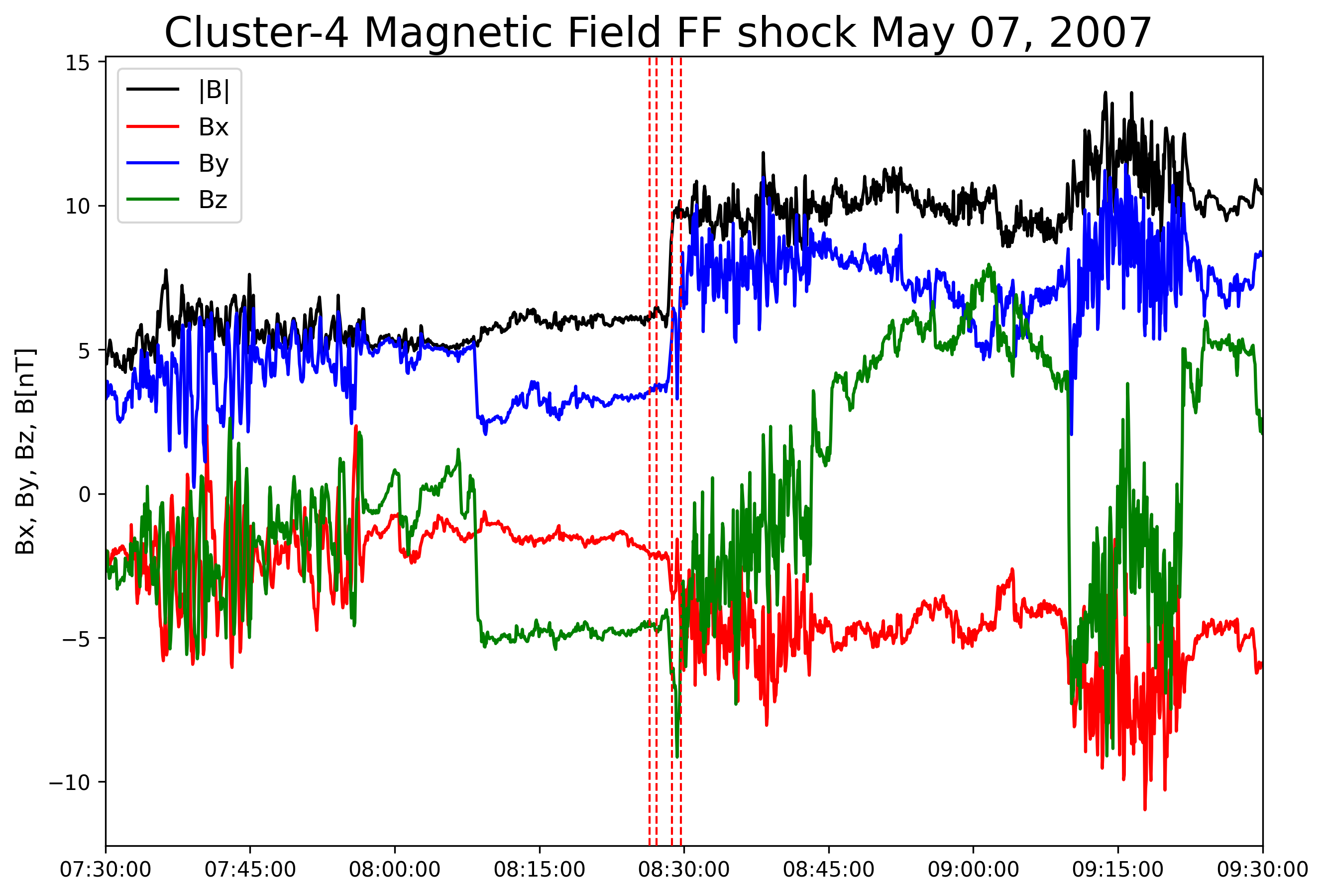}
\caption{The Cluster-4 B magnetic field measurements. The upstream $\Delta t_{up}$ is between \textbf{(08:26:25 - 08:27:10)}, and the downstream $\Delta t_{down}$ is between \textbf{(08:28:45 - 08:29:40)}. The symbols and details of the figures are the same as \ref{fig:WindB}}
\label{fig:cl4b}
\end{figure}

Here I list all magnetic field observations of the spacecraft for the event and apply the minimum variance analysis (MVA) and the magnetic coplanarity (CP) methods on them. 
The upstream and downstream time intervals that most agree between the minimum variance analysis (MVA) and the magnetic coplanarity methods (CP) for all the spacecraft are shown in Table \ref{tab:timeintervalsmay}, and their corresponding Figures are \ref{fig:WindB} for the Wind, \ref{fig:staB} for the STEREO$-$A, \ref{fig:stbB} for the STEREO$-$B, \ref{fig:cl1b} for the Cluster-1, \ref{fig:cl3b} for the Cluster-3, \ref{fig:cl2b} for the Cluster-2, and \ref{fig:cl4b} for the Cluster-4. Table \ref{tab:timeintervalsmay} also shows the ratio between the smallest eigenvalue $\lambda_3$ and the intermediate eigenvalue $\lambda_2$ as well as the angle between the MVA normal and CP normals in the determined upstream and downstream time intervals for each spacecraft. 
\begin{table}[H]
\centering
\caption*{Implementation of the methods on the event May 07, 2007}
\label{tab:sample}
\begin{tabular}{ccccc}
\toprule
\textbf{Spacecraft} & \textbf{$\Delta t_{up}$} & \textbf{$\Delta t_{down}$} & \boldmath$\lambda_2/\lambda_3$ & \boldmath$\Delta\theta_{MVA-CP}$ \\
\midrule
Wind & (06:59:00 - 07:01:47) & (07:03:50 - 07:05:50) & 14.82 & $0.38^{\circ}$\\ 
STEREO$-$A & (08:08:00 - 08:10:05)& (08:12:30 - 08:13:50)& 3.16 & $0.375^{\circ}$\\ 
STEREO$-$B & (09:37:00 - 09:40:35)& (09:42:57 - 09:43:38)& 9.51 & $1.1^{\circ}$\\ 
Cluster-1 & (08:26:10 - 08:27:00)& (08:28:00 - 08:29:00)& 167.9 & $0.21^{\circ}$\\ 
Cluster-3 & (08:26:30 - 08:27:10)& (08:29:00 - 08:29:47)& 22.43 & $1.92^{\circ}$\\ 
Cluster-2 & (08:26:56 - 08:27:40)& (08:28:32 - 08:29:37)& 91.56 & $0.1^{\circ}$\\ 
Cluster-4 & (08:26:25 - 08:27:10)& (08:28:45 - 08:29:40)& 238.63 & $2.49^{\circ}$\\
\bottomrule
\end{tabular}
\caption{Here, \textbf{$\Delta t_{up}$} denotes the defined upstream and \textbf{$\Delta t_{down}$} denotes the defined downstream time intervals with MVA and CP analysis methods. \boldmath$\lambda_2/\lambda_3$ indicates the ratio between the intermediate eigenvalue $\lambda_2$ and the smallest eigenvalue $\lambda_3$, and $\Delta\theta_{MVA-CP}$ is the angle between the MVA and CP normals.}
\label{tab:timeintervalsmay}
\end{table}

\subsection{Analysis of the Event May 07, 2007}
The accepted upstream and downstream time intervals are highly accurate considering the angle difference between the two methods is minimal and the eigenvalue criteria are sufficient for each spacecraft data.
 Using the determined time intervals, the calculations of the ratio between the upstream and downstream magnetic fields, densities, and temperatures as well as the bulk speed, and shock $\theta_{Bn}$ angle are made. These parameters, the minimum variance analysis normal, and the magnetic coplanarity normals are shown in Table \ref{tab:table}, where the solar wind bulk speed and the downstream to upstream ratios for the shock criteria are fulfilled as stated in \ref{eq:criteria1} and \ref{eq:criteria2}, and the results of the additional parameters calculations as stated in \ref{sec:estimatingparameters}
are shown in Table \ref{tab:tableAdd}.
\newsavebox{\corefirst}


\begin{lrbox}{\corefirst}
\Large
\begin{tabular}{|c|c|c|c|c|c|c|c|c|c|}
\hline
Spacecraft & $B_d/B_u$  & $N_d/N_u$ & $T_d/T_u$ & $\Delta V $ & $\theta_{Bn}$ & Spacecraft & MVA & Coplanarity & $\Delta\theta_{MVA-CP}$\\
 \hline
Wind    & $1.86\pm 0.10$   & $1.78 $* & $1.33$*  & $29.6$* & 70.80 & Wind & [-0.763, -0.63, -0.144] & [-0.763, -0.629, -0.15] & 0.37 \\
 \hline
 STEREO$-$A   & $1.89\pm 0.02$     & $2.18\pm 0.9$ & $3.21\pm 2.4$ & $36\pm 6$   &81.83 & STEREO$-$A & [0.889, 0.436, 0.137] & [0.888, 0.436, 0.144] & 0.38 \\
 \hline
STEREO$-$B   & $1.90\pm 0.07$ & $1.90\pm 0.6$ & $2.90\pm 6$ & $40.25\pm 11$ & 59.45 & STEREO$-$B & [-0.855 -0.511 -0.087] &   [-0.864 -0.497 -0.077] & 1.10  \\
\hline
 \multirow{2}{*}{Cluster 1}   & \multirow{2}{*}{$1.53\pm 0.04$} & \multirow{2}{*}{$1.63\pm 0.09$} & $1.10\pm 1.3_{par}$ & \multirow{2}{*}{$31.2\pm 3.1$} & \multirow{2}{*}{86.62} & \multirow{2}{*}{Cluster-1} & \multirow{2}{*}{[0.846 0.532 -0.029]} &   \multirow{2}{*}{[-0.847 -0.530 -0.032]} & \multirow{2}{*}{0.21}  \\
 &&&$1.40\pm 0.32_{perp}$&&&&&&\\
 \hline
 \multirow{2}{*}{Cluster 3}   & \multirow{2}{*}{$1.57\pm 0.05$} & \multirow{2}{*}{$1.69\pm 0.08$} & $1.20\pm 1.5_{par}$ & \multirow{2}{*}{$30.5\pm 2.8$} & \multirow{2}{*}{89.74} & \multirow{2}{*}{Cluster-3} & \multirow{2}{*}{[0.753 0.639 0.153]} &   \multirow{2}{*}{[-0.763 -0.633 -0.121]} & \multirow{2}{*}{1.92}  \\
 &&&$1.40\pm 0.40_{perp}$&&&&&&\\
 \hline
 Cluster 2   & $1.53\pm 0.03$ & $1.42\pm 0.18$ & NaN & NaN & 88.19 & Cluster-2 & [0.838 0.545 0.012] &   [-0.838 -0.544 -0.013] & 0.10  \\
 \hline
 Cluster 4   & $1.57\pm 0.05$ & $1.49\pm 0.19$ & NaN & NaN & 86.55 & Cluster-4 & [0.858 0.506 -0.080] &   [0.838 0.533 0.108] & 2.49  \\
 \hline
 \end{tabular}
 
\end{lrbox}

\scalebox{0}{\usebox{\corefirst}}

\newcommand{\Coefficientfirst}{0.5119}

\begin{table}[H]
   \centering
   \begin{tikzpicture}[scale=0.8, transform shape]
      \clip (0,-\dp\corefirst) rectangle (\Coefficientfirst\wd\corefirst,\ht\corefirst);
      \pgftext[left,base]{\usebox{\corefirst}};
   \end{tikzpicture}
   \caption{\label{tab:table} Resulting core parameters of studying the data of the Wind, STEREO$-$A, STEREO$-$B, Cluster-1, Cluster-3, Cluster-2, and Cluster-4 spacecraft. Here $\theta_{Bn}$ is the shock $\theta$ angle. The values with the asterisk (*) symbol indicate the parameters with indefinite uncertainties because the determined intervals of the upstream and downstream magnetic fields are smaller than the resolution of the plasma data. Due to the CIS experiment is not operational for Cluster-2, and the HIA instrument being switched off for Cluster-4 spacecraft, the plasma parameters are not available for these satellites.}
\end{table}

\begin{table}[H]
   \ContinuedFloat
   \centering
   \begin{tikzpicture}[scale=0.8, transform shape]
      \clip (\Coefficientfirst\wd\corefirst,-\dp\corefirst) rectangle (\wd\corefirst,\ht\corefirst);
      \pgftext[left,base]{\usebox{\corefirst}};
   \end{tikzpicture}
   \caption{(Continued) Resulting core parameters of studying the data of the Wind, STEREO$-$A, STEREO$-$B, Cluster-1, Cluster-3, Cluster-2, and Cluster-4 spacecraft. $\theta_{MVA-CP}$ is the angle difference between the minimum variance normal vectors and the magnetic coplanarity normal vectors.}
\end{table}

\newsavebox{\arfirst}

\begin{lrbox}{\arfirst}
\Large
\begin{tabular}{|c|c|c|c|c|c|c|c|c|}
\hline
Spacecraft & $V_{sh}$ km/s & $C^{up}_s$ km/s  & $V^{up}_A$ km/s & $C^{up}_{ms}$ km/s & Spacecraft & Plasma $\beta_{up}$ & Alfvén Mach & Magnetosonic Mach \\
 \hline
Wind & 313.60 & $47.20\pm 1.60$     & $18.70\pm 6.08$      & $50.78\pm 2.7$      & Wind & $7.6\pm 4.9$      & $4.14\pm 1.72$ & $1.50\pm 0.40$ \\
 \hline
 STEREO$-$A  & 352.41 & $46.27\pm 3.30$     & $34.89\pm 15$      & $57.95\pm 9.41$     & STEREO$-$A & $2.11\pm 1.8$   & $2.19\pm 1.12$ & $1.32\pm 0.42$ \\
 \hline
STEREO$-$B & 354.91 & $46.89\pm 5.85$     & $31.95\pm 7.33$      & $56.74\pm 6.35$     & STEREO$-$B & $2.58\pm 1.34$    & $2.76\pm 1.07$ & $1.55\pm 0.52$ \\
 \hline
  Cluster 1  & 342.02 & $43.88\pm 3.09$     & $25.59\pm 7.70$      & $51.31\pm 3.99$     & Cluster-1 & $3.26\pm 1.89$   & $3.84\pm 1.40$ & $1.99\pm 0.46$ \\
 \hline
  Cluster 3  & 308.11 & $43.88\pm 2.63$     & $26.80\pm 7.64$  & $51.42\pm 3.98$     & Cluster-3 & $3.21\pm 1.83$   & $3.72\pm 1.35$ & $1.93\pm 0.46$ \\
 \hline
\end{tabular}
\end{lrbox}

\scalebox{0}{\usebox{\arfirst}}

\newcommand{\Coefficientsecond}{0.51589}

\begin{table}[H]
   \centering
   \begin{tikzpicture}[scale=0.8, transform shape]
      \clip (0,-\dp\arfirst) rectangle (\Coefficientsecond\wd\arfirst,\ht\arfirst);
      \pgftext[left,base]{\usebox{\arfirst}};
   \end{tikzpicture}
   \caption{\label{tab:tableAdd} Resulting additional parameters of studying the data of the three spacecraft. $V_{sh}$ is the shock speed, $C^{up}_s$ is the upstream sound speed, $V^{up}_A$ is the upstream Alfvén speed and $C^{up}_{ms}$ is the upstream magnetosonic speed.}
\end{table}

\begin{figure}[H]\centering
\subfloat[XY coordinates sketch of the propagation of the IP shock through spacecraft]{\label{fig:xy}\includegraphics[width=.5\linewidth]{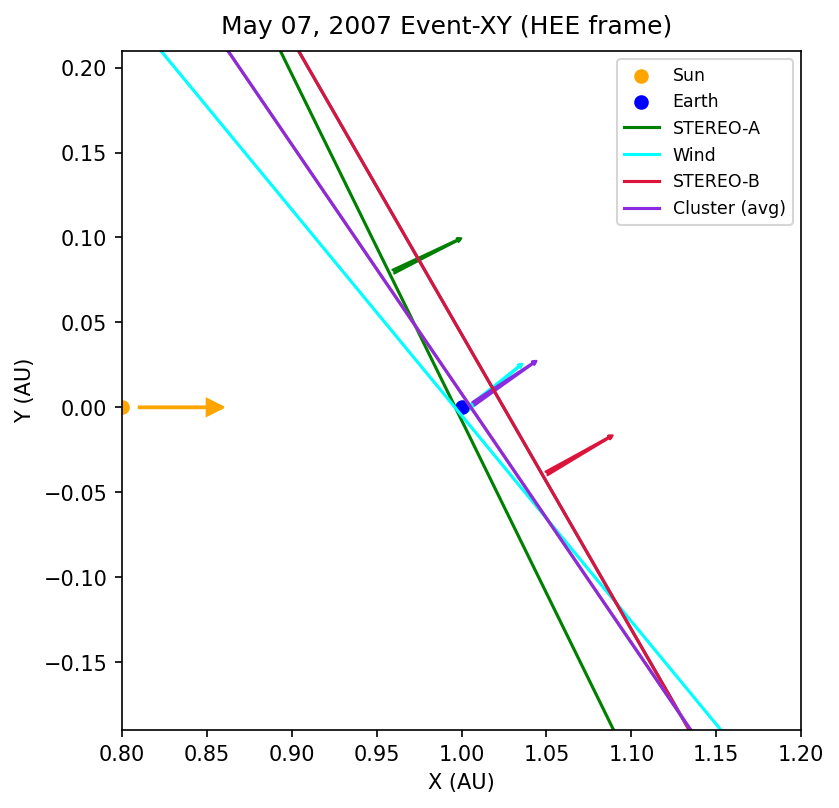}}\hfill
\subfloat[XZ coordinates sketch of the propagation of the IP shock through spacecraft]{\label{fig:xz}\includegraphics[width=.5\linewidth]{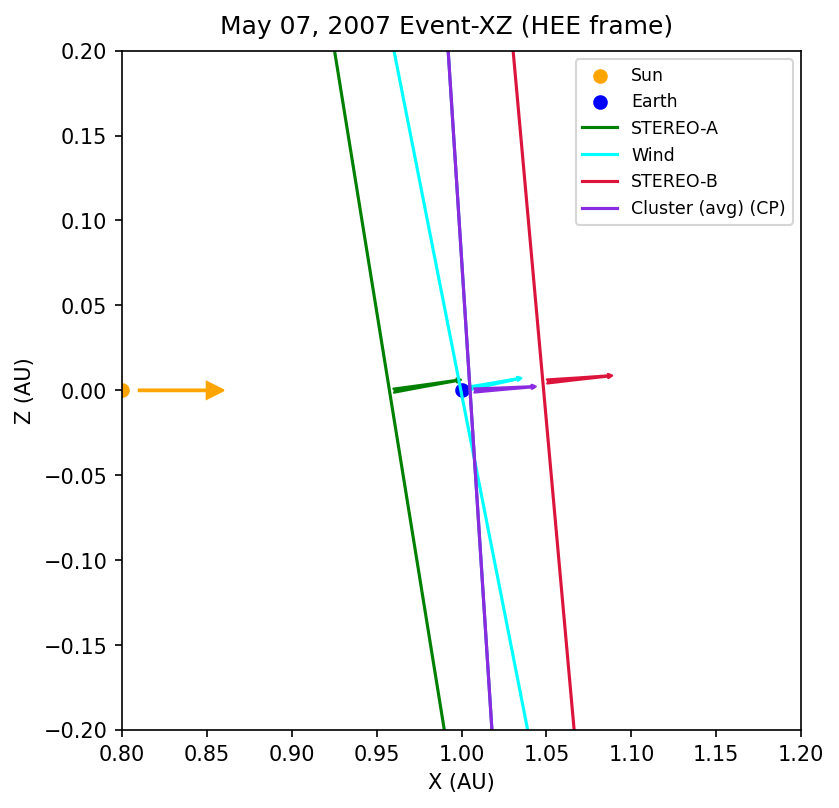}}\par 
\subfloat[YZ coordinates sketch of the propagation of the IP shock through spacecraft.]{\label{fig:yz}\includegraphics[width=.5\linewidth]{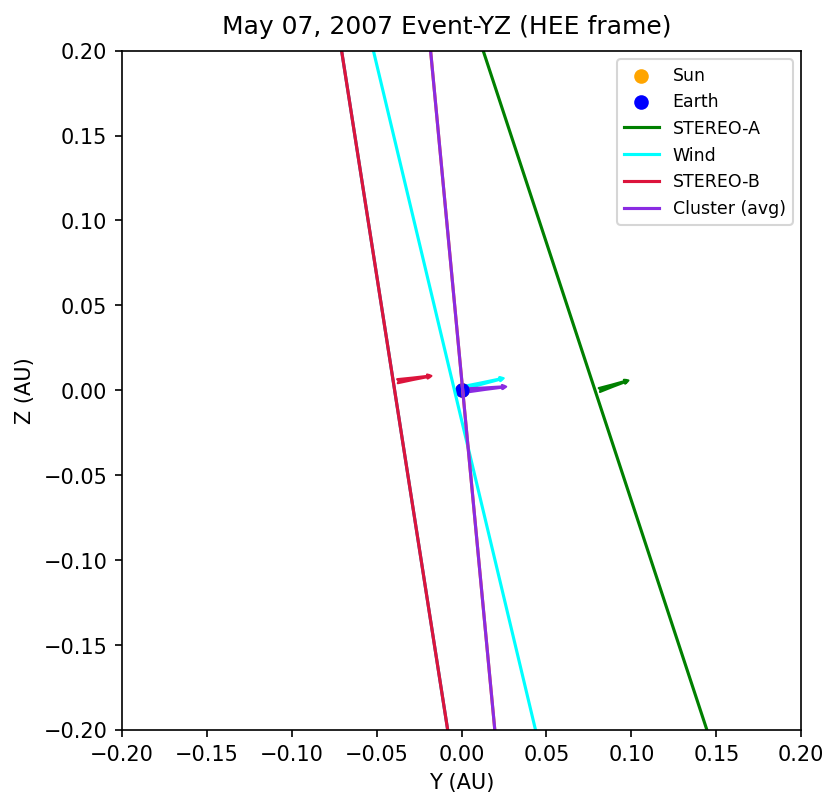}}
\caption{2D XY, XZ, and YZ sketches of the propagation of the IP shock through spacecraft-- Wind, STEREO$-$A and B, and average position of 4 Cluster satellites. The orange arrow represents the Sun-Earth line direction. The arrows on spacecraft positions indicate the normal vector direction and the lines perpendicular to the normal vectors indicate the shock surface orientations. The sizes of the lines are arbitrary. The shock orientations for the 4 Cluster satellites are averaged}
\label{fig:05072D}
\end{figure}

\begin{figure}[H]\centering
\subfloat[XY coordinate 2D sketch of the propagation of the IP shock through 4 Cluster satellites.]{\label{1}\includegraphics[width=.45\linewidth]{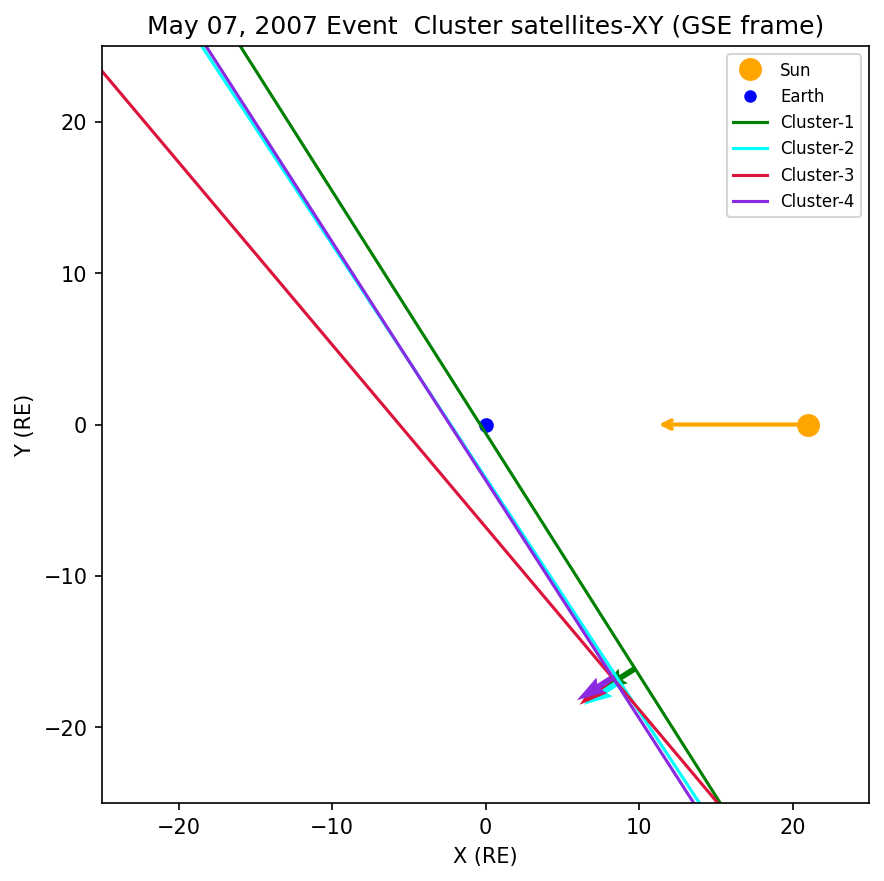}}\hfill
\subfloat[XZ coordinate 2D sketch of the propagation of the IP shock through 4 Cluster satellites.]{\label{2}\includegraphics[width=.45\linewidth]{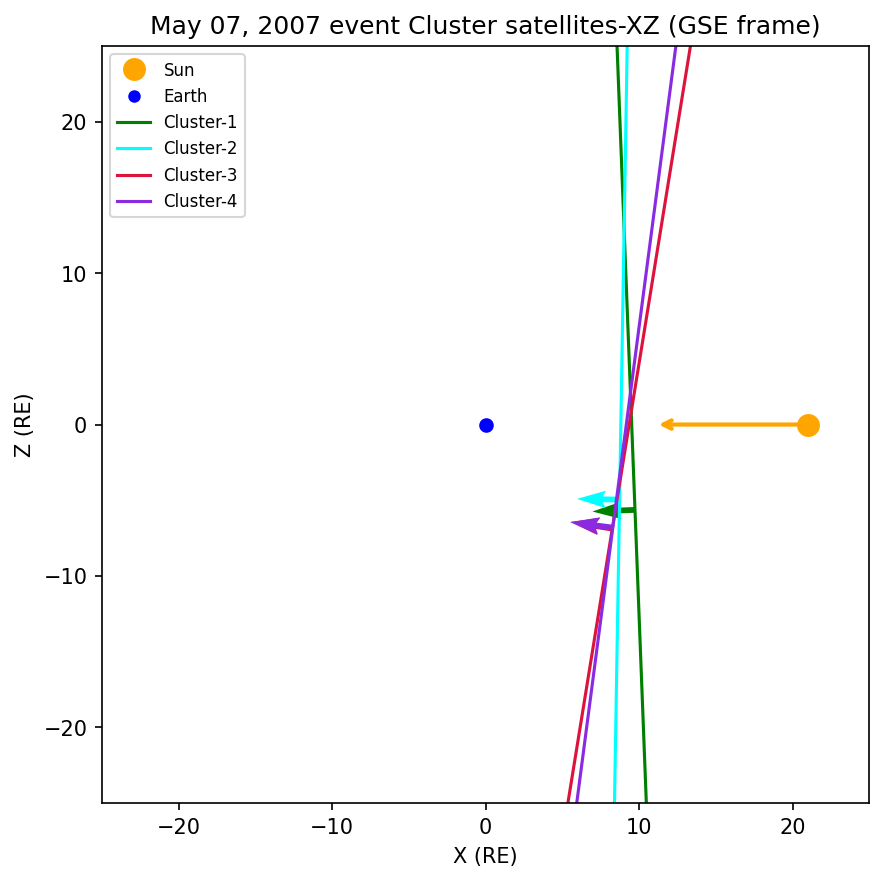}}\par 
\subfloat[YZ coordinate 2D sketch of the propagation of the IP shock through 4 Cluster satellites.]{\label{3}\includegraphics[width=.45\linewidth]{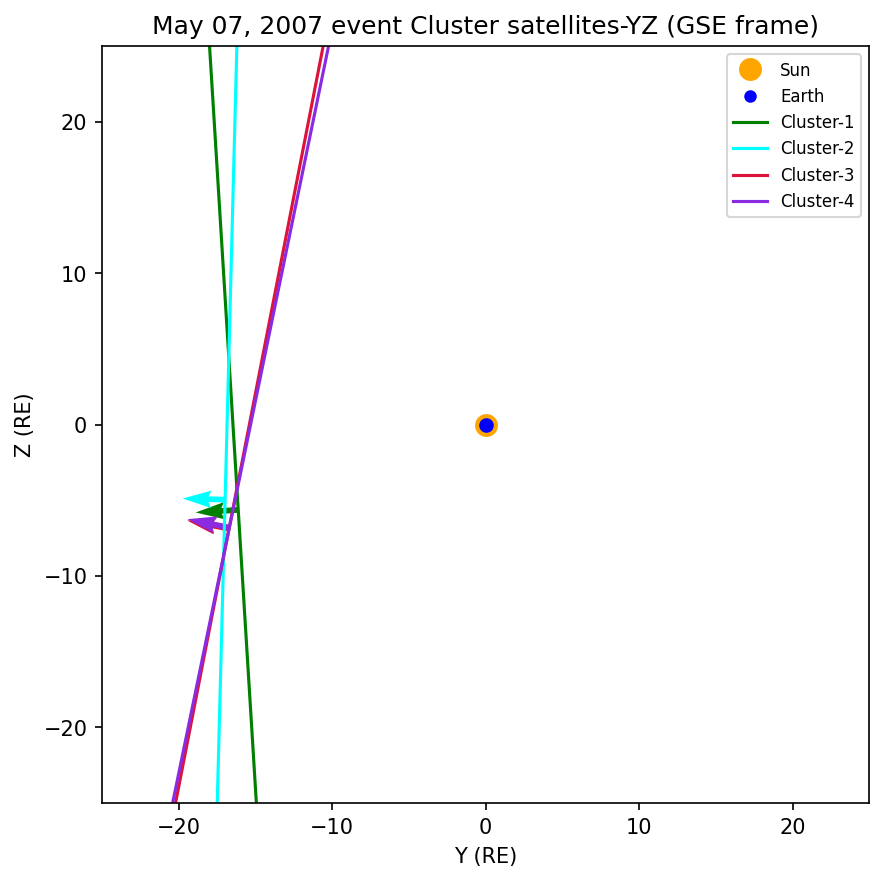}}
\caption{2D sketch of the propagation of the IP shock through 4 Cluster satellites – SC1,
 SC2, SC3, and SC4. The orange arrow represents the Sun-Earth line direction. The arrows on spacecraft positions indicate the normal vector direction and the lines perpendicular to the normal vectors indicate the shock surface orientations.
 The sizes of the lines are arbitrary.}
\label{fig:clustertogether}
\end{figure}

\begin{table}[H]
   \ContinuedFloat
   \centering
   \begin{tikzpicture}[scale=0.8, transform shape]
      \clip (\Coefficientsecond\wd\arfirst,-\dp\arfirst) rectangle (\wd\arfirst,\ht\arfirst);
      \pgftext[left,base]{\usebox{\arfirst}};
   \end{tikzpicture}
   \caption{(Continued) Resulting additional parameters of studying the data of the three spacecraft. Plasma $\beta_{up}$, Alfvén Mach, and Magnetosonic Mach are shown}
\end{table}
Using the results, the 2D sketches of the IP shocks that were detected by the spacecraft are shown in Figure~\ref{fig:xy}, \ref{fig:xz}, \ref{fig:yz}. In these 2D sketches, the shock propagation and normal vector orientation are shown in a temporal development manner. Since 4 Cluster satellites are relatively close to one another, their averaged position as well as normal vectors are shown in the general 2D and 3D sketches. For explicitly showing normal vector directions and positions of all 4 Clusters satellites, it is suitable to change their coordinates and normal vectors into a GSE coordinate system with the positions in the earth radii (RE) unit, see Figure~\ref{fig:clustertogether}.

\subsection{Discussion of the Event May 07, 2007}
\begin{figure}[H]
\centering
\includegraphics[width=1.0\textwidth]{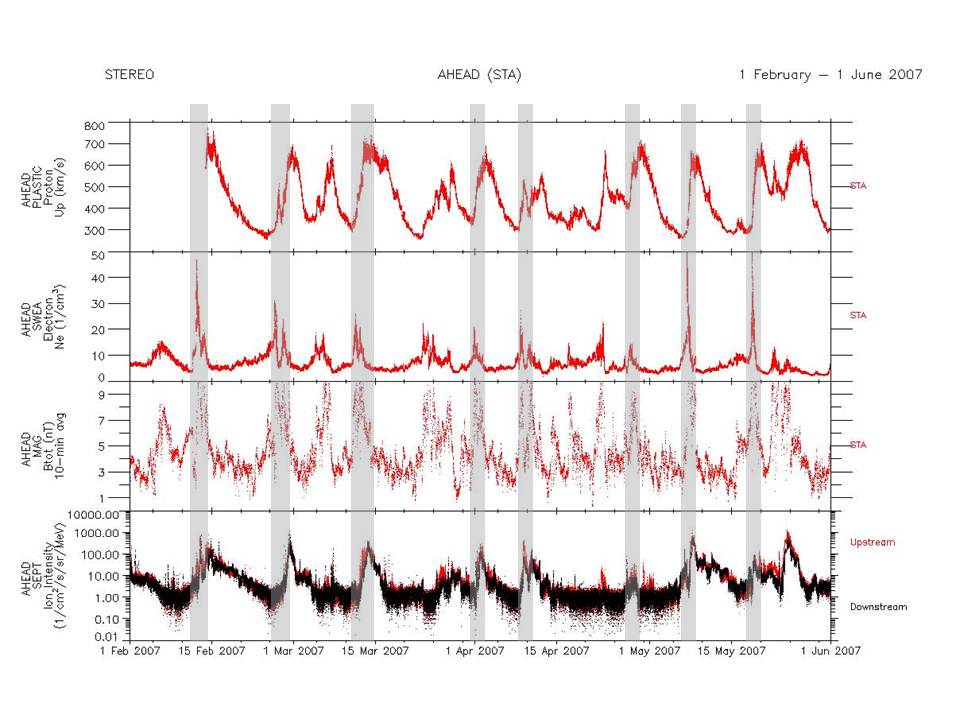}
\caption{The Figure~presents data from STEREO$-$A between February 1, 2007, and June 1, 2007. It illustrates the PLASTIC bulk velocity, the SWEA proxy for electron density, the overall magnetic field strength as measured by MAG, and the SEPT ion intensity ranging from 110 keV to 2200 keV. Periods featuring well-established Co-rotating Interaction Regions (CIRs) with noticeable boosts in ion energy are highlighted with grey shading. \protect\citep[Figure~is from][Figure~2]{opitz2014solar}}
\label{fig:cirpaper}
\end{figure}

\begin{figure}[htp]\centering
\subfloat[]{\label{a}\includegraphics[width=.5\linewidth]{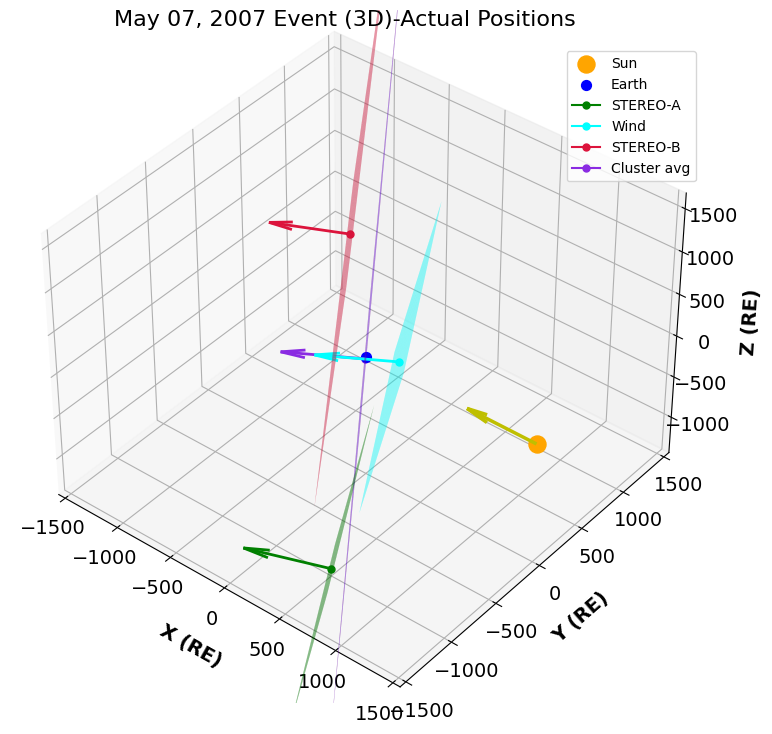}}\hfill
\subfloat[]{\label{b}\includegraphics[width=.5\linewidth]{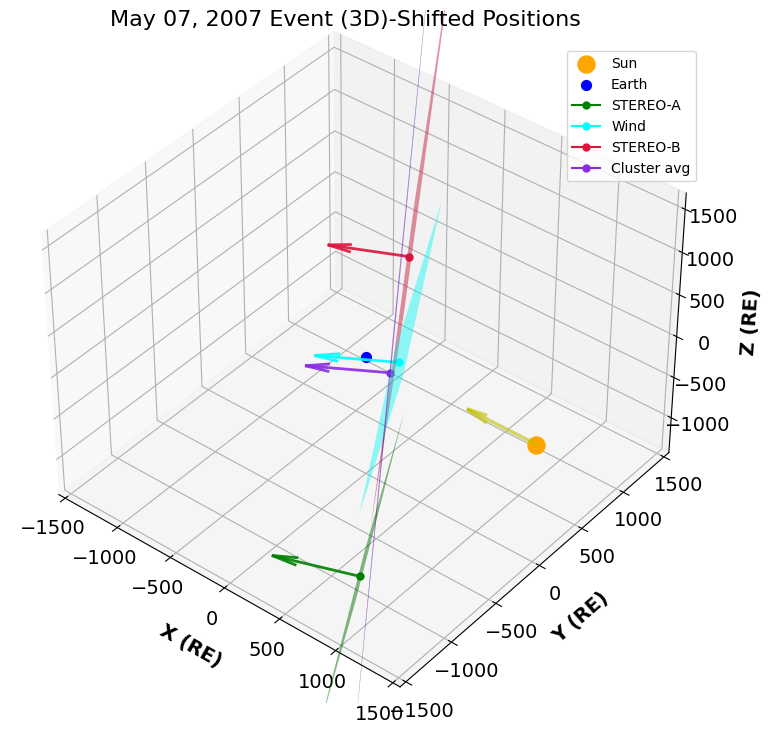}}\par 
\subfloat[]{\label{fig:fittingof0507}\includegraphics[width=.5\linewidth]{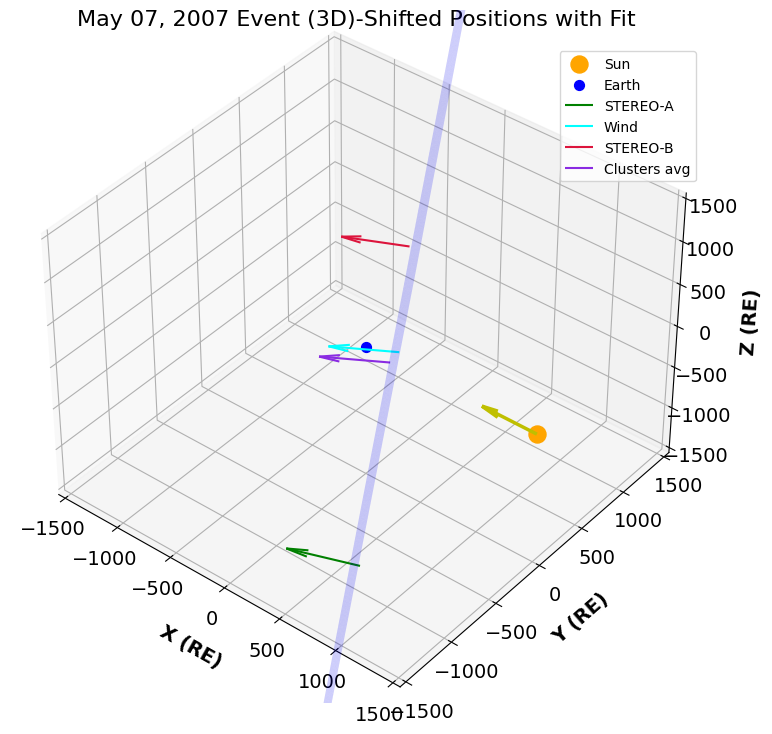}}
\caption{A 3D sketch of the propagation of the IP shock through spacecraft-- Wind, STEREO$-$A, and STEREO$-$B, and average position of 4 Cluster satellites. (a) Actual shock detected positions. (b) The positions of STEREO$-$A, STEREO$-$B, and the average positions of 4 Clusters are shifted back in time to the shock detection time of the Wind spacecraft. (c) The time-shifted positions are fitted with a plane. The arrows indicate the normal vector direction and the planes perpendicular to the normal vectors indicate the shock surface orientations. The sizes of the planes are arbitrary.}
\label{fig:05073D}
\end{figure}

\begin{figure}[H]
\centering
\includegraphics[width=0.6\textwidth]{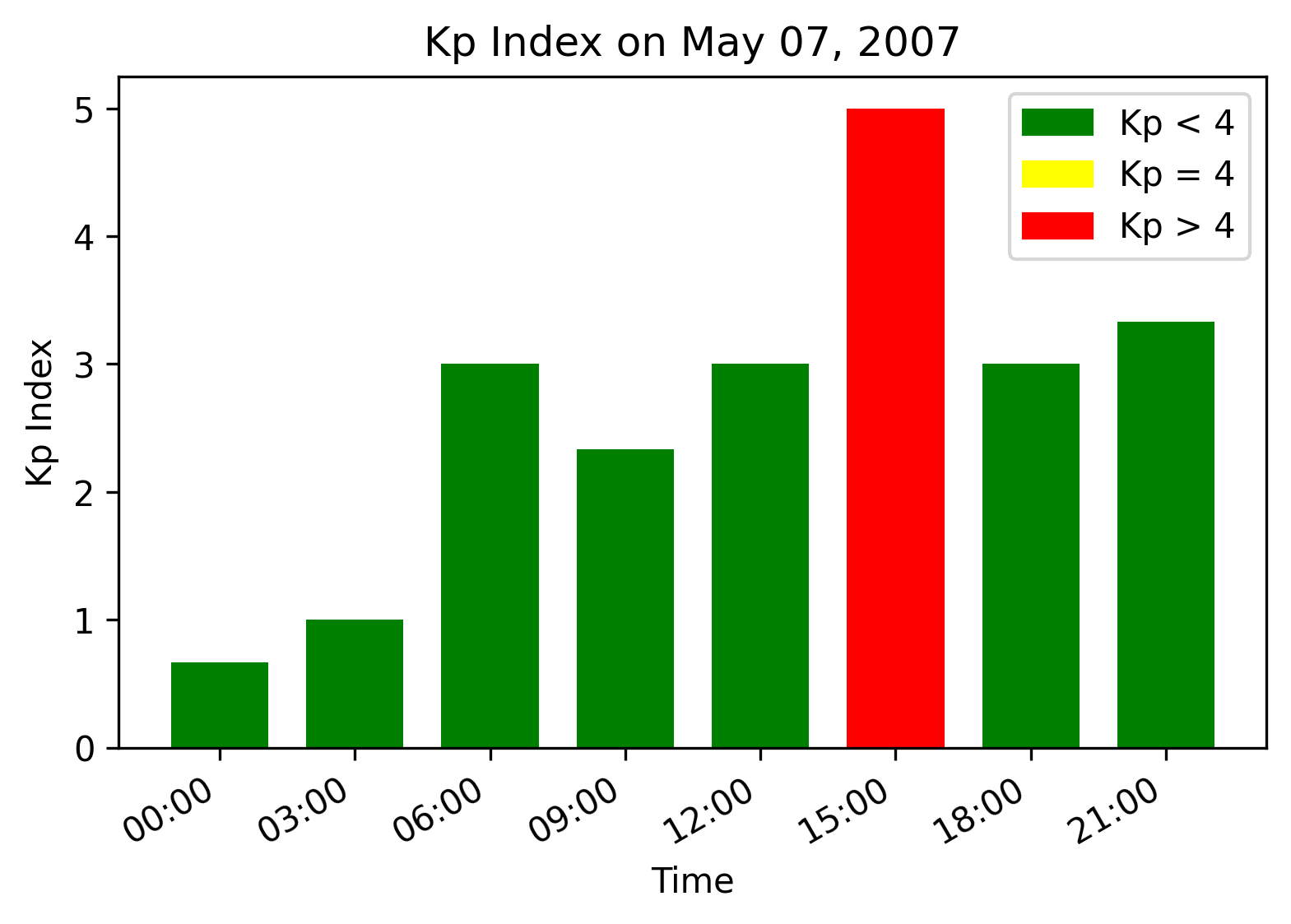}
\caption{Kp-index on May 07, 2007. The green bar indicates moderate geomagnetic activity, while the yellow bar denotes intensifying geomagnetic activity with the red bar being a geomagnetic storm. Here, the red bar indicates a G1-minor geomagnetic storm.}
\label{fig:kp0507}
\end{figure}

The 3D sketch is shown in \ref{fig:05073D}, and in the 3D sketch, the STEREO$-$A, and B, and averaged Clusters positions are time-shifted to the Wind's position to see the overall shape of this IP shock. From the 3D sketch, the overall shape of the IP shock is a planner and can be fitted with a plane. The IP shock fitted plane is tilted $56.42^{\circ}$ with respect to the Sun-Earth line according to Figure~\ref{fig:fittingof0507}. The tilt appears to be almost the same degree as the Parker spiral impacting the Earth from the dawn side. Hence, it is the reason why the Wind detected the shock first even though STEREO$-$A's position is relatively closer to the Sun's direction. This period, 2007, was during the solar minimum phase, and stream interaction region (SIR) or co-rotating stream interactions (CIR) were dominant \citep{opitz2014solar}, and this further proves the result, see Figure~\ref{fig:cirpaper}. Furthermore, to see a correlation between fast-forward shock and geomagnetic activity, I determined the Kp-index of May 07, 2007, as shown in Figure~\ref{fig:kp0507}. It appears that this forward shock-leading event disturbed the magnetosphere, causing a G1-minor geomagnetic storm with the Kp index peaking around 15:00 UTC. The geomagnetic substorm happened about 6 hours after the detection of the shock by the STEREO$-$B spacecraft.    
\subsection{The Event April 23, 2007}
\begin{figure}[H]
\centering
\includegraphics[width=0.8\textwidth]{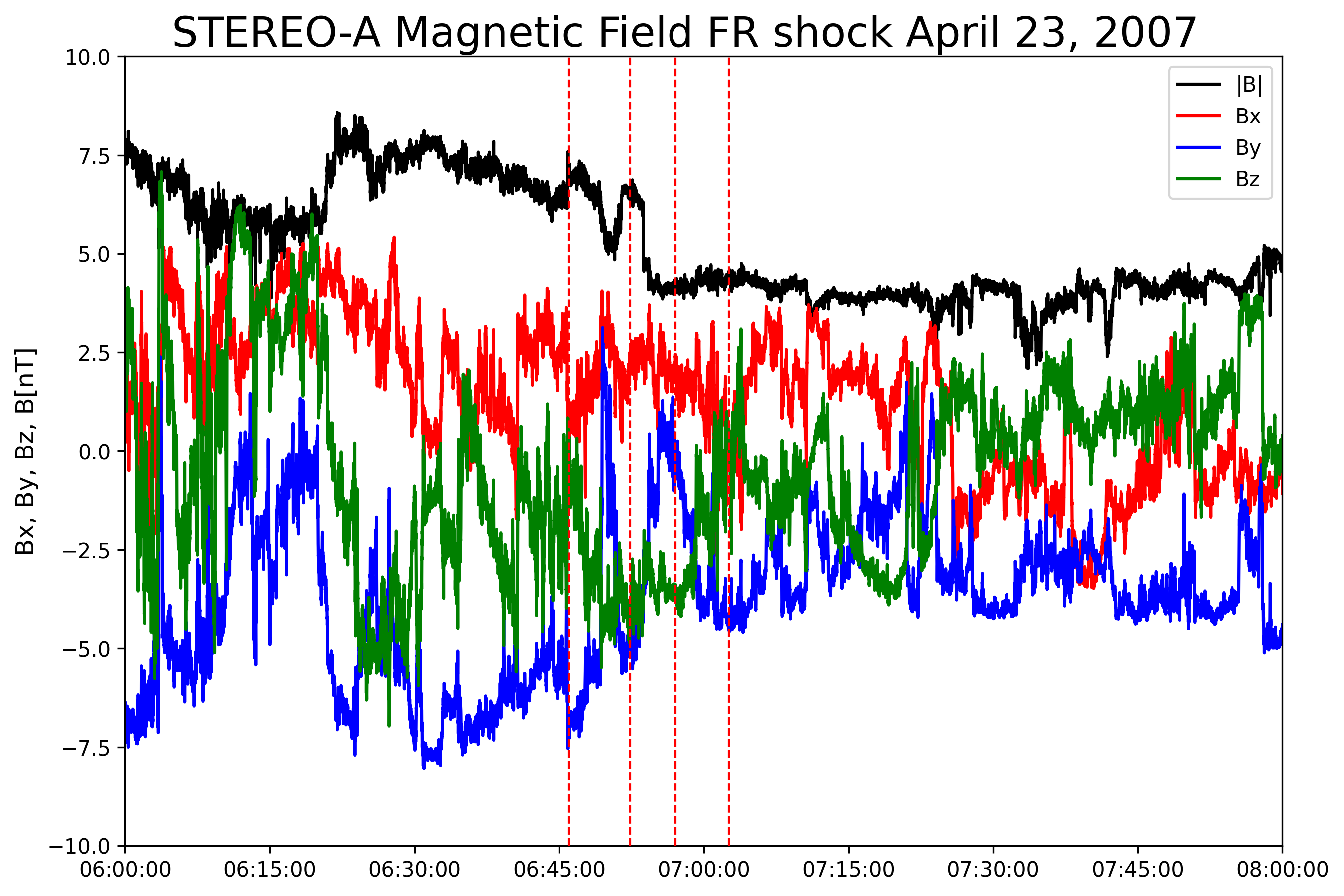}
\caption{The STEREO$-$A magnetic field measurements. The downstream $\Delta t_{down}$ is between \textbf{(06:46:00 - 06:52:23)}, and the upstream $\Delta t_{up}$ is between \textbf{(06:57:03 - 07:02:34)}. FR stands for the fast reverse shock, which means the shock is moving toward its driver. The chosen intervals of the upstream and downstream magnetic field. The red dashed lines each represent the downstream starting time and ending time and the upstream starting time and ending time respectively}
\label{fig:staB0423}
\end{figure}

\begin{figure}[H]
\centering
\includegraphics[width=0.8\textwidth]{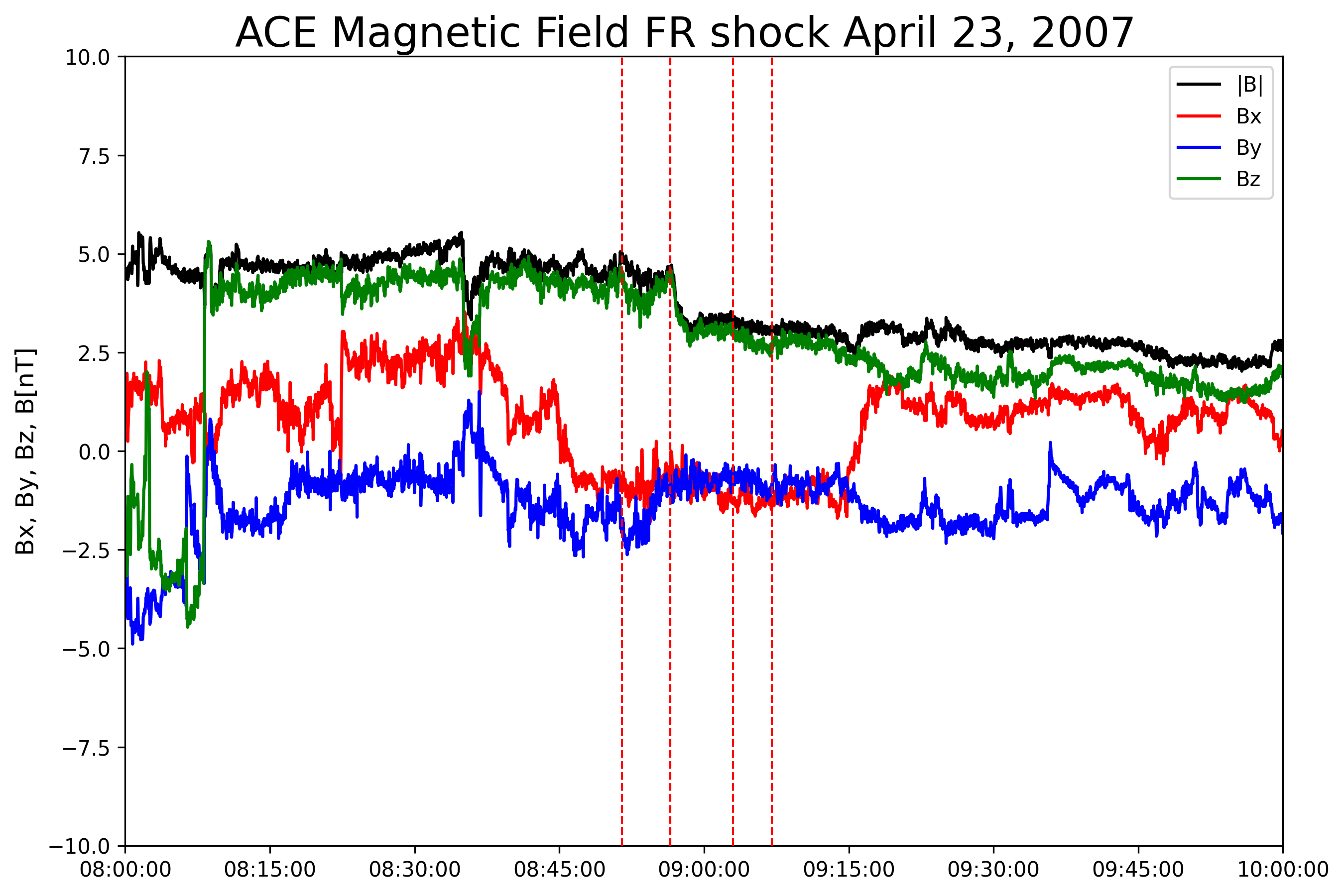}
\caption{ACE magnetic field measurements. The downstream $\Delta t_{down}$ is between \textbf{(08:51:30 - 08:56:30)}, and the upstream $\Delta t_{up}$ is between \textbf{(09:03:00 - 09:07:00)}. The symbols and details of the figures are the same as \ref{fig:staB0423}}
\label{fig:aceB0423}
\end{figure}

\begin{figure}[H]
\centering
\includegraphics[width=0.8\textwidth]{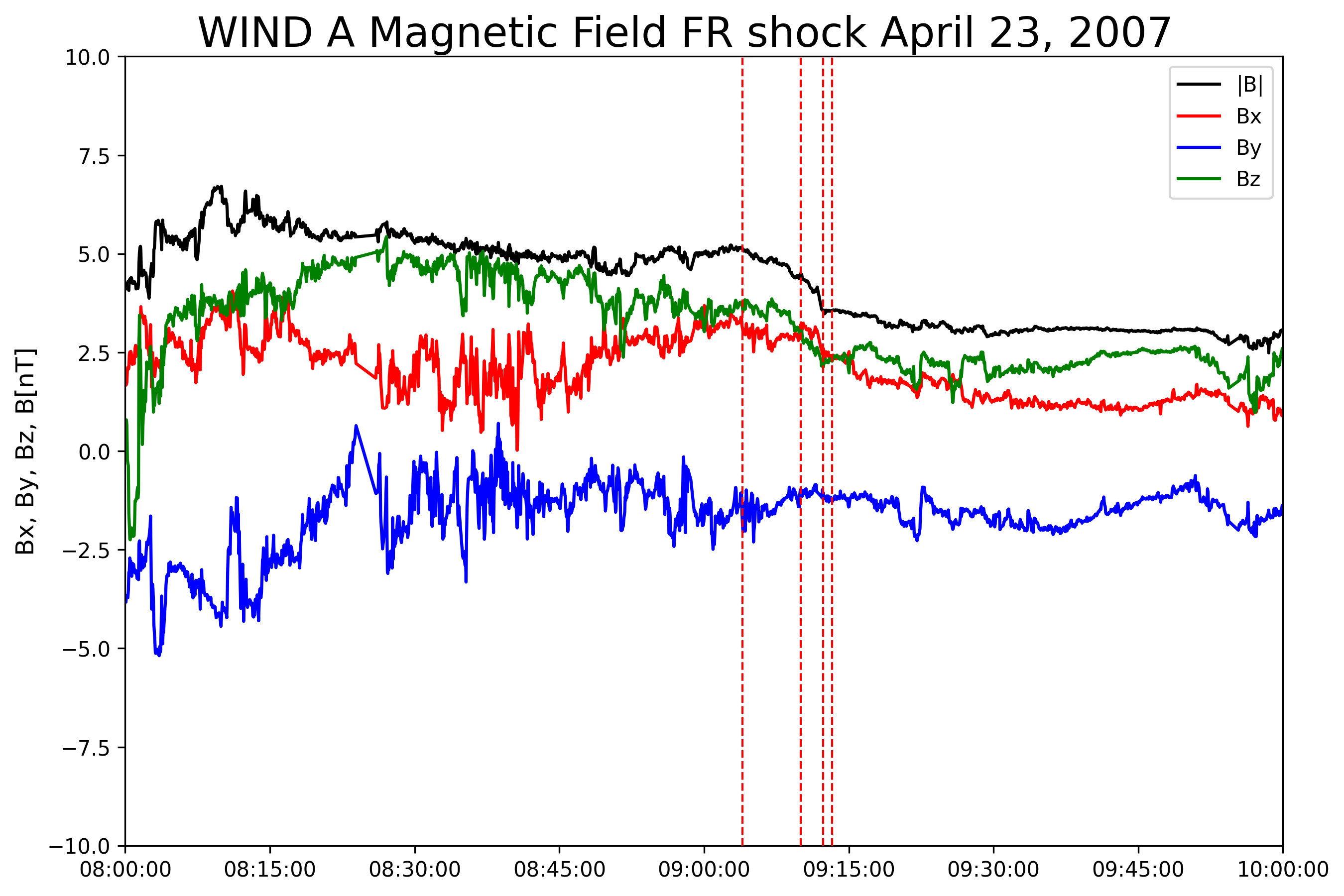}
\caption{The Wind magnetic field measurements. The downstream $\Delta t_{down}$ is between \textbf{(09:04:00 - 09:10:00)}, and the upstream $\Delta t_{up}$ is between \textbf{(09:12:20 - 09:13:15)}. The symbols and details of the figures are the same as \ref{fig:staB0423}}
\label{fig:windB0423}
\end{figure}

\begin{figure}[H]
\centering
\includegraphics[width=0.8\textwidth]{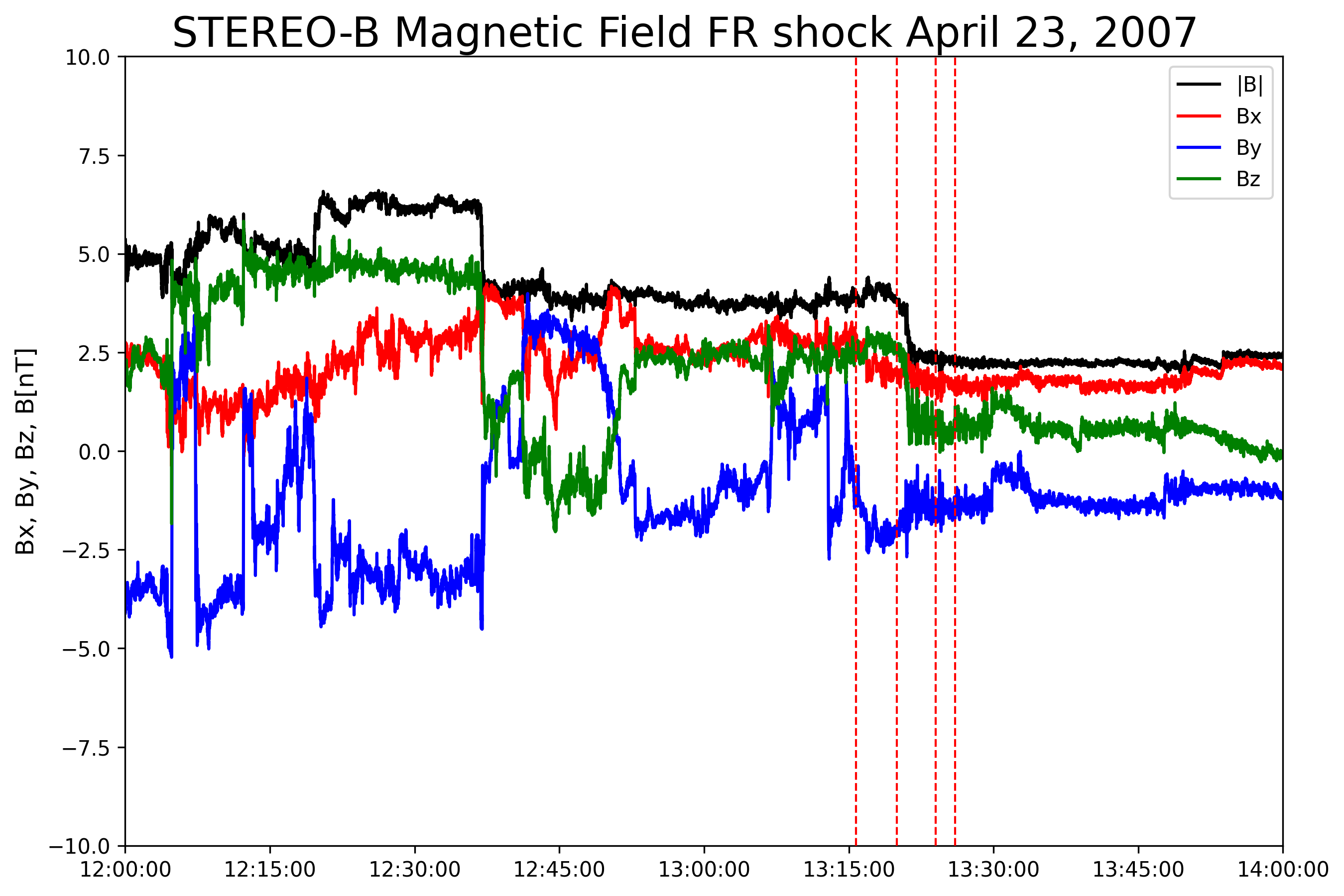}
\caption{STEREO$-$B magnetic field measurements. The downstream $\Delta t_{down}$ is between \textbf{(13:15:45 - 13:20:00)}, and the upstream $\Delta t_{up}$ is between \textbf{(13:24:00 - 13:26:00)}. The symbols and details of the figures are the same as \ref{fig:staB0423}}
\label{fig:stbB0423}
\end{figure}

Here I list all magnetic field observations of the spacecraft for the event and apply the minimum variance analysis (MVA) and the magnetic coplanarity (CP) methods to them. 
In this event, the order between upstream and downstream is swapped because it is a fast-reverse (FR) shock event, which means the shock is up against its driver.
 The upstream and downstream time intervals that most agree between the minimum variance analysis (MVA) and the magnetic coplanarity methods (CP) for all the spacecraft are shown in Table \ref{tab:timeintervalsapril} and their corresponding Figures are \ref{fig:staB0423} for the STEREO$-$A, \ref{fig:aceB0423} for the ACE, \ref{fig:windB0423} for the Wind, and \ref{fig:stbB0423} for the STEREO$-$B. Table \ref{tab:timeintervalsapril} also shows the ratio between the intermediate eigenvalue $\lambda_2$ and the smallest eigenvalue $\lambda_3$ as well as the angle between the MVA normal and CP normals in the determined upstream and downstream time intervals for each spacecraft. 
\begin{table}[H]
\centering
\caption*{Implementation of the methods on the event April 23, 2007}
\label{tab:sample2}
\begin{tabular}{ccccc}
\toprule
\textbf{Spacecraft} & \textbf{$\Delta t_{down}$} & \textbf{$\Delta t_{up}$} & \boldmath$\lambda_2/\lambda_3$ & \boldmath$\Delta\theta_{MVA-CP}$ \\
\midrule
STEREO$-$A & (06:46:00 - 06:52:23) & (06:57:03 - 07:02:34) & 3.29 & $1.07^{\circ}$\\ 
ACE & (08:51:30 - 08:56:30)& (09:03:00 - 09:07:00)& 4.22 & $3.435^{\circ}$\\ 
Wind & (09:04:00 - 09:10:00)& (09:12:20 - 09:13:15)& 3.38 & $1.46^{\circ}$\\ 
STEREO$-$B & (13:15:45 - 13:20:00)& (13:24:00 - 13:26:00)& 8.18 & $1.67^{\circ}$\\ 
\bottomrule
\end{tabular}
\caption{Here, \textbf{$\Delta t_{down}$} denotes the defined downstream and \textbf{$\Delta t_{up}$} denotes the defined upstream time intervals with MVA and CP analysis methods. \boldmath$\lambda_3/\lambda_2$ indicates the ratio between the intermediate eigenvalue $\lambda_2$ and the smallest eigenvalue $\lambda_2$, and $\Delta\theta_{MVA-CP}$ is the angle between the MVA and CP normals.}
\label{tab:timeintervalsapril}
\end{table}

\subsection{Analysis of the Event April 23, 2007}
Similarly to the event about, the upstream and downstream time intervals are highly accurate as well, considering the angle difference between the two methods is minimal and the eigenvalue criteria are fulfilled for each spacecraft data.

Also, using the determined time intervals, the estimations of the ratio between the upstream and downstream magnetic field, densities, and temperatures as well as the bulk speed, and shock $\theta_{Bn}$ angle are made. These parameters, the minimum variance analysis normal and the magnetic coplanarity normal are shown in Table \ref{tab:table2}, where also the solar wind bulk speed and the downstream to upstream ratios for the shock criteria are fulfilled accordingly with \ref{eq:criteria1} and \ref{eq:criteria2}, and the STEREO$-$B ratios are compatible with the shock criteria, proving that the STEREO$-$B did detect the shock. However, the shock detected by the STEREO$-$B appears to become quasi-parallel a few hours after the shock detection times of STEREO$-$A, ACE, and the Wind based on the shock $\theta_{Bn}$ angle, see Table \ref{tab:tableAdd2}, where the calculated results of additional parameters are shown. 

\newsavebox{\coresecond}

\begin{lrbox}{\coresecond}
\Large
\begin{tabular}{|c|c|c|c|c|c|c|c|c|c|}
\hline
Spacecraft & $B_d/B_u$  & $N_d/N_u$ & $T_d/T_u$ & $\Delta V $ & $\theta_{Bn}$ & Spacecraft & MVA & Coplanarity & $\Delta\theta_{MVA-CP}$\\
 \hline
STEREO$-$A   & $1.470\pm 0.16$     & $2.000\pm 0.70$ & $1.480\pm 0.50$ & $46.00\pm 13$ & 88.40 & STEREO$-$A & [-0.86,  0.020, -0.502] & [-0.858,  0.007, -0.51] & 1.070 \\
 \hline
ACE   & $1.410\pm 0.08$  & $1.440\pm 0.16$ & $1.220\pm 0.09$ & $23.720\pm 5.57$ & 46.2 & ACE & [0.872,  -0.189, -0.450] &   [0.869, -0.232, -0.434] & 2.653  \\
\hline
Wind    & $1.349\pm 0.06$ & $1.424\pm 0.40$ & $1.285\pm 0.7$ & $19.26\pm 0.13$ &61.85 & Wind & [-0.823,  0.430,  0.369] & [-0.809,  0.438,  0.390] & 1.349\\
 \hline
STEREO$-$B   & $1.730\pm 0.09$  & $1.980\pm 0.14$ & $1.340\pm 0.15$ & $21.20\pm 3.4$ & 29.39 & STEREO$-$B & [-0.751,  0.591, 0.292] & [-0.731,  0.610, -0.303] & 1.67 \\
 \hline 
\end{tabular}
\end{lrbox}

\scalebox{0}{\usebox{\coresecond}}

\newcommand{\Coefficientthree}{0.5187}

\begin{table}[H]
   \centering
   \begin{tikzpicture}[scale=0.8, transform shape]
      \clip (0,-\dp\coresecond) rectangle (\Coefficientthree\wd\coresecond,\ht\coresecond);
      \pgftext[left,base]{\usebox{\coresecond}};
   \end{tikzpicture}
   \caption{\label{tab:table2} Resulting core parameters of studying the data of the STEREO$-$A, STEREO$-$B, ACE, and Wind spacecraft. Here $\theta_{Bn}$ is the shock $\theta$ angle.}
\end{table}

\begin{table}[H]
   \ContinuedFloat
   \centering
   \begin{tikzpicture}[scale=0.8, transform shape]
      \clip (\Coefficientthree\wd\coresecond,-\dp\coresecond) rectangle (\wd\coresecond,\ht\coresecond);
      \pgftext[left,base]{\usebox{\coresecond}};
   \end{tikzpicture}
   \caption{(Continued) Resulting core parameters of studying the data of the STEREO$-$A, STEREO$-$B, ACE, and Wind spacecrafts. $\theta_{MVA-CP}$ is the angle difference between the minimum variance normal vectors and the magnetic coplanarity normal vectors.}
\end{table}

\newsavebox{\arsecond}

\begin{lrbox}{\arsecond}
\Large
\begin{tabular}{|c|c|c|c|c|c|c|c|c|}
\hline
Spacecraft & $V_{sh}$ km/s & $C^{up}_s$ km/s  & $V^{up}_A$ km/s & $C^{up}_{ms}$ km/s & Spacecraft & Plasma $\beta_{up}$ & Alfvén Mach & Magnetosonic Mach \\
 \hline
STEREO$-$A  & 467.95 & $64.97\pm 7.74$     & $68.715\pm 29.2$      & $94.57\pm 21.85$      & STEREO$-$A & $1.07\pm 0.94$      & $0.998\pm 0.67$ & $0.725\pm 0.41$ \\
 \hline
ACE & 412.36  & $58.28\pm 4.94$     & $52.82\pm 15.90$      & $78.66\pm 11.26$     & ACE & $1.46\pm 0.91$   & $1.49\pm 0.56$ & $1.0\pm 0.27$ \\
 \hline
Wind & 372.59  & $56.14\pm 5.36$     & $56.22\pm 14.71$      & $79.45\pm 11.08$     & Wind & $1.19\pm 0.66$    & $1.24\pm 0.52$ & $0.88\pm 0.31$ \\
 \hline
STEREO$-$B  & 370.31 & $57.31\pm 5.38$     & $29.045\pm 12.28$      & $64.25\pm 07.34$      & STEREO$-$B & $4.67\pm 4.04$      & $1.688\pm 1.08$ & $0.763\pm 0.37$ \\
 \hline
\end{tabular}
\end{lrbox}

\scalebox{0}{\usebox{\arsecond}}

\newcommand{\Coefficientfour}{0.5279}

\begin{table}[H]
   \centering
   \begin{tikzpicture}[scale=0.8, transform shape]
      \clip (0,-\dp\arsecond) rectangle (\Coefficientfour\wd\arsecond,\ht\arsecond);
      \pgftext[left,base]{\usebox{\arsecond}};
   \end{tikzpicture}
   \caption{\label{tab:tableAdd2} Resulting additional parameters of studying the data of the three spacecraft. $V_{sh}$ is the shock speed, $C^{up}_s$ is the upstream sound speed, $V^{up}_A$ is the upstream Alfvén speed and $C^{up}_{ms}$ is the upstream magnetosonic speed.}
\end{table}

\begin{figure}[H]\centering
\subfloat[XY coordinates sketch of the propagation of the IP shock through spacecraft]{\label{fig:xy0423}\includegraphics[width=.5\linewidth]{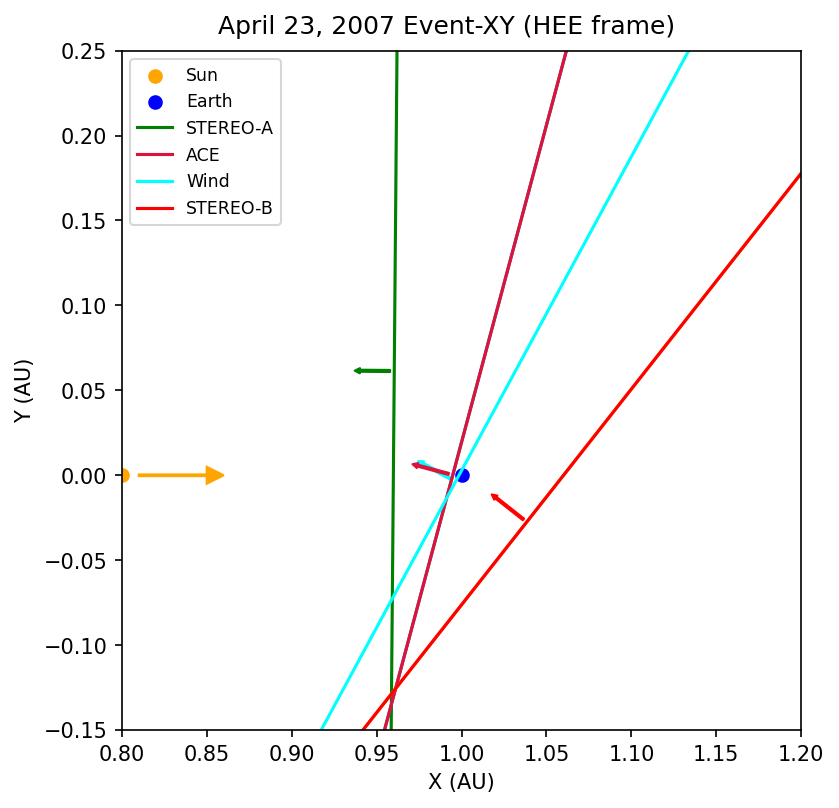}}\hfill
\subfloat[XZ coordinates sketch of the propagation of the IP shock through spacecraft]{\label{fig:xz0423}\includegraphics[width=.5\linewidth]{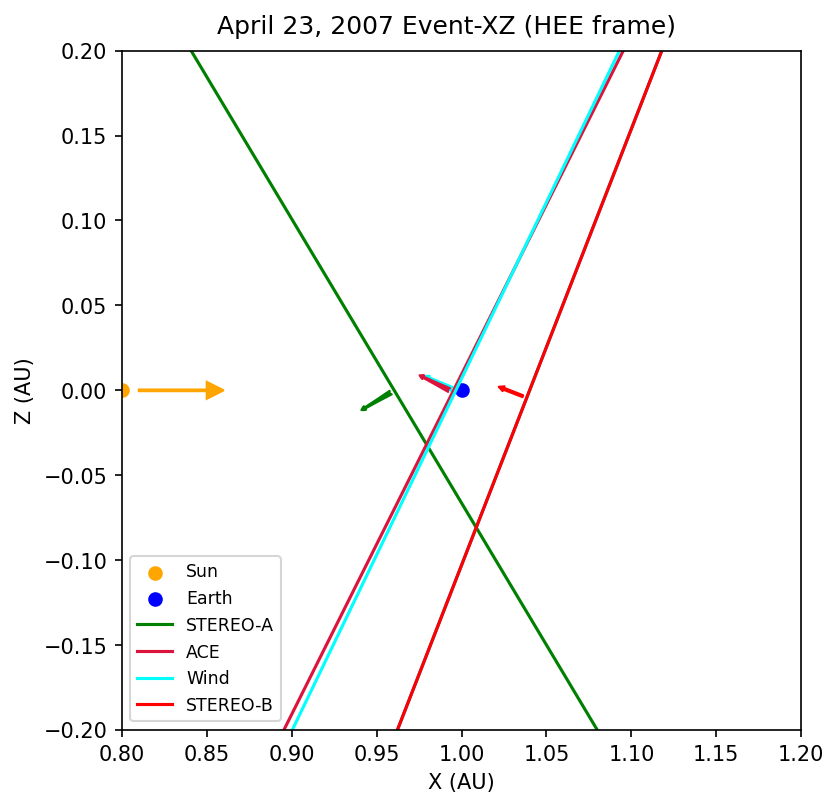}}\par 
\subfloat[YZ coordinates sketch of the propagation of the IP shock through spacecraft.]{\label{fig:yz0423}\includegraphics[width=.5\linewidth]{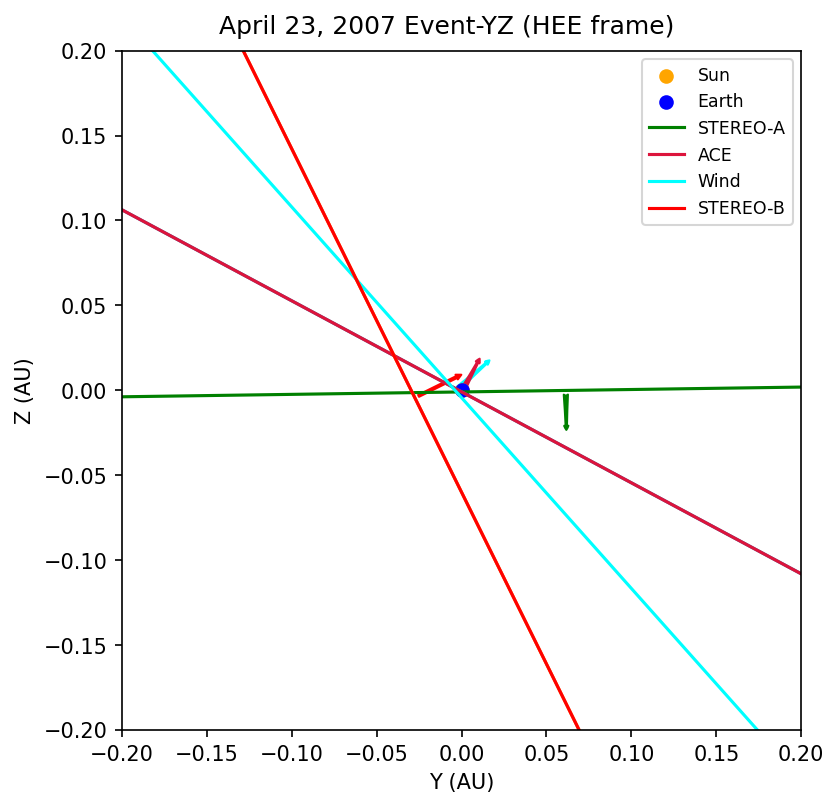}}
\caption{2D XY, XZ, and YZ sketches of the propagation of the IP shock through spacecraft-- STEREO$-$A, ACE, Wind, and STEREO$-$B. The orange arrow represents the Sun-Earth line direction. The arrows on spacecraft positions indicate the normal vector direction and the lines perpendicular to the normal vectors indicate the shock surface orientations. The sizes of the lines are arbitrary.}
\label{fig:04232D}
\end{figure}

\begin{table}[H]
   \ContinuedFloat
   \centering
   \begin{tikzpicture}[scale=0.8, transform shape]
      \clip (\Coefficientfour\wd\arsecond,-\dp\arsecond) rectangle (\wd\arsecond,\ht\arsecond);
      \pgftext[left,base]{\usebox{\arsecond}};
   \end{tikzpicture}
   \caption{(Continued) Resulting additional parameters of studying the data of the three spacecraft. Plasma $\beta_{up}$, Alfvén Mach, and Magnetosonic Mach are shown}
\end{table}
Similarly, the sketches of the IP shocks that were detected by the spacecraft are shown in \ref{fig:xy0423}, \ref{fig:xz0423}, \ref{fig:yz0423}. In these 2D sketches, the shock propagation and normal vector orientation are shown in a temporal development manner. 
\subsection{Discussion of the Event April 23, 2007}
\begin{figure}[H]
\centering
\includegraphics[width=0.6\textwidth]{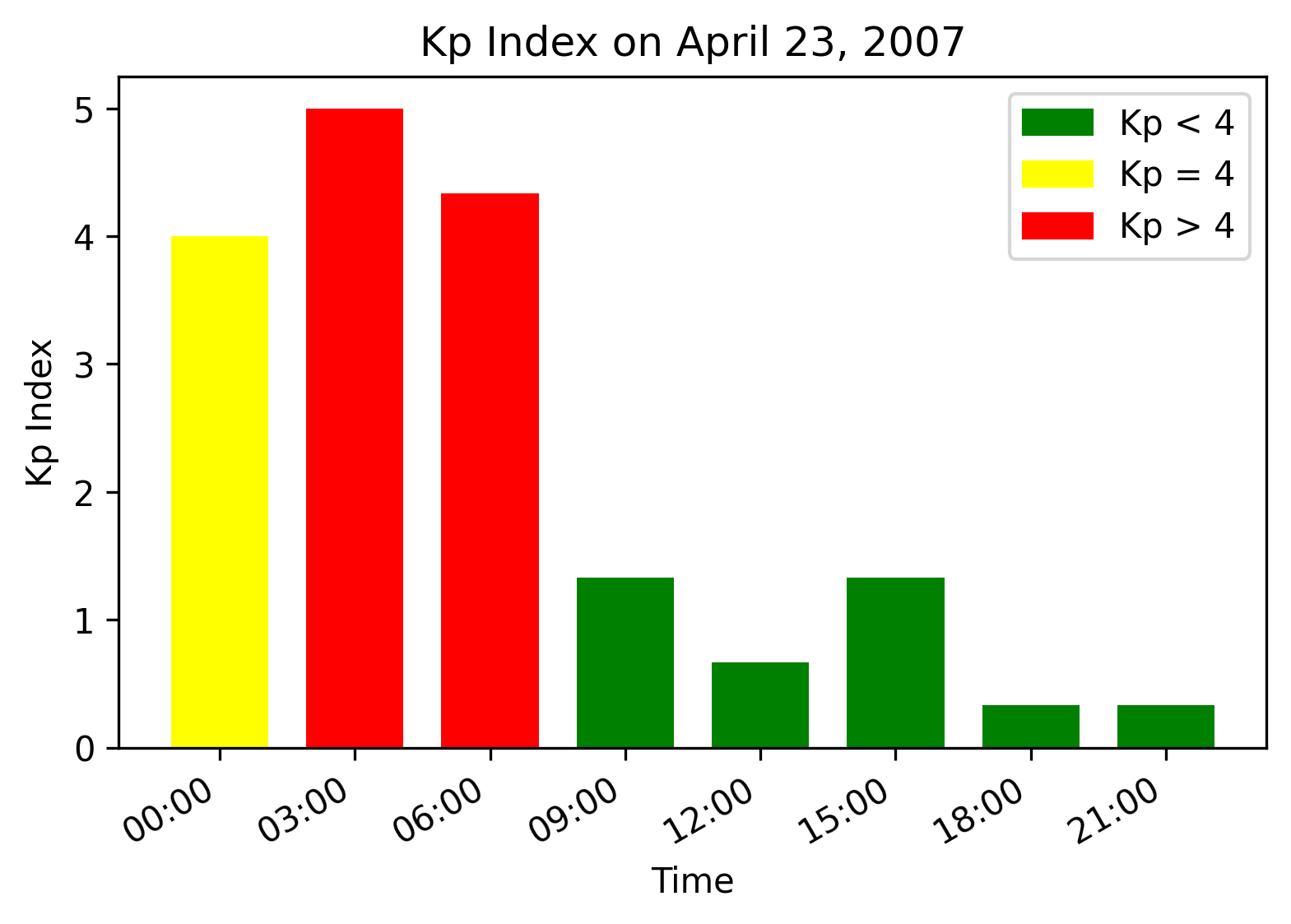}
\caption{The Kp-index on April 23, 2007. The symbols and details of the Figure~is the same as \ref{fig:kp0507}}
\label{fig:kp0423}
\end{figure}

\begin{figure}[htp]\centering
\subfloat[]{\label{fig:04233Dactualposition}\includegraphics[width=.5\linewidth]{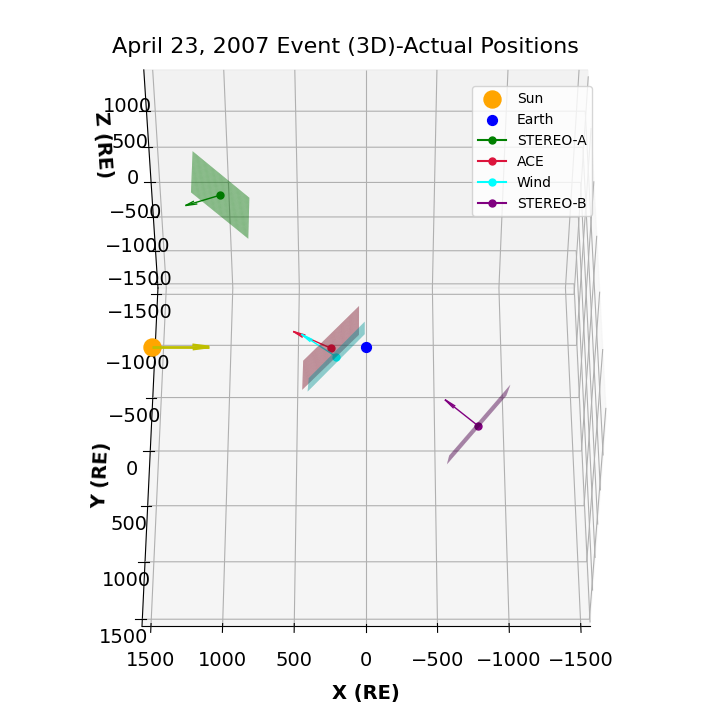}}\hfill
\subfloat[]{\label{fig:04233Dsub}\includegraphics[width=.5\linewidth]{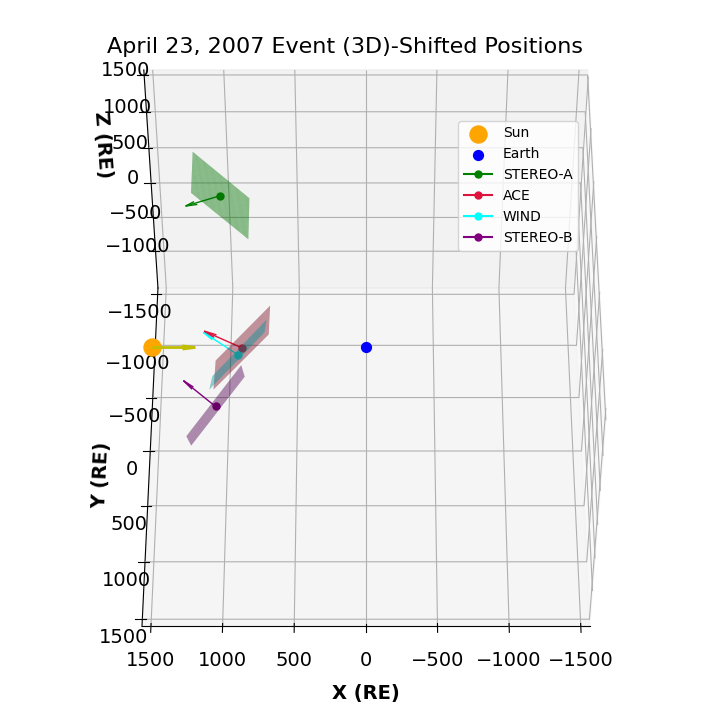}}\hfill
\subfloat[]{\label{fig:04233Danother}\includegraphics[width=.5\linewidth]{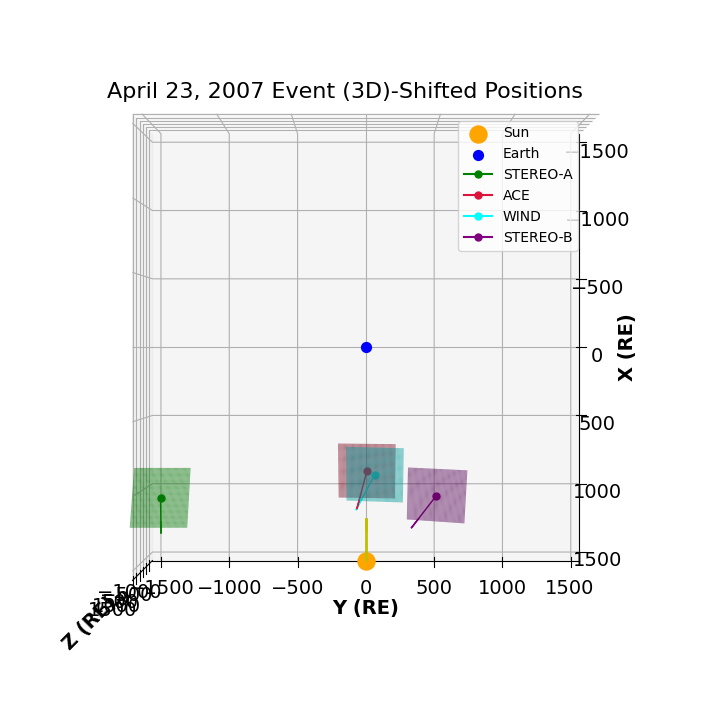}}
\caption{ (a) A 3D sketch of the propagation of the IP shock through 3 spacecraft-- STEREO$-$A, Wind, ACE, and STEREO$-$B. (b) The positions of Wind, ACE, and STEREO$-$B are shifted back in time to the shock detection time of STEREO$-$A. (C) The shifted positions are shown from the top view. The arrows indicate the normal vector direction and the planes perpendicular to the normal vectors indicate the shock surface orientations. The sizes of the planes are arbitrary.}
\label{fig:04233D}
\end{figure}

The 3D sketch is shown in \ref{fig:04233D}, and in the 3D sketch, the Wind, ACE, and STEREO$-$B positions are time-shifted to the STEREO$-$A's position to see the overall shape of this IP shock. It appears the shape of the shock is not uniform and is twisted from the STEREO$-$A spacecraft to the other three spacecraft in the Z-axis along the transverse direction (Y-axis) to the Sun-Earth line, see Figure~\ref{fig:xz0423}. As seen from Figure~\ref{fig:04233Danother}, the STEREO$-$A alone is on the dusk (left) side of the Sun-Earth line while the Wind, the ACE, and the STEREO$-$B are on the dawn (right). In this XY-plane view, the shock normal vectors appear to be changing or slightly rotating their direction from the STEREO$-$A to the Wind, the ACE, and the STEREO$-$B along Y-axis, and the shock is changing from the quasi-perpendicular to quasi-parallel based on the shock $\theta_{Bn}$ angle. This IP shock is a fast reverse shock, meaning it travels toward the Sun even though the shock propagates away from the Sun with the solar wind. The shock detection time between the STEREO$-$A and the Wind/ACE is about two hours while the between STEREO$-$A and B is almost six hours, yet the shock orientation is significantly changed from the STEREO$-$A to the other three spacecraft indicating the change is spatial, not temporal. So, due to this nature, the IP shock could be a local ripple. The ripples on the shock surface are known for being caused by ICME (interplanetary coronal mass ejections) shock drivers as they do not propagate into homogeneous interplanetary medium \citep{acuna2008stereo}. However, the source of this event is a stream-interaction region (SIR) as listed here \url{https://stereo-ssc.nascom.nasa.gov/pub/ins_data/impact/level3/STEREO_Level3_Shock.pdf}. 
I determined the Kp-index of this event as seen in Figure~\ref{fig:kp0423}. Similar to the fast-forward shock event of May 07, 2007, a G1-minor geomagnetic storm occurred on this day, but the beginning of the geomagnetic storm happened three hours before the shock detection time of the STEREO$-$A while the ending of the storm was partially at the same time as the STEREO$-$A. As stated before, SIR/CIRs form when the fast-moving solar wind catches the slow-moving solar wind \citep{ richardson2018solar}. This sometimes forms a pair of shocks, one leading as a fast-forward shock while a rarefied shock trails the solar wind as a fast-reverse shock but oftentimes as just sole fast-forward or fast reverse shocks \citep{jian2019solar}. Nevertheless, since it always involves the fast-moving solar wind, it makes total sense why the G1-minor geomagnetic storm with a Kp-5 index happened before the detection of the shock because it seems the fast-moving solar wind caused the minor storm before the fast-reverse shock finally arrived at the spacecraft positions.
 The normal orientation inconsistency can also be explained by the argument that this shock is a fast-reverse shock because SIRs have characteristic tilts such that forward waves direct towards the solar equatorial plane while the reverse waves tend to move in the direction of the solar poles \citep{kilpua2015properties}. Hence, this tilting nature may explain the XZ component tilted orientation between the STEREO$-$A and the other three spacecraft -- the Wind, the ACE, and the STEREO$-$B. 
\chapter{Summary and conclusions}
In this thesis, a single spacecraft study of the interplanetary (IP) shock is presented. This thesis aims to study and determine how IP shocks are developing and evolving in spatial and temporal propagation. 
For this purpose, two methods are implemented, namely the minimum variance analysis of the magnetic field (MVA) and the magnetic coplanarity methods (CP) for the determination of upstream and downstream time intervals. The two methods use magnetic field measurement data, which is known for its high resolutions compared to those of plasma data concerning their respective heliospheric variational rates. To acquire data, I used the Coordinated Data Analysis Web (CDAWeb) of NASA for each spacecraft. Data are in the first event case – from early morning to noon in 2007-05-07 for seven different spacecraft, namely Wind, Stereo A and B, and 4 Cluster satellites. 

The second event case, similarly, - from early morning to noon in 2007-04-23 for four different spacecraft, namely Wind, STEREO$-$A and B, and ACE.
 After acquiring data, I changed every coordinate system of the data to the HEE coordinate system so that it is easy to compare the data and how the shock parameters change when propagating through space. However, near Earth, vectors are changed to the GSE coordinate system. For accepting the upstream and downstream time intervals, two criteria are made such as the $\theta$ angle differences between MVA and the magnetic coplanarity must be considerably small, $<15^{\circ}$, and the ratio between the intermediate eigenvalue to the smallest eigenvalue must be greater than 3. For both case studies, I found such upstream and downstream intervals, indicating the results are highly accurate. Hence, pinpointing the shock formations has been very successful due to using the two methods to determine the shock normal vectors.

The two event studies are not timely ordered. First, the Event May 07, 2007, is studied, and then the Event April 23, 2007, is followed. Hence the Event May 07 is referred to as the first event and the Event April 23 is the second event. The first event is a fast-forward (FF) shock event, the shock propagating away from the Sun in the solar wind frame of reference in addition to the solar wind on which it propagates is also traveling away from the Sun. The shock parameters agree within the errors. From the 2D sketches as well as the 3D sketch, on which the positions of the STEREO$-$A and B, and 4 Clusters spacecraft are time-shifted to the shock detection time of the Wind spacecraft to see the geometry of the shock, the shock appears planar and tilted $56.42^{\circ}$ to the Sun-Earth line, explaining why the Wind first detected the IP shock even though its location is behind the STEREO$-$A in respect to the Sun-Earth line. The tilt of this planar shock surface is almost identical to the usual propagation of the Parker spiral impacting the Earth. Consequently, this indicates the origin of the shock is co-rotating stream interactions (CIRs), which agrees with the detected CIR on that day.
 There is no sign of temporal change in this scale. In the spatial range of 40 million km, no change was observable.
 The G1-minor geomagnetic storm happened after this shock was detected, indicating this shock event caused the geomagnetic storm on that day.
 
 The second event is the fast reverse (FR) shock event, the shock propagates toward the Sun in the solar wind frame of reference although the shock propagates away from the Sun with its carrier solar wind. There are several peculiarities with this shock event, it appears the shape of the shock is not uniform and twisted from the STEREO$-$A spacecraft to the other spacecraft along the transverse direction to the Sun-Earth line. The shock normal vectors also were observed to be changing along Y-axis. 
A G1-minor geomagnetic storm occurred on this day, similar to the fast-forward shock event of May 07, 2007. Intriguingly, the onset of this storm took place three hours prior to the shock detection time recorded by STEREO$-$A. The termination of the storm partially coincided with the STEREO$-$A detection time.
 The formation of Stream Interaction Regions/Corotating Interaction Regions (SIR/CIRs) occurs when a faster solar wind stream encounters a slower one. The source of this shock is SIR/CIR. 
Given the involvement of fast-moving solar wind, it's logical that the G1-minor geomagnetic storm, with a Kp-5 index, began before the shock detection. It appears that the fast-moving solar wind instigated the minor storm prior to the arrival of the fast-reverse shock at the spacecraft's position.
 The orientation irregularity can be potentially accounted for by the characteristic tilts of SIRs. The forward waves of SIRs tend to propagate towards the solar equatorial plane, while the reverse waves lean towards the solar poles. This inherent tilting could explain the tilted orientation of the XZ component observed between the STEREO$-$A and the other three spacecraft – Wind, ACE, and STEREO$-$B. Even without this characteristic tilting, the fast-reverse shock is being compressed, which, given its defining nature, could explain the irregularities.  
\newpage
\bibliographystyle{plainnat}
\bibliography{References}
\end{document}